\documentclass{article}%
\usepackage[a4paper,left=1.1in,right=1.1in,top=1.1in,bottom=1.1in]{geometry}
\usepackage{amssymb}
\usepackage{graphicx}
\usepackage{amsmath}
\usepackage{amsfonts}%
\providecommand{\U}[1]{\protect\rule{.1in}{.1in}}

\begin{document}

\begin{center}
\bigskip

\bigskip

${\LARGE MANY-BODY\,EFFECTS\,AND\,QUANTUM\,\ FLUCTUATIONS\bigskip}$

${\LARGE \ FOR\ DISCRETE\,\,TIME\,CRYSTALS\bigskip}$

$\,{\LARGE IN\,}\ {\LARGE BOSE-EINSTEIN\,CONDENSATES}$

\bigskip

Jia Wang$^{1}$, Peter Hannaford$^{2}$ and Bryan J. Dalton$^{1}$\bigskip

$^{1}$Centre for Quantum Science and Technology Theory, Swinburne University
of Technology, Melbourne, Victoria 3122, Australia

$^{2}$Optical Sciences Centre, Swinburne University of Technology, Melbourne,
Victoria 3122, Australia

\bigskip
\end{center}

\textbf{Abstract }We present a fully comprehensive multi-mode quantum
treatment based on the truncated Wigner approximation (TWA) to study many-body
effects and effects of quantum fluctuations on the formation of a discrete
time crystal (DTC) in a Bose-Einstein condensate (BEC) bouncing resonantly on
a periodically driven atom mirror. Zero-range contact interactions between the
bosonic atoms are assumed. Our theoretical approach avoids the restrictions
both of mean-field theory, where all bosons are assumed to remain in a single
mode, and of time-dependent Bogoliubov theory, which assumes boson depletion
from the condensate mode is small. We show that the mean-field and
time-dependent Bogoliubov approaches can be derived as approximations to the
TWA treatment. Differing initial conditions, such as a finite temperature BEC,
can also be treated. For realistic initial conditions corresponding to a
harmonic trap condensate mode function, our TWA calculations performed for
period-doubling agree broadly with recent mean-field calculations for times
out to at least $2000$ mirror oscillations, except at interaction strengths
very close to the threshold value for DTC formation where the position
probability density differs significantly from that determined from mean-field
theory. For typical attractive interaction strengths above the threshold value
for DTC formation and for the chosen trap and driving parameters, the TWA
calculations indicate a quantum depletion due to quantum many-body
fluctuations of less than about two atoms out of a total of $600$ atoms at
times corresponding to $2000$ mirror oscillations, in agreement with
time-dependent Bogoliubov theory calculations. On the other hand, for
interaction strengths very close to the threshold value for DTC formation, the
TWA calculations predict a large quantum depletion - as high as about $260$
atoms out of $600$. We also show that the mean energy per particle of the DTC
does not increase significantly for times out to at least $2000$ mirror
oscillations and typically oscillates around an average value close to its
initial value; so TWA theory predicts the absence of thermalisation. Finally,
we find that the dynamical behaviour of our system is largely independent of
whether the boson-boson interaction is attractive or repulsive, and that it is
possible to create a stable DTC based on repulsive interactions.\pagebreak

\section{Introduction}

\label{Section - Introduction}

Discrete time crystals (DTC) are periodically driven non-equilibrium states of
quantum many-body systems that spontaneously break discrete time-translation
symmetry due to particle interactions and start to evolve with a period $s$
times $(s=2,3,..)$ longer than the period $T\ $of the drive \cite{Sacha15a},
\cite{Khemani16a}, \cite{Else16a}, \cite{Yao17a}, \cite{Sacha20a}. Such time
crystals are predicted to be robust against external perturbations and to
persist indefinitely in the thermodynamic limit of large $N$, in analogy with
space crystals. A number of different platforms for creating discrete time
crystals have been proposed \cite{Sacha15a}, \cite{Khemani16a}, \cite{Else16a}%
, \cite{Yao17a}. Preliminary experimental evidence for the realization of DTCs
has been reported for a number of platforms, including a spin chain of
interacting ions \cite{Zhang17a}, nitrogen-vacancy spin impurities in diamond
\cite{Choi17a}, nuclear spins in organic molecules \cite{Pal18a} and ordered
ADP crystals \cite{Rovny18a}, \cite{Rovny18b}, and a superfluid Bose-Einstein
condensate of ultra-cold atoms \cite{Smits18a}, \cite{Liao18a},
\cite{Smits20a}. Overviews of the topic of time crystals are given in
\cite{Sacha18b}, \cite{Yao18a}, \cite{Khemani19a}, \cite{Sacha20a}.

Another platform that has been proposed for a discrete time crystal and which
is yet to be realized experimentally involves the use of a Bose Einstein
condensate (BEC) of weakly interacting bosonic atoms bouncing resonantly on a
periodically driven mirror \cite{Sacha15a}, \cite{Giergel18a},
\cite{Giergiel20a}, \cite{Kuros20a}. A many-body mean-field approach with the
condensate wave-function expanded as a linear combination of $s$ Wannier
functions (which are defined in the time domain) has been employed to study
such discrete time crystals in the case of period-doubling $(s=2)$
\cite{Sacha15a}, \cite{Kuros20a}, period-quadrupling $(s=4)$ \cite{Kuros20a}
and higher periodicities $s=10-100$ \cite{Giergel18a}, \cite{Giergiel20a}.
Also, a time-dependent Bogoliubov theory has been used to treat the $s=2$ case
\cite{Kuros20a} .

A concern in applying a mean-field (single-mode) or a few-mode approach to
study time crystals and discrete time-translation symmetry breaking is whether
the lack of thermalisation and decay of the condensate in such studies is an
artefact imposed by the adopted approximations \cite{Khemani19a}.
Thermalisation refers to a common phenomenon that many-body systems subjected
to periodic driving eventually reach infinite temperature, where no time
crystal can exist. On the other hand, describing a many-body system with no
more than a few modes precludes thermalisation and quantum depletion of the
condensate by assumption, raising concerns that the predicted time crystal may
not exist once one considers a more rigorous multi-mode approach.

Features that a theory of DTC should take into account include: the system
being many body, the possible effects of there being many modes that the
bosons could occupy and whether one or two modes could dominate, the roles of
both interactions and driving, the evolution dynamics being quantum and
allowing for a steady periodic non-equilibrium state to appear with period a
multiple of the drive period, whether or not thermalisation occurs, and the
sensitivity of DTC behaviour to changes in the initial conditions - including
the effect of the initial BEC temperature. The truncated Wigner approximation
(TWA)\ can take into account all of these features - whereas (for example) a
mean-field theory only allows for the bosons to occupy one mode, and
time-dependent Bogoliubov theory does not allow for substantial depletion of
the condensate mode.

In this paper we present a full theoretical many-body study of conditions for
creating discrete time crystals in a Bose-Einstein condensate which is allowed
to bounce resonantly on an oscillating mirror in a gravitational field as
proposed in \cite{Sacha15a}. We use the TWA to predict/ confirm the presence
of the non-trivial quantum phenomena of DTC using a more advanced and accurate
method than in previous studies \cite{Sacha15a}, \cite{Kuros20a}. The main
preparation process considered is that of a weakly interacting BEC in a
harmonic trap, strongly confined in the transverse direction. We study
many-body effects and effects of quantum fluctuations on the evolution for the
case of period-doubling $(s=2)$\ and for a zero temperature BEC. Processes
that could prevent a DTC being created, such as depletion of bosonic atoms
from the condensate mode, are allowed for, and the TWA approach is used to
examine whether thermalisation - which also destroys a DTC - occurs in
experimentally accessable time periods. We compare our results with previous
studies \cite{Sacha15a}, \cite{Kuros20a} where the theoretical approach is
based on mean-field theory and the Gross-Pitaevskii equation (GPE), and hence
assumes that the bosonic atoms remain in a single condensate mode, and on
time-dependent Bogoliubov theory, where the depletion from the condensate mode
is assumed to be small.

The theoretical framework outlined in the present paper is intended to provide
the basis for more extensive future studies, including for other periodicities
$sT$\ and for non-zero temperature. We also wish to explore more fully the
conditions for $sT$\ periodicity than is possible in this initial paper, such
as choices of the boson-boson interaction strength, different initial
conditions, the stability of the time crystal, etc, but using a theory which
is not restricted to assuming all bosons remain in a single condensate mode
(such as in mean-field theory) or that the quantum fluctuations from the
condensate field are small (as in time-dependent Bogoliubov theory). For these
reasons, and because other expositions of some of the theoretical background
do not cover in one paper all the background material, we have set out a full
description of our approach - including the description of the prepared
initial state via Bogoliubov theory as well as the evolution of the bouncing,
driven BEC. Our approach assumes that the total boson number $N$\ is large, as
is usual for a BEC.

The truncated Wigner approximation (TWA)\ \cite{Steel98}, \cite{Blakie08},
\cite{Dalton15a}, \cite{Gardiner17a}, \cite{King19}, is a phase-space theory
method which is a now well-established approach for treating dynamical
behaviour in cold bosonic gases. We expect it to be more reliable than either
the mean-field theory or time-dependent Bogoliubov theory, since these
approaches can be derived from TWA as approximations (see Appendix
\ref{Appendix - Mean Field Theory and Time-Dependent Bogoliubov Theory}). The
TWA involves both a field-theory approach, where the BEC is treated as a
single quantum field, and a mode-theory approach, where the separate quantum
modes or single-particle states that the modes could occupy, are used. The
modes that describe the evolution of the BEC are either Floquet modes, which
describe single particles moving in a time-periodic potential, or
gravitational modes, which describe single particles moving in a static
gravitational potential. Both in the single quantum field and the separate
mode approaches, observable quantities at each time can be expressed as
phase-space averages involving a Wigner distribution functional (function) or
as a stochastic average involving stochastic fields or mode amplitudes. For
phase-space averages the observable quantities are first expressed in terms of
symmetrically ordered combinations of the quantum operators (see Ref.
\cite{Dalton15a}). Similarly, evolution equations for the quantum state can be
described by Fokker-Planck equations for the Wigner functional (function), or
as in the present paper by equivalent Ito stochastic field equations in the
field-theory version or Ito stochastic differential equations for stochastic
phase-space variables in the mode-theory version \cite{Blakie08},
\cite{Dalton15a}. This equivalence ensures that phase space and stochastic
averages for observable quantities are equal. The Ito equations are
deterministic in the present case, but depend on initial values for the
phase-space variables - which are specified by a stochastic distribution
determined from the initial state of the BEC just after it is prepared. The
validity of the TWA is discussed in Refs. \cite{Blakie08}, \cite{Gardiner17a},
as well as in Section \ref{SubSubSection - Validity and Reliability of TWA}
and Appendix \ref{Appendix - Details re TWA Validity}.

The preparation of the initial state of the BEC is described by
time-independent Bogoliubov theory \cite{Dalton15a}, \cite{King19},
\cite{Morgan00}, \cite{Lewenstein}, \cite{Proukakis08} in which the BEC is
formed in a harmonic trap potential and the condensate wave function is
obtained from a time-independent Gross-Pitaevskii equation (GPE). The
condensate is assumed to be in a Fock state. The non-condensate field is
required to be orthogonal to the condensate mode, and non-condensate modes are
described both via an arbitrary set of mode functions chosen to be orthogonal
to the condensate mode and also via Bogoliubov mode functions determined from
Bogoliubov-de Gennes equations \cite{Morgan00}. There is a pair of Bogoliubov
mode functions for each mode frequency, which satisfy biorthogonality
conditions. Separate Wigner distribution functions for the Bogoliubov modes
and for the condensate mode are obtained when the density operator for the
initial state factorises into condensate and non-condensate density operators
- as in the Bogoliubov approximation. Stochastic phase-space variables are
introduced for both the condensate mode and the Bogoliubov modes, and the
stochastic properties of these phase-space variables are determined for the
case where the Bogoliubov modes are unoccupied and all bosons are in the
condensate mode in a phase-invariant state - such as would occur at zero
temperature. The initial stochastic phase-space variables for the Floquet or
gravitational modes (and their statistical properties) are determined from
stochastic phase-space variables for the condensate and Bogoliubov modes via
matching the stochastic field functions expanded either in terms of Floquet
(or gravitational) modes or in terms of the condensate and (non-condensate)
Bogoliubov modes. To help understand the dynamical processes involved, we also
apply the TWA treatment to a different initial condition, where the condensate
mode function is described as a linear combination of Wannier-like states.
Here, the non-condensate modes are all orthogonal to the condensate mode, but
not chosen as Bogoliubov modes.

Expressions for various quantities of physical interest are obtained as a
function of evolution time, including the position probability density (PPD);
the spatial position quantum correlation function (QCF) - which describes the
spatial coherence of the BEC; the mean values of one-body projectors (OBP) -
which describe the probabilities of finding a boson in the original condensate
mode or Wannier modes; and the quantum depletion - which describes the loss of
bosons from the condensate mode. The TWA treatment is compared with that based
entirely on mean-field theory as treated via the time-dependent version of the
Gross-Pitaevskii equation. Numerical calculations are performed for evolution
times out to $2000$ mirror oscillations and for a wide range of interaction
strengths as well as the two different initial conditions.

We find that for attractive interactions the combination of boson-boson
interaction and periodic driving enables a stable discrete time crystal to be
created for a realistic harmonic trap initial condition when the magnitude of
the interaction parameter exceeds a \emph{threshold} value, $|gN|>0.012$ for
our choice of parameters (see Table 1). In the absence of driving, we find
that the interaction strength corresponding to that required to create a DTC
is strong enough to couple many modes, which is indicated by a substantial
quantum depletion. However, with driving, the quantum depletion is strongly
suppressed and the thermalisation is quenched, implying that the absence of
our system's thermalisation is a genuine many-body effect and a direct
consequence of driving in the presence of a sufficiently strong interaction.
We also find that the dynamical behaviour of our system is largely independent
of whether the boson-boson interaction is attractive or repulsive.

We also find that our TWA results agree well with the mean-field
GPE\ calculations except at interaction strengths very close to the threshold
value for creation of a single stable wave-packet and formation of a discrete
time crystal. We show in Appendix
\ref{Appendix - Mean Field Theory and Time-Dependent Bogoliubov Theory} that
the mean-field and time-dependent Bogoliubov approaches are derivable as
approximations to the TWA treatment.\textbf{ }For realistic initial conditions
corresponding to the harmonic trap condensate mode function and typical
interaction strengths required to produce a DTC, our TWA calculations indicate
a quantum depletion less than about two atoms out of a total of $600$ atoms at
times corresponding to $2000$ mirror oscillations, in agreement with recent
many-body calculations based on a time-dependent Bogoliubov approach
\cite{Kuros20a}. However, for interaction strengths very close to the
threshold value for DTC formation, the quantum depletion due to the quantum
fluctuations can be as high as about $260$ atoms out of a total of $600$ atoms
at times corresponding to $2000$ mirror oscillations. Here, the time-dependent
Bogoliubov approach would break down, since Bogoliubov theory assumes that the
non-condensate field is relatively small. In the absence of driving and for a
relatively strong interaction ($0.05\leq|gN|\leq0.1$), we find that the
quantum depletion can be more than $220$ atoms out of a total of $600$ atoms
at times corresponding to $2000$ mirror oscillations, indicating that driving
in the presence of a sufficiently strong interaction needed to form a DTC
supresses the quantum depletion associated with the quantum many-body fluctuations.

In Section \ref{Section - Phase Space Theory for Evolution}, we describe the
phase-space theory used to study the evolution of the periodically driven
many-body system, which includes details of the truncated Wigner approximation
(Sect. \ref{SubSection - TWA}), the periodically driven BEC with stochastic
phase amplitudes for separate modes (Sect.
\ref{SubSection - Stochastic Phase Amplitudes}), the preparation of a BEC with
condensate and Bogoliubov modes (Sect. \ref{SubSection - Preparation of BEC}),
and the quantum state for the initial BEC (Sect.
\ref{SubSection - Quantum State for Initial BEC}). In Section
\ref{Section - Numerical Results}, we present numerical results of the TWA
calculations; in Section \ref{Section - Discussion} we discuss the
significance of our results in comparison with previous calculations; and in
Section \ref{Section - Conclusions and Summary} we present our conclusions and
suggest future directions. Details of the theory are set out in Appendices,
which are included as Online Supplementary Material.

\pagebreak

\section{Phase Space Theory for Evolution of a Periodically Driven BEC}

\label{Section - Phase Space Theory for Evolution}

The effect of quantum fluctuations due to many-body effects on the predictions
of time crystal behaviour in a BEC system may be treated using phase-space
methods. If the truncated Wigner $W^{+}$ distribution functional
\cite{Steel98}, \cite{Blakie08}, \cite{King19}, \cite{Dalton15a} is used, the
resulting Ito stochastic field equations are similar to the Gross-Pitaevski
equation that arises using the mean-field approximation, but now allow for
quantum fluctuations due to the fields being stochastic rather than
deterministic. The treatment outlined below is based on the $1D$
approximation, as in the papers of Sacha et al. \cite{Sacha15a},
\cite{Giergel18a}, \cite{Giergiel20a}, \cite{Kuros20a}.\smallskip

\subsection{Hamiltonian}

\label{SubSection - Hamiltonian}

The \emph{Hamiltonian} for the $1D$ model for the periodically driven BEC is
given in terms of the field operators $\hat{\Psi}(z)$, $\hat{\Psi}(z)^{\dag}$
for the annihilation, creation of a bosonic atom of mass $m$ at position $z$
as \cite{Giergel18a}%

\begin{align}
\hat{H}\mbox{\rule{-0.5mm}{0mm}}  &  =\mbox{\rule{-1mm}{0mm}}\int%
\mbox{\rule{-1mm}{0mm}}dz\left(  \frac{\hbar^{2}}{2m}\frac{\partial}{\partial
z}\hat{\Psi}(z)^{\dag}%
\mbox{\rule{-0.5mm}{0mm}}\,\mbox{\rule{-0.5mm}{0mm}}\frac{\partial}{\partial
z}\hat{\Psi}(z)+\hat{\Psi}(z)^{\dag}V(z,t)\hat{\Psi}(z)\right. \nonumber\\
&  \hspace{2cm}\left.  +\frac{g}{2}\hat{\Psi}(z)^{\dag}\hat{\Psi}(z)^{\dag
}\hat{\Psi}(z)\hat{\Psi}(z)\right) \label{Eq.HamiltonianFieldModel}\\
&  =\widehat{{\small K}}+\widehat{{\small V}}+\widehat{{\small U}}
\label{Eq.HamiltTerms}%
\end{align}
and is the sum of kinetic energy, time-dependent potential energy and
interaction energy terms.

The potential energy contains a driving term at frequency $\omega=2\pi/T$ and
is given by%
\begin{equation}
V(z,t)=mg_{E}\,z(1-\lambda\cos\omega t) \label{Eq.PotentialTerm}%
\end{equation}
This corresponds to an initially prepared BEC being allowed to fall in a
vertical gravitational field ($g_{E}$\ is the gravitational acceleration) and
allowed to bounce off an atom mirror oscillating with a period $T$ and
amplitude $\lambda$. The system is described in the oscillating frame.

The interaction energy is based on a zero-range contact interaction between
the bosonic atoms. The $1D$ coupling constant \textbf{is} $g=2\hslash
\omega_{\bot}a_{s}$ - which is proportional to the s-wave scattering length
$a_{s}$ and the oscillation frequency $\omega_{\bot}$ for the BEC atoms in a
transverse trap - may be tuned via a Feshbach resonance. \smallskip

\subsection{Truncated Wigner Approximation}

\label{SubSection - TWA}

In this section we outline the \emph{truncated Wigner approximation}
(TWA)\ for phase-space field theory and set out basic expressions for some
quantities of physical interest such as the position probability density and
the first order quantum correlation function. \smallskip

\subsubsection{Functional Fokker-Planck Equation - W$^{+}$ Distribution}

The \emph{functional Fokker-Planck equation} (FFPE) for the $W^{+}$
\emph{Wigner distribution functional }$W[\mathbf{\psi}(z)]$ which represents
the quantum density operator $\widehat{\rho}$ can be obtained via applying the
correspondence rules in conjunction with the Liouville-von Neumann equation
for the density operator. The contributions can be written as a sum of terms
from $\widehat{{\small K}},\widehat{{\small V}},\widehat{{\small U}}$. A
derivation for $3D$ is presented in Ref \cite{Dalton15a} (see Section 15.1.6
and Appendix I). The field operators $\hat{\Psi}(z)$, $\hat{\Psi}(z)^{\dag}%
$\ are represented by two unrelated\textbf{ }time-independent c-number fields
$\psi(z)$\ and $\psi^{+}(z)$,\ respectively. For short, $\mathbf{\psi
}(z)\equiv\psi(z),\psi^{+}(z)$, and $W[\mathbf{\psi}(z)]$ is time dependent.

The kinetic energy term is
\begin{align}
&  \left(  \frac{\partial}{\partial t}W[\mathbf{\psi}(z)]\right)
_{K}\nonumber\\
&  =\frac{i}{\hslash}\int dz\,\left[  \left\{  \frac{\delta}{\delta\psi
(z)}\left(  -\frac{\hbar^{2}}{2m}\frac{\partial^{2}}{\partial z^{2}%
}{\small \psi(z)\,}W[\ \mathbf{\psi}(r)]\right)  \right\}  \right. \nonumber\\
&  \left.  +\left\{  \frac{\delta}{\delta\psi^{+}(z)}\left(  \frac{\hbar^{2}%
}{2m}\frac{\partial^{2}}{\partial z^{2}}{\small \psi}_{u}^{+}{\small (z)\,}%
W[\ \mathbf{\psi}(r)]\right)  \right\}  \right]
\label{Eq.FFPEKineticFermiFldModel}%
\end{align}
and only contributes to the drift term in the FFPE.

The potential energy term is
\begin{align}
&  \left(  \frac{\partial}{\partial t}W[\mathbf{\psi}(z)]\right)
_{V}\nonumber\\
&  =\frac{i}{\hslash}\int{\small d}z\,\left[  \left\{  \frac{\delta}%
{\delta\psi(z)}\left(  (V(z,t))\psi(z)\,W[\ \mathbf{\psi}(r)]\right)
\right\}  -\left\{  \frac{\delta}{\delta\psi^{+}(z)}\left(  (V(z,t))\psi
^{+}(z)\,W[\ \mathbf{\psi}(z)]\right)  \right\}  \right] \nonumber\\
&  \label{Eq.FFPEPotentialFermiFldModel}%
\end{align}
and also only contributes to the drift term in the FFPE.

The boson-boson interaction term is
\begin{align}
&  \left(  \frac{\partial}{\partial t}W[\mathbf{\psi}(z)]\right)
_{U}\nonumber\\
&  {\small =}\frac{i}{\hslash}g\int{\small d}z\,\left[  \left\{  \frac{\delta
}{\delta\psi(z)}(\psi^{+}(z)\psi(z)-\delta_{C}(z,z))\,\psi(z)\,W[\mathbf{\psi
}(z)]\right\}  \right. \nonumber\\
&  \left.  -\left\{  \frac{\delta}{\delta\psi^{+}(z)}(\psi^{+}(z)\psi
(z)-\delta_{C}(z,z))\,\psi^{+}(z)\,W[\mathbf{\psi}(z)]\right\}  \right.
\nonumber\\
&  \left.  -\frac{1}{4}\left\{  \frac{\delta}{\delta\psi(z)}\frac{\delta
}{\delta\psi(z)}\frac{\delta}{\delta\psi^{+}(z)}\,\psi(z)\,W[\mathbf{\psi
}(z)]\right\}  \right. \nonumber\\
&  \left.  +\frac{1}{4}\left\{  \frac{\delta}{\delta\psi^{+}(z)}\frac{\delta
}{\delta\psi^{+}(z)}\frac{\delta}{\delta\psi(z)}\,\psi^{+}(z)\,W[\mathbf{\psi
}(z)]\right\}  \right]  \label{Eq.FFPEInteractFermiFldModel}%
\end{align}
where for two position coordinates $z$ and $z^{\#}$
\begin{equation}
\delta_{C}(z,z^{\#})=%
{\displaystyle\sum\limits_{k}}
\phi_{k}(z)\phi_{k}(z^{\#})^{\ast} \label{Eq.DefnDelta}%
\end{equation}
with the $\phi_{k}(z)$ being any set of suitable orthonormal mode functions
satisfying the condition $%
{\displaystyle\int}
dz\,\phi_{k}(z)^{\ast}\phi_{l}(z)=\delta_{k,l}$. A unitary change in these
modes does not change $\delta_{C}(z,z^{\#})$.

The exact FFPE involves only first and third-order functional derivatives -
there are no second-order diffusion terms. In the truncated Wigner $W^{+}$
treatment, the third-order derivatives are discarded so only the first-order
drift terms remain. This proceedure is justisfied on the basis that the region
of phase space that is most important is where $\psi(z),\psi^{+}(z)\sim
\sqrt{N}$, where $N$ is the number of bosons present. As applying a further
derivative gives a contribution of order $1/\sqrt{N}$ smaller than the
previous term, discarding the third-order derivative terms is justified when
$N$\smallskip$\gg1$. The validity of the TWA is discussed in Refs.
\cite{Blakie08}, \cite{Gardiner17a}, as well as in Section
\ref{SubSubSection - Validity and Reliability of TWA} and Appendix
\ref{Appendix - Details re TWA Validity}.\smallskip\ 

\subsubsection{Ito Stochastic Field Equations for Time Evolution}

The \textit{Ito stochastic field equations} (SFE) Ito equations are equivalent
to the functional Fokker-Planck equation for the distribution functional
assuming the latter only involve up to second-order functional derivatives.
There are no further approximations involved. This equivalence is demonstrated
in Chapter 14 of Ref \cite{Dalton15a} and is based on the requirement that the
phase space average of any function of the phase space field functions
obtained from the Wigner distribution functional is the same as the stochastic
average of the same function of stochastic field functions. The form of the
Ito stochastic equations that enables this equivalence is obtained from the
terms in the FFPE.\textbf{ }

The Ito (SFE)\textbf{ }corresponding to the functional Fokker-Planck equation
for the $W^{+}$ distribution functional are as follows: (see Section 15.1.8 in
Ref. \cite{Dalton15a})%
\begin{align}
\frac{\partial}{\partial t}\widetilde{\psi}(z,t)  &  =-\frac{i}{\hslash
}\left[  -\frac{\hslash^{2}}{2m}\frac{\partial^{2}}{\partial z^{2}%
}\widetilde{\psi}(z,t)+V(z,t)\widetilde{\psi}(z,t)+g\{\widetilde{\psi}%
^{+}(z,t)\widetilde{\psi}(z,t)-\delta_{C}(z,z)\}\widetilde{\psi}(z,t)\right]
\nonumber\\
&  \label{Eq.ItoSFEA}%
\end{align}
and
\begin{align}
\frac{\partial}{\partial t}\widetilde{\psi}^{+}(z,t)  &  =+\frac{i}{\hslash
}\left[  -\frac{\hslash^{2}}{2m}\frac{\partial^{2}}{\partial z^{2}%
}\widetilde{\psi}^{+}(z,t)+V(z,t)\widetilde{\psi}^{+}(z,t)+g\{\widetilde{\psi
}^{+}(z,t)\widetilde{\psi}(z,t)-.\delta_{C}(z,z)\}\widetilde{\psi}%
^{+}(z,t)\right] \nonumber\\
&  \label{Eq.ItoSFAB}%
\end{align}
where the $\widetilde{\psi}(z,t),\widetilde{\psi}^{+}(z,t)$ are time-dependent
stochastic field functions that now replace the original phase-space field
functions $\psi(z),\psi^{+}(z)$. These are functions of position which may be
expanded in terms of an independent set of (non-stochastic) basis functions or
modes and where the amplitudes (or stochastic phase variables) for the mode
expansion are treated as stochastic quantities. Stochasticity in the context
of phase space theory is discussed in Appendix G of Ref. \cite{Dalton15a}. As
there are no diffusion terms in the FFPE, there are no Gaussian-Markov noise
terms satisfying the standard results $\overline{\Gamma_{a}\,(t)}=0$,
$\overline{\Gamma_{a}\,(t_{1})\Gamma_{b}\,(t_{2})}$ $=\delta_{a,b}\delta
(t_{1}-t_{2})$ etc, in the Ito SFE for the truncated Wigner $W^{+}$ treatment
(the bar denotes a stochastic average). The solutions to the Ito SFE are
uniquely determined from the initial functions $\widetilde{\psi}(z,0)$ and
$\widetilde{\psi}^{+}(z,0)$, as if the fields were like classical fields.
However, $\widetilde{\psi}(z,t),\widetilde{\psi}^{+}(z,t)$ are non-classical
and stochastic\textbf{ }because the initial functions $\widetilde{\psi}(z,0)$
and $\widetilde{\psi}^{+}(z,0)$ are members of an ensemble of stochastic
fields, with a distribution chosen to represent the features of the initial
quantum state $\widehat{\rho}(0)$. This in turn leads to an ensemble of
time-dependent stochastic fields $\widetilde{\psi}(z,t),\widetilde{\psi}%
^{+}(z,t)$ representing the time evolution of the quantum state, this being
equivalent to the time evolution in the Wigner distribution functional
$W[\mathbf{\psi}(z)].$\smallskip

\subsubsection{Mean-Field Theory and Gross-Pitaevski Equation}

The Ito stochastic field equations resemble the \emph{Gross-Pitaevskii
equation} (GPE) that applies for the mean-field approximation in which all
bosons are assumed to occupy a single condensate mode. The GPE would represent
another approach to treating the periodically driven BEC, but as it does not
allow for the presence of other modes this mean-field approach cannot allow
for quantum fluctuation effects that are treated via the TWA. The
time-dependent GPE for the condensate wave function $\Phi_{c}(z,t)$ -
normalised as $%
{\displaystyle\int}
dz\,\Phi_{c}(z,t)^{\ast}$ $\Phi_{c}(z,t)$ $=N$ - is given by%
\begin{equation}
\frac{\partial}{\partial t}\Phi_{c}(z,t)=-\frac{i}{\hslash}\left[
-\frac{\hslash^{2}}{2m}\frac{\partial^{2}}{\partial z^{2}}\Phi_{c}%
(z,t)+V(z,t)\Phi_{c}(z,t)+g\{\Phi_{c}(z,t)^{\ast}\Phi_{c}(z,t)\}\Phi
_{c}(z,t)\right]  \label{Eq.GPE}%
\end{equation}
If we choose $\widetilde{\psi}^{+}(z,t)=\widetilde{\psi}(z,t)^{\ast}$ and
$\widetilde{\psi}(z,t)=\Phi_{c}(z,t)$ the Ito SFE has almost the same form as
the GPE. However, as well as $\Phi_{c}(z,t)$ being non-stochastic, the
quantity $\delta_{C}(z,z)$ is absent from the non-linear term in the GPE that
allows for boson-boson interactions in the mean-field approximation. As $%
{\displaystyle\int}
dz\,\Phi_{c}(z,t)^{\ast}\Phi_{c}(z,t)=N$ and $%
{\displaystyle\int}
dz\,\delta_{C}(z,z)=n_{M}$ (where $n_{M}$ is the number of modes that need to
be used to treat the physics), we see that the $\delta_{C}(z,z)$ term in the
Ito SFE \smallskip will be relatively unimportant compared to the
$\widetilde{\psi}^{+}(z,t)\widetilde{\psi}(z,t)$ term in the usual case for a
BEC system, where the number of bosons considerably exceeds the number of
modes that need to be considered. The stochasticity of the initial
$\widetilde{\psi}(z,0)$ and $\widetilde{\psi}^{+}(z,0)$ should be more
important than the $\delta_{C}(z,z)$ term in treating quantum effects for
interacting bosons.

We define throughout a unity normalised \emph{condensate mode function}
$\psi_{c}(z,t)$\ as\textbf{\ }%
\begin{equation}
\psi_{c}(z,t)=\Phi_{c}(z,t)/\sqrt{N} \label{Eq.CondModeFn}%
\end{equation}
\smallskip

\subsubsection{Validity and Reliability of Truncated Wigner Approximation}

\label{SubSubSection - Validity and Reliability of TWA}

Since mean-field theory or time-dependent Bogoliubov theory can be derived
from the TWA as approximations (as shown in Appendix
\ref{Appendix - Mean Field Theory and Time-Dependent Bogoliubov Theory}), we
expect the TWA to be more reliable than both of these approximate approaches -
which themselves have been successfully used to treat cold bosonic gases.
Indeed, the TWA has already been extensively used to describe cold bosonic
gases. However, as explained above, the TWA approach involves a key
approximation -- the neglect of third-order derivative terms in the FFPE.
Ultimately, the reliability of the TWA rests on whether these terms can be
neglected, but as far as we are aware no calculations of the size of these
terms has been carried out. So, at present, the only tests of the reliabilty
of the TWA are: (a) whether its predictions agree with experiment - and the
relevant experiment has not yet been carried out; and/or (b) whether it agrees
with exact full FFPE calculations -- and these have not been done. Detailed
issues are discussed in Appendix \ref{Appendix - Details re TWA Validity}.
\smallskip

\subsubsection{Mean Energy}

The expression for the \emph{mean energy}
\begin{equation}
\left\langle \widehat{H}\right\rangle =Tr(\widehat{H}\,\widehat{\rho})
\label{Eq.MeanEnergy}%
\end{equation}
in terms of the stochastic field functions $\widetilde{\psi}%
(z,t),\widetilde{\psi}^{+}(z,t)$ is
\begin{align}
\left\langle \widehat{H}\right\rangle  &  =%
{\displaystyle\int}
dz\,\left(  \frac{\hbar^{2}}{2m}\overline{\frac{\partial}{\partial
z}\widetilde{\psi}^{+}(z,t)\,\frac{\partial}{\partial z}\widetilde{\psi}%
(z,t)}\right)  \,\nonumber\\
&  +%
{\displaystyle\int}
dz\,\left(  V(z,t)\,\overline{\widetilde{\psi}^{+}(z,t)\,\,\widetilde{\psi
}(z,t)}\right) \nonumber\\
&  +%
{\displaystyle\int}
dz\,\left(  \frac{g}{2}\;\overline{\widetilde{\psi}^{+}(z,t)\,\widetilde{\psi
}^{+}(z,t)\,\widetilde{\psi}(z,t)\,\widetilde{\psi}(z,t)}\right) \nonumber\\
&  -%
{\displaystyle\int}
dz\,\left(  g\,\delta_{C}(z,z)\;\overline{\widetilde{\psi}^{+}%
(z,t)\,\,\widetilde{\psi}(z,t)}\right) \nonumber\\
&  -%
{\displaystyle\int}
dz\,\left(  \frac{\hbar^{2}}{4m}(\triangle\delta_{C}(z,z))\right)  -%
{\displaystyle\int}
dz\,\left(  \frac{1}{2}\delta_{C}(z,z)\;V(z,t)\right) \nonumber\\
&  +%
{\displaystyle\int}
dz\,\left(  \frac{g}{4}\,\delta_{C}(z,z)^{2}\right)
\label{Eq.MeanEnergyStochastic}%
\end{align}
where
\begin{equation}
\triangle\delta_{C}(z,z)=%
{\displaystyle\sum\limits_{k}}
\left(  \frac{\partial}{\partial z}\phi_{k}(z)\right)  \left(  \frac{\partial
}{\partial z}\phi_{k}(z)\right)  ^{\ast} \label{Eq.DefnDzDelta}%
\end{equation}
and $\phi_{k}(z)$ being any set of suitable orthonormal mode functions. \ The
proof is given in Appendix \ref{Appendix - Mean Energy}.

For the present situation where the number of bosons considerably exceeds the
number of modes that need to be considered, only the first three terms are
important. The last three terms are non-stochastic. \ An expression
for\textbf{ }the mean energy is given below in Eq.
(\ref{Eq.MeanEnergyGravModes}) in terms of gravitational modes.\smallskip

\subsubsection{Position Probability Density}

The \emph{number operator} is given by
\begin{equation}
\widehat{N}=%
{\displaystyle\int}
dz\,\hat{\Psi}^{\dag}(z)\hat{\Psi}(z) \label{Eq.NumberOpr}%
\end{equation}
which is the integral over an operator $\widehat{N}(z)=\hat{\Psi}^{\dag
}(z)\hat{\Psi}(z)$ associated with the position probability density for the
bosons in the system. We consider states which are eigenstates of
$\widehat{N}$ with eigenvalue $N$.

The \emph{position probability density} (PPD)\ is defined as the mean value of
$\widehat{N}(z)$
\begin{equation}
F(z,t)=Tr(\hat{\Psi}^{\dag}(z)\hat{\Psi}(z)\rho(t))
\label{Eq.PositionProbDefn}%
\end{equation}
This essentially specifies the relative numbers of bosons found at various
positions $z$ as a function of $t$.

As indicated in the Introduction, to employ the $W^{+}$ distribution we must
express $\hat{\Psi}^{\dag}(z)\hat{\Psi}(z)$ in the symmetric form $\{\hat
{\Psi}^{\dag}(z)\hat{\Psi}(z)\}=\frac{1}{2}(\hat{\Psi}^{\dag}(z)\hat{\Psi
}(z)+\hat{\Psi}(z)\hat{\Psi}^{\dag}(z))$ via $\hat{\Psi}^{\dag}(z)\hat{\Psi
}(z)=\{\hat{\Psi}^{\dag}(z)\hat{\Psi}(z)\}-\frac{1}{2}\delta_{C}(z,z)$, so
that the position number density is then given either in terms of a functional
integral or equivalently via a stochastic average%
\begin{align}
F(z,t)  &  =%
{\textstyle\int}
{\textstyle\int}
D^{2}\psi D^{2}\psi^{+}\,\psi(z)W[\mathbf{\psi}(z)]\psi^{+}(z)-\frac{1}%
{2}\delta_{C}(z,z)\label{Eq.PosiProbPhaseSpaceIntegal}\\
&  =\overline{\widetilde{\psi}(z,t)\widetilde{\psi}^{+}(z,t)}-\frac{1}%
{2}\delta_{C}(z,z) \label{Eq.PositProbStochasAver}%
\end{align}
The stochastic average is the expression used in numerical calculations. Note
the presence of the $\frac{1}{2}\delta_{C}(z,z)$, but as the spatial integral
of this term is of order $\frac{1}{2}n_{M}$, whilst that of $\overline
{\widetilde{\psi}(z,t)\widetilde{\psi}^{+}(z,t)}$ is of order $N_{c}\gg n_{M}%
$, the second term in Eq. (\ref{Eq.PositProbStochasAver}) should not be important.

It should be noted that other distribution functionals can also be used to
represent the density operator, and these have their own functional
Fokker-Planck equations and Ito stochastic field equations. These include the
positive $P+$ distribution functional, where here the FFPE includes a
diffusion term and the Ito SFE involve Gaussian-Markoff noise terms. The $P+$
case is discussed in Ref \cite{Dalton15a} (see Sections 15.1.5 and 15.1.7).

A direct numerical solution of the Ito stochastic field equations for each
initial $\widetilde{\psi}(z,0)$ and $\widetilde{\psi}^{+}(z,0)$ is one
possible approach to calculating the position probability density via Eq
(\ref{Eq.PositProbStochasAver}), with an ensemble of different initial
stochastic fields chosen to represent the properties of the initial quantum
state. In such a treatment the underlying presence of mode functions and their
frequencies is not made explicit. Our approach however will involve mode
expansions.\smallskip

\subsubsection{Quantum Correlation Function}

As well as the position probability density, there is a further quantity that
is of interest in describing BECs. This is the first-order \emph{quantum
correlation function} (QCF)\ which is defined \cite{Leggett01a} as%
\begin{equation}
P(z,z^{\#},t)=Tr(\hat{\Psi}^{\dag}(z^{\#})\hat{\Psi}(z)\rho(t)) \label{Eq.QCF}%
\end{equation}
where $z,z^{\#}$ are two spatial positions. Since $\hat{\Psi}^{\dag}%
(z^{\#})\hat{\Psi}(z)=\{\hat{\Psi}^{\dag}(z^{\#})\hat{\Psi}(z)\}-\frac{1}%
{2}\delta_{C}(z,z^{\#})$, the QCF can also be expressed as a stochastic
average via
\begin{equation}
P(z,z^{\#},t)=\overline{\widetilde{\psi}(z,t)\widetilde{\psi}^{+}(z^{\#}%
,t)}-\frac{1}{2}\delta_{C}(z,z^{\#}) \label{Eq.QCFStochAver}%
\end{equation}

This QCF can be expressed in terms of \emph{natural orbitals} $\chi_{i}(z,t)$
and their \emph{occupation numbers }$p_{i}$, as will now be shown. First, we
see that $P(z,z^{\#},t)=P(z^{\#},z,t)^{\ast}$. Then if we introduce an
orthonormal set of mode functions $\phi_{k}(z)$ we can define a matrix $P$
with elements $P_{k,l}=%
{\displaystyle\iint}
dz_{1}\,dz_{1}^{\#}\,\phi_{k}^{\ast}(z)\,P(z,z^{\#},t)\,\phi_{l}(z^{\#})$. It
is easy to show that $P$ is Hermitian, so it can be diagonalised via a unitary
matrix $U$ such that $P=U\Delta U^{\dag}$, where $\Delta$ is a diagonal matrix
containing the eigenvalues $p_{i}$ of $P$. These eigenvalues are real and
positive. Substituting for $P$ we find that the QCF\ can be written in terms
of natural orbitals $\chi_{i}(z,t)$ and their occupation numbers $p_{i}$ (see
Eq.(\ref{Eq.QCFNaturalOrbitals})) as
\begin{equation}
P(z,z^{\#},t)=%
{\textstyle\sum\limits_{i=0}}
p_{i}\,\chi_{i}(z,t)\,\chi_{i}(z^{\#},t)^{\ast} \label{Eq.QCFNaturalOrbitals}%
\end{equation}
where the natural orbitals are
\begin{equation}
\chi_{i}(z,t)=%
{\textstyle\sum\limits_{k}}
U_{k,i}\,\phi_{k}(z,t) \label{Eq.NaturalOrb}%
\end{equation}
Situations where almost all bosons occupy one natural orbital are an indicator
of a BEC. By convention the natural orbital with the largest occupancy is
listed as $i=0$. The QCF\ essentially specifies the long-range spatial
coherence that applies in BECs.\smallskip

\subsubsection{One-Body Projector}

\label{SubSubSection - One Body Projector}

We can define a \emph{one-body projection }operator (OBP) onto any single
particle state $\left\vert \phi\right\rangle $ for an $n$ body system in first
quantisation as
\begin{equation}
\widehat{M}=%
{\displaystyle\sum\limits_{i=1}^{n}}
(\left\vert \phi\right\rangle \left\langle \phi\right\vert )_{i}
\label{Eq.OneBodyProjectorOpr}%
\end{equation}
the sum being over \emph{all} the identical bosons. Note that for a system of
identical bosons the operator must be \emph{symmetric} under particle
permutations. Typically in a BEC with $N\gg1$ bosons in the condensate mode
the single-particle state would be taken as the initial\textbf{\ }condensate
mode function\textbf{\ }with $\left\langle z|\psi_{c}\right\rangle =$
$\psi_{c}(z,0)$. This quantity is a Hermitian operator for the many-body
system and can be regarded as an observable. Its mean value could therefore be
measured as a function of time.

The operator $\widehat{M}$ can be written in second quantisation form using an
orthonormal set of mode functions $\phi_{k}(z)$ and their annihilation,
creation operators $\widehat{a}_{k}$, $\widehat{a}_{k}^{\dag}$ using standard
procedures in which the field operators are expanded in terms of a set of
orthonormal mode functions $\phi_{k}(z)$ as $\widehat{\Psi}(z)=%
{\displaystyle\sum\limits_{k}}
\phi_{k}(z)\,\widehat{a}_{k}$ and $\widehat{\Psi}(z)^{\dag}=%
{\displaystyle\sum\limits_{k}}
\phi_{k}(z)^{\ast}\,\widehat{a}_{k}^{\dag}$. We have for the case of the
condensate mode function
\begin{align}
\widehat{M}_{c}  &  =%
{\displaystyle\sum\limits_{k,l}}
\widehat{a}_{k}^{\dag}\,\widehat{a}_{l}\,\left\langle \phi_{k}%
(1)|\,(\left\vert \psi_{c}\right\rangle \left\langle \psi_{c}\right\vert
)_{1}\,|\,\phi_{l}(1)\right\rangle \nonumber\\
&  =%
{\displaystyle\sum\limits_{k,l}}
\widehat{a}_{k}^{\dag}\,\widehat{a}_{l}\,%
{\displaystyle\int}
dz_{1}\,\,\phi_{k}^{\ast}(z_{1})\psi_{c}(z_{1},0)%
{\displaystyle\int}
dz_{1}^{\#}\,\,\psi_{c}^{\ast}(z_{1}^{\#},0)\phi_{l}(z_{1}^{\#})\nonumber\\
&  =%
{\displaystyle\sum\limits_{k,l}}
\left(
{\displaystyle\int}
dz^{\#}\,\phi_{k}(z^{\#})\widehat{\Psi}(z^{\#})^{\dag}\right)  \left(
{\displaystyle\int}
dz\,\phi_{l}^{\ast}(z)\widehat{\Psi}(z)\right)  \left(
{\displaystyle\int}
{\displaystyle\int}
dz_{1}\,dz_{1}^{\#}\,\phi_{k}^{\ast}(z_{1})\psi_{c}(z_{1},0)\psi_{c}^{\ast
}(z_{1}^{\#},0)\phi_{l}(z_{1}^{\#})\right) \nonumber\\
&  =%
{\displaystyle\int}
{\displaystyle\int}
dz\,dz^{\#}\,\psi_{c}(z^{\#},0)\,\widehat{\Psi}(z^{\#})^{\dag}\,\widehat{\Psi
}(z)\,\psi_{c}^{\ast}(z,0) \label{Eq.OneBodyProjSecondQn}%
\end{align}
where the completeness results $%
{\displaystyle\sum\limits_{k}}
\phi_{k}(z^{\#})\,\phi_{k}^{\ast}(z_{1})=\delta(z^{\#}-z_{1})$ and $%
{\displaystyle\sum\limits_{k}}
\phi_{l}(z_{1}^{\#})\,\phi_{l}^{\ast}(z)=\delta(z-z_{1}^{\#})$ are used and
the integrals over $z_{1}$ and $z_{1}^{\#}$ are carried out. Thus, the
one-body projector operator involves an integral of the position correlation
operator $\widehat{\Psi}(z^{\#})^{\dag}\,\widehat{\Psi}(z)$ times $\psi
_{c}(z^{\#},0)\psi_{c}^{\ast}(z,0)$.

Hence, we see that the mean value for the one-body projector operator is
\begin{equation}
M_{c}(t)=%
{\displaystyle\int}
{\displaystyle\int}
dz\,dz^{\#}\,\psi_{c}(z^{\#},0)\,\,\psi_{c}^{\ast}(z,0)\,P(z,z^{\#},t)
\label{Eq.MeanOneBodyProjector}%
\end{equation}
which is the double space integral of $\psi_{c}(z^{\#},0)\psi_{c}^{\ast}(z,0)$
multiplied by the time dependent \emph{first order quantum correlation
function} $P(z,z^{\#},t)$, which is defined and its expression given in terms
of stochastic field functions as in Eqs. (\ref{Eq.QCF}) and
(\ref{Eq.QCFStochAver}).

If we substitute for the QCF\ in terms of the stochastic field functions
$\widetilde{\psi}(z,t)$ and $\widetilde{\psi}^{+}(z,t)$ we get%
\begin{equation}
M_{c}(t)=%
{\displaystyle\int}
{\displaystyle\int}
dz\,dz^{\#}\,\psi_{c}(z^{\#},0)\,\psi_{c}^{\ast}(z,0)\left\{  \overline
{\widetilde{\psi}(z,t)\widetilde{\psi}^{+}(z^{\#},t)}-\frac{1}{2}\delta
_{C}(z,z^{\#})\right\}  \label{Eq.ObservableStochResult}%
\end{equation}
Note that the result involves the condensate mode function $\psi_{c}(z,0)$\ at
time $0$, rather than at time $t$, as in the expression in Eq.
(\ref{Eq.CondModeOccupation}) for the occupancy $N_{C}(t)$ of the
time-dependent condensate mode. Thus, the mean value of the one-body projector
is a different quantity to the quantum depletion - defined below in Section
\ref{Quantum Depletion}. \ We can use the mean value of the\ one-body
projector $M_{c}(t)$\ as an observable. This can obviously be calculated from
the time-dependent stochastic field functions $\widetilde{\psi}(z,t)$ and
$\widetilde{\psi}^{+}(z^{\#},t)$ and the condensate mode function $\psi
_{c}(z,0)$ as a function of time. The term $\frac{1}{2}\delta_{C}(z,z^{\#})$
would be small compared to the $\overline{\widetilde{\psi}(z,t)\widetilde{\psi
}^{+}(z^{\#},t)}$ term for $N_{c}\gg1$. The quantity $M_{c}(t)$\ has the
desired feature of revealing the periodicity of the QCF\ via its Fourier
transform (FT).

Also we have
\begin{equation}
M_{c}(0)=N \label{Eq.InitialM}%
\end{equation}
assuming the initial condition of all bosons in mode $\psi_{c}$ and using Eq.
(\ref{Eq.FockStateT0Props}) (see below). It is straightforward to rephrase the
above treatment in a normalised form $M_{c}(t)/M_{c}(0)$ using Eq.
(\ref{Eq.InitialM}) for the initial value of $M_{c}(t)$

The Fourier transform of the position probability density reveals the
frequencies at which the PPD is oscillating. By taking the FT of $M_{c}(t)$ we
have
\begin{equation}
FT\{M_{c}(t)\}=%
{\displaystyle\int}
{\displaystyle\int}
dz\,dz^{\#}\,\psi_{c}(z^{\#},0)\,\,\psi_{c}^{\ast}(z,0)\,\;FT\{P(z,z^{\#},t)\}
\label{Eq.FTResult}%
\end{equation}
since the initial condensate mode function $\,\psi_{c}(z,0)$ is
time-independent. Thus the FT of $M_{c}(t)$ would reveal the \emph{same
frequencies} as the FT of the QCF.

Hence, the one-body projector approach yields a useful indicator for
describing the behaviour of the QCF (and also of its diagonal terms when
$z=z^{\#}$ - the position probability density), whose periodicities determine
whether time crystal behaviour is occurring. The FT of $M_{c}(t)$\ does not of
course explain the periodicity of the QCF or the PPD - that explanation
requires a consideration of the numerous time scales involved in the system
(the drive period $T$, the gravitational mode frequencies, the time scale
associated with boson-boson interaction, the bounce time of the BEC, etc.).

The mean value of the one body projector $M_{c}(t)$ for a many-body system is
similar to the autocorrelation function or the \emph{fidelity} $F_{K}(t)$
introduced by Kuros et. al. \cite{Kuros20a} (see Eq. (6) therein) in a
mean-field theory. The fidelity was defined by\textbf{\ }%
\begin{equation}
F_{K}(t)=|\left\langle \Phi_{c}(0)|\Phi_{c}(t)\right\rangle |^{2}=|%
{\displaystyle\int}
dz\;\Phi_{c}(z,0)\Phi_{c}^{\ast}(z,t)\,|^{2} \label{Eq.KurosFidelity}%
\end{equation}
where $\Phi_{c}(z,t)$\ is the condensate wave-function, given by the solution
of the GPE. This specifies how similar the time-dependent condensate
wave-function is to its initial form.

This can be written as (note the position orders)
\begin{equation}
F_{K}(t)=%
{\displaystyle\int}
{\displaystyle\int}
dz\,dz^{\#}\,P_{K}(z^{\#},z,0)\,P_{K}(z,z^{\#},t)
\label{Eq.KurosFidelityResult}%
\end{equation}
where
\begin{equation}
P_{K}(z,z^{\#},t)\,=\Phi_{c}(z,t)\Phi_{c}^{\ast}(z^{\#},t) \label{Eq.KurosQCF}%
\end{equation}
is the QCF in the single-mode mean-field approximation. The similarity of
Eq.(\ref{Eq.KurosFidelityResult}) to Eq.(\ref{Eq.MeanOneBodyProjector}) for
the mean value of the condensate mode function OBP $M_{c}(t)$ is clear. The
Fourier transform of $F_{K}(t)$ is related to the FT of the QCF $P_{K}%
(z,z^{\#},t)$ via a similar relationship to that in Eq. (\ref{Eq.FTResult}).
Note that $F_{K}(t)$ is not proportional to $%
{\displaystyle\int}
dz\,\rho(z,0)\,\rho(z,t)$\ - as stated in Ref. \cite{Kuros20a}, where
$\rho(z,t)=|\Phi_{c}(z,t)|^{2}$ is the mean-field position probability
density.\smallskip

\subsubsection{Other One-Body Projectors}

The one-body projector concept can be extended to cases where the mode
function chosen is different to the condensate mode $\psi_{c}(z,0).$ One case
of particular interest is where the mode function \ is one of the two
Wannier-like modes $\Phi_{i}(z,t)$ $(i=1,2)$ (see Sect
\ref{SubSubSection Wannier Modes}\ and Eq (\ref{Eq.ApproxWannier}) below).

In this case we would have for the OBP operator $\widehat{M}_{Wi}(t)$ and its
mean value $M_{Wi}(t)\equiv N_{i}(t)$
\begin{align}
\widehat{M}_{Wi}(t)  &  =%
{\displaystyle\int}
{\displaystyle\int}
dz\,dz^{\#}\,\Phi_{i}(z^{\#},t)\,\widehat{\Psi}(z^{\#})^{\dag}\,\widehat{\Psi
}(z)\,\Phi_{i}^{\ast}(z,t)\nonumber\\
M_{Wi}(t)  &  =Tr\left(  \widehat{M}_{Wi}(t)\widehat{\rho}(t)\right)  \equiv
N_{i}(t)\nonumber\\
N_{i}(t)  &  =%
{\displaystyle\int}
{\displaystyle\int}
dz\,dz^{\#}\,\Phi_{i}(z^{\#},t)\,\,\Phi_{i}^{\ast}(z,t)\left\{  \overline
{\widetilde{\psi}(z,t)\widetilde{\psi}^{+}(z^{\#},t)}-\frac{1}{2}\delta
_{C}(z,z^{\#})\right\}  \qquad i=1,2\nonumber\\
&  \label{Eq.OtherOBP}%
\end{align}

\subsubsection{Quantum Depletion from Condensate Mode}

\label{Quantum Depletion}

We can expand the field operators, field functions and the stochastic field
functions in terms of a time-dependent condensate mode function $\psi
_{c}(z,t)=\Phi_{c}(z,t)/\sqrt{N}$ obtained from the solution of the
time-dependent GPE (\ref{Eq.GPE}) and any set of orthogonal non-condensate
modes $\psi_{k}(z,t)$ as
\begin{align}
\widehat{\Psi}(z)  &  =\psi_{c}(z,t)\widehat{c}_{0}(t)+%
{\displaystyle\sum\limits_{k\neq0}}
\psi_{k}(z,t)\widehat{c}_{k}(t)\qquad\widehat{\Psi}(z)^{\dag}=\psi
_{c}(z,t)^{\ast}\widehat{c}_{0}^{\dag}(t)+%
{\displaystyle\sum\limits_{k\neq0}}
\psi_{k}(z,t)^{\ast}\widehat{c}_{k}^{\dag}(t)\label{Eq.FieldOprsCondNonCond}\\
\widetilde{\psi}(z,t)  &  =\psi_{c}(z,t)\widetilde{\gamma}_{0}(t)+%
{\displaystyle\sum\limits_{k\neq0}}
\psi_{k}(z,t)\widetilde{\gamma}_{k}(t)\qquad\widetilde{\psi}(z,t)^{+}=\psi
_{c}(z,t)^{\ast}\widetilde{\gamma}_{0}^{+}(t)+%
{\displaystyle\sum\limits_{k\neq0}}
\psi_{k}(z,t)^{\ast}\widetilde{\gamma}_{k}^{+}(t)\nonumber\\
&  \label{Eq.StochFieldFnsCondNonCond}%
\end{align}
where $\widehat{c}_{0}(t)$, $\widehat{c}_{0}^{\dag}(t)$ are condensate mode
annihilation, creation operators and $\widetilde{\gamma}_{0}(t)$,
$\widetilde{\gamma}_{0}^{+}(t)$ are the related stochastic amplitudes. Similar
operators $\widehat{c}_{k}(t)$, $\widehat{c}_{k}^{\dag}(t)$ and stochastic
amplitudes $\widetilde{\gamma}_{k}(t)$, $\widetilde{\gamma}_{k}^{+}(t)$ apply
for the non-condensate modes.

The number of bosons left in the original condensate mode $N_{C}(t)$ is given
by%
\begin{equation}
N_{C}(t)=\left\langle \widehat{c}_{0}^{\dag}\widehat{c}_{0}\right\rangle .
\label{Eq.CondensateNumber}%
\end{equation}
We can now derive an expression for the \textit{quantum depletion} (QD)
\begin{equation}
N_{D}(t)=N-N_{C}(t) \label{Eq.QuantumDepletion}%
\end{equation}
which specifies the loss of bosons from the condensate mode and is another
physical quantity of interest. The quantum depletion gives a measure of the
failure of the mean-field theory

As $\{\widehat{c}_{0}^{\dag}\widehat{c}_{0}\}=\widehat{c}_{0}^{\dag
}\widehat{c}_{0}+1/2$ we can use the Wigner distribution approach to replace
$\left\langle \{\widehat{c}_{0}^{\dag}\widehat{c}_{0}\}\right\rangle $ by
$\overline{\widetilde{\gamma}_{0}^{+}(t)\widetilde{\gamma}_{0}(t)}$ so that we
have
\begin{equation}
N_{C}(t)=\overline{\widetilde{\gamma}_{0}^{+}(t)\widetilde{\gamma}_{0}(t)}-1/2
\label{Eq.QuantumRetention}%
\end{equation}

By using Eq. (\ref{Eq.StochFieldsCondNonCondModes}) for the stochastic fields
$\widetilde{\Psi}(z,t)$, $\widetilde{\Psi}^{+}(z,t)$\ the stochastic phase
amplitudes for the condensate mode are given by
\begin{equation}
\widetilde{\gamma}_{0}(t)=\int dz\,\psi_{c}^{\ast}(z,t)\widetilde{\Psi
}(z,t)\qquad\widetilde{\gamma}_{0}^{+}(t)=\int dz^{\#}\,\psi_{c}%
(z^{\#},t)\widetilde{\Psi}^{+}(z^{\#},t) \label{Eq.StochAmpCondModes}%
\end{equation}
Hence, using Eqs. (\ref{Eq.QCF}) and (\ref{Eq.QCFStochAver}) and $\delta
_{C}(z,z^{\#})=\psi_{c}(z,t)\psi_{c}(z^{\#},t)^{\ast}+%
{\displaystyle\sum\limits_{k\neq c}}
\psi_{k}(z,t)\psi_{k}(z^{\#},t)^{\ast}$ we can then show that
\begin{align}
N_{C}(t)  &  =%
{\displaystyle\int}
{\displaystyle\int}
dz\,dz^{\#}\,\psi_{c}(z^{\#},t)\left\langle \widehat{\Psi}(z^{\#})^{\dag
}\,\widehat{\Psi}(z)\right\rangle \psi_{c}(z,t)^{\ast}\nonumber\\
&  =%
{\displaystyle\int}
{\displaystyle\int}
dz\,dz^{\#}\,\psi_{c}(z^{\#},t)\psi_{c}(z,t)^{\ast}P(z,z^{\#},t)
\label{Eq.CondModeOccupation}%
\end{align}
where $P(z,z^{\#},t)$ is the first-order quantum correlation function. Noting
the result (\ref{Eq.MeanOneBodyProjector}) for the mean value of the one-body
projector, we see that the quantum depletion has a similar relation to the
QCF, but now involving the time-dependent condensate mode function. Taking the
Fourier transform of each side of Eq. (\ref{Eq.CondModeOccupation}) results in
the FT of the quantum depletion not reflecting the same periodicities as in
the QCF due to the presence of the $\psi_{c}(z^{\#},t)\psi_{c}(z,t)^{\ast}%
$factor. However, as noted above the quantum depletion has a different role,
namely indicating whether or not the mean-field theory still applies.

By expressing $P(z,z^{\#},t)$ in terms of natural orbitals via
(\ref{Eq.QCFNaturalOrbitals}) we see that the condensate mode occupancy is
also given by
\begin{equation}
N_{C}(t)=%
{\displaystyle\sum\limits_{i=0}}
p_{i}\times\left\vert
{\displaystyle\int}
dz\,\psi_{c}^{\ast}(z,t)\chi_{i}(z,t)\right\vert ^{2}
\label{Eq.CondModeOccNatModes}%
\end{equation}

Subsituting for the QCF in terms of stochastic fields gives
\begin{align}
N_{C}(t)  &  =%
{\displaystyle\int}
{\displaystyle\int}
dz\,dz^{\#}\,\psi_{c}(z^{\#},t)\psi_{c}(z,t)^{\ast}\left(  \overline
{\widetilde{\psi}(z,t)\widetilde{\psi}^{+}(z^{\#},t)}-\frac{1}{2}\delta
_{C}(z,z^{\#})\right) \nonumber\\
&  =%
{\displaystyle\int}
{\displaystyle\int}
dz\,dz^{\#}\,\psi_{c}(z^{\#},t)\psi_{c}(z,t)^{\ast}\left(  \overline
{\widetilde{\psi}(z,t)\widetilde{\psi}^{+}(z^{\#},t)}\right)  \;-\frac{1}{2}
\label{Eq.StochResultCondModeOccupn}%
\end{align}
where the result $\delta_{C}(z,z^{\#})=\psi_{c}(z,t)\psi_{c}(z^{\#},t)^{\ast}+%
{\displaystyle\sum\limits_{k\neq c}}
\psi_{k}(z,t)\psi_{k}(z^{\#},t)^{\ast}$ has again been used.\smallskip

\subsection{Periodic Driven BEC- Stochastic Phase Amplitudes for Separate
Modes}

\label{SubSection - Stochastic Phase Amplitudes}

In this section we set out the dynamical equations for the stochastic
phase-space amplitudes associated with separate modes that are involved when
the field operator, field function and stochastic field are expanded in terms
of a suitable set of orthogonal mode functions. These stochastic amplitude
equations describe the time evolution of the periodically driven BEC. Later,
we will consider other mode functions (Bogoliubov modes) that are more
suitable for describing the preparation of the BEC in a trap potential before
it is released and subjected to periodic driving. There are, however, several
alternative choices that are all suitable for treating the driven
BEC.\smallskip

\subsubsection{Floquet Modes}

\label{SubSubSection - Floquet Modes}

As we are considering the effect of a periodic driving field on the BEC one
possibility is to choose Floquet modes \cite{Floquet}, \cite{Sacha18b}, which
are essentially single boson wave functions for evolution due to the periodic
field. They satisfy the equation
\begin{equation}
-\frac{\hslash^{2}}{2m}\frac{\partial^{2}}{\partial z^{2}}\phi_{k}%
(z,t)+V(z,t)\phi_{k}(z,t)-i\hbar\frac{\partial}{\partial t}\phi_{k}%
(z,t)=\hslash\nu_{k}\phi_{k}(z,t) \label{Eq.FloquetModes}%
\end{equation}
and are periodic $\phi_{k}(z,t)=\phi_{k}(z,t+T)$ as well as being orthonormal
$%
{\displaystyle\int}
dz\,\phi_{k}(z,t)^{\ast}\phi_{l}(z,t)=\delta_{k,l}$ at all times $t$. The
Floquet frequencies $\nu_{k}$ are time independent, and form zones analogous
to the Brillouin zones that occur for particles moving in a spatially periodic potential.

Thus, we have
\begin{align}
\widehat{\psi}(z)  &  =%
{\displaystyle\sum\limits_{k}}
\widehat{a}_{k}(t)\,\phi_{k}(z,t)\qquad\widehat{\psi}^{\dag}(z)=%
{\displaystyle\sum\limits_{k}}
\widehat{a}_{k}^{\dag}(t)\,\phi_{k}^{\ast}(z,t)\label{Eq.FieldOprs}\\
\psi(z)  &  =%
{\displaystyle\sum\limits_{k}}
\alpha_{k}(t)\,\phi_{k}(z,t)\qquad\psi^{+}(z)=%
{\displaystyle\sum\limits_{k}}
\alpha_{k}^{+}(t)\,\phi_{k}^{\ast}(z,t)\label{Eq.FieldFns}\\
\widetilde{\psi}(z,t)  &  =%
{\displaystyle\sum\limits_{k}}
\widetilde{\alpha}_{k}(t)\,\phi_{k}(z,t)\qquad\widetilde{\psi}^{+}(z,t)=%
{\displaystyle\sum\limits_{k}}
\widetilde{\alpha}_{k}^{+}(t)\,\phi_{k}^{\ast}(z,t) \label{Eq.StochFldFns}%
\end{align}
are the expansions for the field operators, field functions and stochastic
field that occur in the Hamiltonian, the FFPE and the Ito SFE, respectively.
The mode annihilation, creation operators for the Floquet modes are time
dependent, but still satisfy the standard Bose commutation rules
$[\widehat{a}_{k},\widehat{a}_{l}]=$ $[\widehat{a}_{k}^{\dag},\widehat{a}%
_{l}^{\dag}]=0,$ $[\widehat{a}_{k},\widehat{a}_{l}^{\dag}]=\delta_{k,l}$. The
phase-space variables $\alpha_{k},\alpha_{k}^{+}$ that represent the mode
annihilation, creation operators are also time dependent - as are their
related stochastic phase-space variables $\widetilde{\alpha}_{k}%
,\widetilde{\alpha}_{k}^{+}$.

However, using Floquet modes in the numerics has the disadvantage that they
are time dependent, so sets of such modes must be stored on the computer as
functions of time. Instead the numerics are more conveniently carried out
using time-independent gravitational modes. For completeness, we have set out
the equations that would be used for the evolution of stochastic phase
variables for Floquet modes together with Floquet-based expressions for the
position probability density and QCF in Appendix
\ref{Appendix - Floquet Mode Treatment}.\smallskip

\subsubsection{Gravitational Modes}

\label{SubSubSect - Gravit Modes}

Rather than Floquet modes - since we are considering the effect of a static
gravitational field on the BEC - it is convenient to choose gravitational
modes, which are essentially single boson wave-functions for evolution due to
the static gravitational field. These have the advantage of being time
independent as well as being orthogonal. They satisfy the equation%
\begin{equation}
-\frac{\hslash^{2}}{2m}\frac{\partial^{2}}{\partial z^{2}}\xi_{k}%
(z)+mg_{E}z\,\xi_{k}(z)=\hslash\epsilon_{k}\xi_{k}(z) \label{Eq.GravModes}%
\end{equation}

Expansion of the field function and stochastic field function in terms of
these modes gives%
\begin{align}
\psi(z) &  =%
{\displaystyle\sum\limits_{k}}
\eta_{k}\,\xi_{k}(z)\qquad\psi^{+}(z)=%
{\displaystyle\sum\limits_{k}}
\eta_{k}^{+}\,\xi_{k}^{\ast}(z)\label{Eq.FieldFnGravModes}\\
\widetilde{\psi}(z,t) &  =%
{\displaystyle\sum\limits_{k}}
\widetilde{\eta}_{k}(t)\,\xi_{k}(z)\qquad\widetilde{\psi}^{+}(z,t)=%
{\displaystyle\sum\limits_{k}}
\widetilde{\eta}_{k}^{+}(t)\,\xi_{k}^{\ast}(z)\label{Eq.StochFieldGravModes}%
\end{align}
The phase-space variables $\eta_{k},\eta_{k}^{+}$\ that represent the mode
annihilation, creation operators are time independent - whereas their related
stochastic phase-space variables $\widetilde{\eta}_{k},\widetilde{\eta}%
_{k}^{+}$ are time dependent.\smallskip

\paragraph{Position ProbabilityDensity, QCF and Mean Energy - Gravitational
Modes}

The position probability density in Eq. (\ref{Eq.PositProbStochasAver}) can be
expressed in terms of gravitational mode functions as
\begin{equation}
F(z,t)=%
{\displaystyle\sum\limits_{k,l}}
\xi_{k}(z)\,\xi_{l}^{\ast}(z)\,\left[  \overline{\widetilde{\eta}%
_{k}(t)\,\widetilde{\eta}_{l}^{+}(t)}\,-\frac{1}{2}\delta_{k,l}\right]
\label{Eq.PositProbGravModes}%
\end{equation}
and involves the stochastic average of products of stochastic phase-space variables.

Similarly, the QCF in Eq. (\ref{Eq.QCFStochAver}) can also be expressed in
terms of gravitational mode functions (see Eq.(\ref{Eq.QCFNaturalOrbitals}))
as%
\begin{equation}
P(z,z^{\#},t)=%
{\displaystyle\sum\limits_{k,l}}
\xi_{k}(z)\,\xi_{l}^{\ast}(z^{\#})\,\left[  \overline{\widetilde{\eta}%
_{k}(t)\,\widetilde{\eta}_{l}^{+}(t)}\,-\frac{1}{2}\delta_{k,l}\right]
\label{Eq.QCFGravModes}%
\end{equation}

Furthermore, the mean energy in Eq. (\ref{Eq.MeanEnergyStochastic}) can be
simplified by using the expansion of the stochastic field functions
(\ref{Eq.StochFieldGravModes}) in terms of gravitational modes along with the
defining Eq. (\ref{Eq.GravModes}) for these modes. We have%
\begin{align}
\left\langle \widehat{H}\right\rangle  &  =%
{\displaystyle\sum\limits_{k=0}}
\hslash\epsilon_{k}\,\left(  \overline{\widetilde{\eta}_{k}%
(t)\,\widetilde{\eta}_{k}^{+}(t)}\,-\frac{1}{2}\right) \nonumber\\
&  -mg_{E}\lambda\cos\omega t\,%
{\displaystyle\int}
dz\,z\,\left(  \overline{\widetilde{\psi}^{+}(z,t)\,\widetilde{\psi}%
(z,t)}-\frac{1}{2}\delta_{C}(z,z)\right) \nonumber\\
&  +\frac{g}{2}%
{\displaystyle\int}
dz\,\left(  \overline{\widetilde{\psi}^{+}(z,t)\,\widetilde{\psi}%
^{+}(z,t)\,\widetilde{\psi}(z,t)\,\widetilde{\psi}(z,t)}-2\delta
_{C}(z,z)\,\overline{\widetilde{\psi}^{+}(z,t)\,\widetilde{\psi}(z,t)}%
+\frac{1}{2}\delta_{C}(z,z)^{2}\right) \nonumber\\
&  \label{Eq.MeanEnergyGravModes}%
\end{align}
The derivation is set out in Appendix \ref{Appendix - Mean Energy}. \smallskip

\paragraph{Evolution of Stochastic Phase Variables for Gravitational Modes}

Coupled equations for the stochastic phase-space variables $\widetilde{\eta
}_{k},\widetilde{\eta}_{k}^{+}$ can be obtained allowing for the effect of the
oscillating potential $mg_{E}z\,\lambda\sin\omega t$ and the boson-boson
interactions. They can be written in the form
\begin{align}
\frac{\partial}{\partial t}\widetilde{\eta}_{k}  &  =-i\epsilon_{k}%
\widetilde{\eta}_{k}-i\frac{g}{\hbar}%
{\displaystyle\sum\limits_{n}}
E_{k,n}\,\widetilde{\eta}_{n}-i\frac{g}{\hbar}%
{\displaystyle\sum\limits_{n}}
F_{k,n}\,\widetilde{\eta}_{n}\label{Eq.StochPhaseGravA}\\
\frac{\partial}{\partial t}\widetilde{\eta}_{k}^{+}  &  =+i\epsilon
_{k}\widetilde{\eta}_{k}^{+}+i\frac{g}{\hbar}%
{\displaystyle\sum\limits_{n}}
E_{k,n}^{+}\,\widetilde{\eta}_{n}^{+}+i\frac{g}{\hbar}%
{\displaystyle\sum\limits_{n}}
F_{k,n}^{\ast}\,\widetilde{\eta}_{n}^{+} \label{Eq.StochPhaseGravB}%
\end{align}
where
\begin{align}
E_{k,n}  &  =%
{\displaystyle\int}
dz\,\xi_{k}^{\ast}(z)\left(  \widetilde{\psi}^{+}(z,t)\,\widetilde{\psi
}(z,t)\,-\delta_{C}(z,z)\right)  \,\xi_{n}(z)\nonumber\\
E_{k,n}^{+}  &  =%
{\displaystyle\int}
dz\,\xi_{k}(z)\left(  \widetilde{\psi}^{+}(z,t)\,\widetilde{\psi
}(z,t)\,-\delta_{C}(z,z)\right)  \,\xi_{n}^{\ast}(z)\label{Eq.EMatrices}\\
F_{k,n}  &  =%
{\displaystyle\int}
dz\,\xi_{k}^{\ast}(z)\left(  mg_{e}z\,\lambda\sin\omega t\right)  \,\xi_{n}(z)
\label{Eq.FMatrix}%
\end{align}
Note that $F_{k,n}$ is periodic with period $T$. The time dependence of the
$E_{k,n}$ and $E_{k,n}^{+}$ will arise from the time dependence of the
$\widetilde{\eta}_{k}$ and $\widetilde{\eta}_{k}^{+}$ via the stochastic field
functions, and their time dependence will ultimately depend on $T$ and the
$\epsilon_{k}$. In this method the stochastic fields are determined at each
time point from Eq. (\ref{Eq.StochFieldGravModes}), which then can also be
used to determine the quantum depletion (see Eq. (\ref{Eq.StochAmpCondModes}%
)).\smallskip

\paragraph{Initial Conditions - Gravitational Modes}

Initial conditions in terms of gravitational modes are
\begin{equation}
\widetilde{\psi}(z,0)=%
{\displaystyle\sum\limits_{k}}
\widetilde{\eta}_{k}(0)\,\xi_{k}(z)\qquad\widetilde{\psi}^{+}(z,0)=%
{\displaystyle\sum\limits_{k}}
\widetilde{\eta}_{k}^{+}(0)\,\xi_{k}^{\ast}(z) \label{Eq.InitCondGravModes}%
\end{equation}
This requires a consideration of how to treat the preparation process, as will
be explained below.\smallskip

\subsubsection{Wannier Modes}

\label{SubSubSection Wannier Modes}

Yet another choice for orthonormal mode functions that could be used to treat
the time crystal topic is the Wannier modes. These are defined as linear
combinations of the Floquet modes with frequencies in the first Floquet zone,
and are analogous to the Wannier functions defined for spatially periodic
potentials as linear combinations of Bloch functions associated with the first
Brillouin zone. Spatial Wannier functions are spatially localised around
different spatial lattice points; temporal Wannier functions are localised in
time around different time lattice points $nT$ (see Ref \cite{Sacha15c}). The
Wannier modes are
\begin{equation}
\Phi_{nT}(z,t)=\mathcal{N}%
{\textstyle\sum\limits_{\{\nu_{k}\}}}
\,\exp(-i\nu_{k}(t-nT))\;\phi_{k}(z,t) \label{Eq.WannierModes}%
\end{equation}
where the sum is over Floquet modes in the first Floquet zone and $N$ is a
normalizing factor, and $n$ is an integer. If $m$ is another integer the
periodic properties of the Floquet modes leads to the result $\Phi
_{(n+m)T}(z,t+mT)=\Phi_{nT}(z,t)$, which shows that for a given position $z$
the Wannier mode $\Phi_{nT}(z,t)$ is a function of $(t-nT)$. This indicates
that the Wannier mode is centred in time around $nT$. It may be temporally
localised.\textbf{ }

The Wannier modes satisfy the approximate\textbf{ }orthogonality condition
\begin{equation}%
{\textstyle\int}
dz\,\Phi_{nT}(z,t)^{\ast}\,\Phi_{mT}(z,t)=\mathcal{N}^{2}%
{\textstyle\sum\limits_{\{\nu_{k}\}}}
\exp(-i\nu_{k}(m-n)T)\approx\delta_{n,m} \label{Eq.OrthogWannier}%
\end{equation}
if the first Floquet zone is broad.

Approximate versions of Wannier modes are also used (Ref. \cite{Sacha15a}),
based on a two-mode approximation to the solutions of the Gross-Pitaevski
equation. They are two linear combinations of two Floquet modes designated as
$\phi_{1}(z,t)$ and $\phi_{2}(z,t)$. For effects involving a period $sT$ we
choose
\begin{align}
\Phi_{1}(z,t)  &  =\frac{1}{\sqrt{2}}(\phi_{1}(z,t)+\exp(-i\frac{2\pi}%
{sT}t)\,\phi_{2}(z,t))\nonumber\\
\Phi_{2}(z,t)  &  =\frac{1}{\sqrt{2}}(\phi_{1}(z,t)-\exp(-i\frac{2\pi}%
{sT}t)\,\phi_{2}(z,t)) \label{Eq.ApproxWannier}%
\end{align}
These functions repeat over a period $sT$ and are orthogonal. The two Floquet
modes may be chosen to be similar to the condensate mode function at $t=0,$
enabling the initial conditions to be described in terms of these approximate
Wannier modes plus non-condensate modes such as Bogoliubov modes.\smallskip

\subsection{Preparation of the BEC - Condensate and Bogoliubov Modes}

\label{SubSection - Preparation of BEC}

As described in Sect. \ref{Section - Introduction}, the BEC is prepared in the
standard way in a trap potential before it is allowed to bounce on the
oscillating mirror. The description of the states for the initially prepared
BEC can be treated as a time-independent problem via variational methods based
on minimising the mean value $\left\langle \widehat{H}\right\rangle $ of the
energy, subject to constraints such as the mean boson number $\left\langle
\widehat{N}\right\rangle $ being $N$. Here, we are only interested in being
able to represent the initial BEC state in terms of the Wigner distribution
just after the trap is switched off. The initial state of the BEC for which
$N_{c}$ bosonic atoms mainly occupy a single condensate mode may be described
theoretically in several ways,\ including the single-mode approximation. In
the single-mode approximation all the bosons are assumed to occupy just one
mode. Here the condensate bosons are described via a mean-field approach, with
the condensate wave function $\Phi_{c}(z)$ obtained from the variational
approach as the solution of a time-independent GPE, but with the potential
energy term now given by the time-independent trapping potential $V_{trap}(z)$
and including the chemical potential $\mu$ to allow for the constraint that
$\left\langle \widehat{N}\right\rangle =N$ (see Eqs (\ref{Eq.GPECondensPrepnA}%
), (\ref{Eq.FockStateT0Props}) below). However, collisional interactions cause
bosons to be transferred from the condensate mode to non-condensate modes, so
the treatment of the initial state should allow for this.

One widely used approach is to use Bogoliubov modes \cite{Bogoliubov},
\cite{Fetter72}, \cite{Morgan00}, \cite{Dalton15a}, \cite{King19},
\cite{Lewenstein}, \cite{Proukakis08} to describe the non-condensate modes.
The field operator is written as the sum of condensate and fluctuation fields.
For the fluctuation field Bogoliubov mode operators and their mode functions
are introduced to enable an approximation to the Hamiltonian in Eq
(\ref{Eq.HamiltonianFieldModel}) that is correct to second-order in the
fluctuation field to be expressed as the sum of a condensate mode term and
terms describing non-interacting quantum harmonic oscillators (see Appendix
\ref{Appendix - Bogoliubov Theory} for details). First-order terms in the
fluctuation field are eliminated since the condensate field satisfies the
time-independent GPE. The mode functions are determined from Bogoliubov-
de-Gennes (BDG) equations based on the condensate wave function $\Phi_{c}$.
Bogoliubov modes thus allow for quantum fluctuations from the mean-field
theory described by the GPE and the condensate wave-function. This model will
be used to determine the initial stochastic\textbf{\ }quantities. The
description in terms of Bogoliubov modes for the BEC preparation is then
matched to the initial conditions for the BEC falling onto the oscillating
mirror, where the periodically driven BEC is treated as above in terms of
Floquet or gravitational modes.\smallskip

\subsubsection{Condensate Wave Function, Condensate and Non-Condensate Fields}

The condensate wave-function $\Phi_{c}(z)$ is determined from the
time-independent GPE
\begin{equation}
\left[  -\frac{\hslash^{2}}{2m}\frac{\partial^{2}}{\partial z^{2}}%
+V_{trap}(z)+g\{\Phi_{c}(z)^{\ast}\Phi_{c}(z)\}\right]  \,\Phi_{c}%
(z)=\mu\,\Phi_{c}(z) \label{Eq.GPECondensPrepnA}%
\end{equation}
where $\mu$ is the chemical potential. The condensate wave-function is
normalised in terms of the number $N$ of bosons in the condensate, as $%
{\displaystyle\int}
dz\,\Phi_{c}(z)^{\ast}\Phi_{c}(z)=N$.

In Bogoliubov theory the field operator $\hat{\Psi}(z)$ is first written as
the sum of the condensate field $\widehat{\Psi}_{c}(z)$ and a fluctuation (or
non-condensate) field $\delta\widehat{\Psi}(z)$ as
\begin{equation}
\hat{\Psi}(z)=\widehat{\Psi}_{c}(z)+\delta\widehat{\Psi}(z)
\label{Eq.FieldOpr1}%
\end{equation}
where the condensate field and the fluctuation field are given by
\begin{align}
\widehat{\Psi}_{c}(z)  &  =\widehat{c}_{0}\psi_{c}(z)\label{Eq.CondFieldOpr}\\
\delta\widehat{\Psi}(z)  &  =\sum_{i\neq C}\widehat{c}_{i}\psi_{i}(z)
\label{Eq.FluctFieldOpr3}%
\end{align}

The condensate mode annihilation operator is $\widehat{c}_{0}$ and the
condensate mode function is
\begin{equation}
\psi_{c}(z)=\Phi_{c}(z)/\sqrt{N} \label{Eq.CondensateModeFn}%
\end{equation}
which is normalised to unity \cite{Fetter72}. The equation for the condensate
mode function is
\begin{equation}
\left[  -\frac{\hslash^{2}}{2m}\frac{\partial^{2}}{\partial z^{2}}%
+V_{trap}(z)+gN\{\psi_{c}(z)^{\ast}\psi_{c}(z)\}\right]  \,\psi_{c}%
(z)=\mu\,\psi_{c}(z) \label{Eq.CondensateModeEqn}%
\end{equation}
This equation is used to eliminate terms in the Hamiltonian that are linear in
the fluctuation field. Note that the time variable $t$ \ is no longer present.

The fluctuation field may be expanded in terms of any suitable standard set of
normalised mode functions $\psi_{i}(z)$, where the corresponding
non-condensate mode annihilation operator is $\widehat{c}_{i}$. The
non-condensate modes are required to be orthogonal to $\psi_{c}(z)$, so we
have the constraint%

\begin{equation}
\int dz\,\psi_{c}(z)^{\ast}\delta\widehat{\Psi}(z)=0 \label{Eq.CondModeOrthog}%
\end{equation}
It follows that the non-zero commutation result for the fluctuation field is
\begin{equation}
\lbrack\delta\widehat{\Psi}(z),\delta\widehat{\Psi}(z^{\#})^{\dag}%
]=\delta(z-z^{\#})-\psi_{c}(z)\psi_{c}(z^{\#})^{\ast} \label{Eq.CommRule}%
\end{equation}

The choice of the non-condensate mode functions is arbitrary so far, but one
choice would be the normalised eigenfunctions of the \emph{Gross-Pitaevskii
operator} $\left(  -\frac{\hslash^{2}}{2m}\frac{\partial^{2}}{\partial z^{2}%
}+V_{trap}(z)+g\{\Phi_{c}(z)^{\ast}\Phi_{c}(z)\}\right)  $\ with eigenvalues
$\mu_{i}\neq\mu$. However, in the present paper we will use Bogoliubov modes
$u_{k}(z),v_{k}(z)$ rather than Gross-Pitaevski modes.\smallskip

\subsubsection{Bogoliubov Modes}

The fluctuation field operator is expanded in terms of annihilation, creation
operators $\widehat{b}_{k},\widehat{b}_{k}^{\dag}$ for Bogoliubov modes and
associated mode functions $u_{k}(z),v_{k}(z)$ as
\begin{equation}
\delta\widehat{\Psi}(z)=%
{\displaystyle\sum\limits_{k\neq0}}
\left[  u_{k}(z)\,\widehat{b}_{k}-v_{k}(z)^{\ast}\,\widehat{b}_{k}^{\dag
}\right]  \qquad\delta\widehat{\Psi}(z)^{\dag}=%
{\displaystyle\sum\limits_{k\neq0}}
\left[  -v_{k}(z)\,\widehat{b}_{k}+u_{k}(z)^{\ast}\,\widehat{b}_{k}^{\dag
}\right]  \label{Eq.FluctFieldOpr}%
\end{equation}
The summation over $k\neq0$ is to indicate that the condensate mode
$\widehat{c}_{0}$ is excluded, since the fluctuation field is only intended to
treat non-condensate modes. The orthogonality condition
(\ref{Eq.CondModeOrthog}) then leads to the following orthogonality conditions
between $u_{k}(z),v_{k}(z)$ and the condensate mode function.
\begin{align}
\int dz\,\psi_{c}(z)^{\ast}u_{k}(z)  &  =0\nonumber\\
\int dz\,\psi_{c}(z)\,v_{k}(z)  &  =0 \label{Eq.BogolCondModeOrthog}%
\end{align}
The commutation rules (\ref{Eq.CommRule}) for $\delta\widehat{\Psi}%
(z),\delta\widehat{\Psi}(z^{\#})^{\dag}$ together with the requirement that
$\widehat{b}_{k},\widehat{b}_{k}^{\dag}$ satisfy standard Bose commutation
rules lead to the following conditions for the $u_{k}(z),v_{k}(z)$%
\begin{align}%
{\displaystyle\sum\limits_{k\neq0}}
\left[  u_{k}(z)\,v_{k}(z^{\#})^{\ast}-v_{k}(z)^{\ast}\,u_{k}(z^{\#})\right]
&  =0\nonumber\\%
{\displaystyle\sum\limits_{k\neq0}}
\left[  u_{k}(z)\,u_{k}(z^{\#})^{\ast}-v_{k}(z)^{\ast}\,v_{k}(z^{\#})\right]
&  =\delta(z-z^{\#})-\psi_{c}(z)\psi_{c}(z^{\#})^{\ast} \label{Eq.CondUVMkA}%
\end{align}
The inverse relations giving the $\widehat{b}_{k},\widehat{b}_{k}^{\dag}$ in
terms of the fluctuation fields are
\begin{align}
\widehat{b}_{k}  &  =%
{\displaystyle\int}
dz\,\left[  u_{k}(z)\,\delta\widehat{\Psi}(z)+v_{k}(z)^{\ast}\,\delta
\widehat{\Psi}(z)^{\dag}\right]  \qquad k\neq0\nonumber\\
\widehat{b}_{k}^{\dag}  &  =%
{\displaystyle\int}
dz\,\left[  v_{k}(z)\,\delta\widehat{\Psi}(z)+u_{k}(z)^{\ast}\,\delta
\widehat{\Psi}(z)^{\dag}\right]  \qquad k\neq0 \label{Eq.BogolModeOprs}%
\end{align}
which is shown using Eqs (\ref{Eq.FluctFieldOpr}), (\ref{Eq.CondUVMkA}) and
(\ref{Eq.CondModeOrthog}). Further conditions on $u_{k}(z),v_{k}(z)$ follow
using the commutation rules for $\widehat{b}_{k},\widehat{b}_{k}^{\dag}$ and
$\delta\widehat{\Psi}(z),\delta\widehat{\Psi}(z^{\#})^{\dag}$ and the
orthogonality conditions (\ref{Eq.BogolCondModeOrthog}) in conjunction with
Eq. (\ref{Eq.BogolModeOprs}). These are%
\begin{align}%
{\displaystyle\int}
dz\,\left[  u_{k}(z)\,v_{l}(z)\,-v_{k}(z)\,u_{l}(z)\right]   &  =0\qquad
k,l\neq0\nonumber\\%
{\displaystyle\int}
dz\,\left[  u_{k}(z)^{\ast}\,u_{l}(z)\,-v_{k}(z)^{\ast}\,v_{l}(z)\right]   &
=\delta_{k,l}\qquad k,l\neq0 \label{Eq.BiorthogCond}%
\end{align}
This shows that the Bogoliubov mode functions $u_{k}(z),v_{k}(z)$ do not
satisfy standard orthogonality and normalisation conditions -
\emph{biorthogonality} conditions apply instead \cite{Lewenstein}. For each
Bogoliubov frequency $\omega_{k}$ we note that there are two mode functions
involved.\smallskip

\subsubsection{Generalised Bogoliubov-de Gennes Equations}

The Bogoliubov mode functions $u_{k}(z),v_{k}(z)$ and the mode frequencies
$\omega_{k}$ are chosen to satisfy a\emph{\ generalised} form \cite{Morgan00}
of the Bogoliubov-de Gennes (BDG) equations%
\begin{align}
(%
\mathcal{L}%
+g\,n_{C}(z))\,u_{k}(z)-g\,\Phi_{c}(z)^{2}\,v_{k}(z)  &  =\hslash\omega
_{k}\,u_{k}(z)+C_{k}\psi_{c}(z)\nonumber\\
-g\,\Phi_{c}^{\ast}(z)^{2}\,u_{k}(z)+(%
\mathcal{L}%
+g\,n_{C}(z))\,v_{k}(z)  &  =-\hslash\omega_{k}\,v_{k}(z)-C_{k}\psi_{c}^{\ast
}(z) \label{Eq. BogolDeGennes}%
\end{align}
where $k\neq0$. The differential operator $%
\mathcal{L}%
$ is defined by
\begin{equation}%
\mathcal{L}%
=\left[  -\frac{\hslash^{2}}{2m}\frac{\partial^{2}}{\partial z^{2}}%
+V_{trap}(z)+g\,n_{C}(z)-\mu\right]  \label{Eq.DiflferentialOprL}%
\end{equation}
and the quantity $C_{k}$ is given by%
\begin{equation}
C_{k}=C_{k}[u_{k}(z),v_{k}(z)]=\int dz\,g\,n_{C}(z)\,(\psi_{c}^{\ast
}(z)\,u_{k}(z)-\psi_{c}(z)\,v_{k}(z)) \label{Eq.C}%
\end{equation}
Note that the GPE (\ref{Eq.GPECondensPrepnA}) can be written $%
\mathcal{L}%
\,\Phi_{c}(z)=%
\mathcal{L}%
\,\psi_{c}(z)=0$. The generalised BDG equations are integro-differential
equations, since $C_{k}$ is a functional of $u_{k}(z),v_{k}(z)$. The equations
depend only on quantities obtained from the condensate wave function $\Phi
_{c}(z)$, such as $n_{C}(z)$, $\Phi_{c}(z)^{2}$ and $\psi_{c}(z)$.
Fortunately, the $u_{k}(z),v_{k}(z)$ and the Bogoliubov frequencies
$\omega_{k}$ can be obtained from eigenmodes of the standard BDG equations,
which are eigenvalue equations only involving differential operators. Note
that the sign convention differs from that in Ref \cite{Morgan00}.

It can be shown from the generalised BDG equations (\ref{Eq. BogolDeGennes})
that the eigenvalues $\omega_{k}$ are real, and that the $u_{k}(z),v_{k}(z)$
satisfy the required orthogonality conditions (\ref{Eq.BogolCondModeOrthog})
with the condensate mode function $\psi_{c}(z)$. The inclusion of the term
involving $C_{k}$ is necessary to ensure orthogonality. By convention
$\omega_{k\text{ }}$is taken to be positive. The proofs make use of the
Hermitian properties of $%
\mathcal{L}%
$, where $%
{\displaystyle\int}
dz\,U^{\ast}\,(%
\mathcal{L}%
\,V)=%
{\displaystyle\int}
dz\,(%
\mathcal{L}%
\,U^{\ast})\,V$ and $U,V$ are arbitrary functions. Furthermore, if
$u_{k},v_{k}$ are solutions for $\omega_{k}$, then $v_{k}^{\ast},u_{k}^{\ast}$
are solutions for $-\omega_{k}$. The proof involves the property of $C_{k}$
that $C_{k}[u_{k}(z),v_{k}(z)]^{\ast}=-C_{k}[v_{k}^{\ast}(z),u_{k}^{\ast}%
(z)]$. The generalised BDG\ equations seem to have a solution for $\omega
_{0}=0$ with $u_{0}(z)=\psi_{c}(z)$ and $v_{0}(z)=\psi_{c}^{\ast}(z)$.
However, this is inconsistent with Eq. (\ref{Eq.BogolCondModeOrthog}), based
on the non-condensate field being required to be orthogonal to the condensate
mode (see Eq.(\ref{Eq.CondModeOrthog})).

The second-order approximation to the Hamiltonian
(\ref{Eq.HamiltonianFieldModel}) can be written in terms of the Bogoliubov
mode operators $\widehat{b}_{k},\widehat{b}_{k}^{\dag}$ as the sum of
uncoupled harmonic oscillators, one for each Bogoliubov mode. The form is
given in Appendix \ref{Appendix - Bogoliubov Theory} as
Eq.(\ref{Eq.BogolGrandCanHamilt}).\smallskip

\subsubsection{Standard Bogoliubov-de Gennes Equations}

The standard Bogoliubov-de Gennes equations can be obtained by replacing the
modes $u_{k},v_{k}$ by new modes $\overline{u}_{k},\overline{v}_{k}$ via the
expressions
\begin{equation}
u_{k}=\overline{u}_{k}-\frac{C_{k}\psi_{c}}{\hslash\omega_{k}}\qquad\qquad
v_{k}=\overline{v}_{k}-\frac{C_{k}\psi_{c}^{\ast}}{\hslash\omega_{k}}
\label{Eq.StandardModesB}%
\end{equation}
Substituting into the generalised BDG equations we then obtain the
\emph{standard} BDG equations \cite{Blakie08}
\begin{align}
(%
\mathcal{L}%
+g\,n_{C}(z))\,\overline{u}_{k}(z)-g\,\Phi_{c}(z)^{2}\,\overline{v}_{k}(z)  &
=\hslash\omega_{k}\,\overline{u}_{k}(z)\nonumber\\
-g\,\Phi_{c}^{\ast}(z)^{2}\,\overline{u}_{k}(z)+(%
\mathcal{L}%
+g\,n_{C}(z))\,\overline{v}_{k}(z)  &  =-\hslash\omega_{k}\,\overline{v}%
_{k}(z) \label{Eq.StandardBogDeGennes}%
\end{align}
where the orthogonality conditions (\ref{Eq.BogolCondModeOrthog}) have been
used. \ The quantity $C_{k}$ is now also given by
\begin{equation}
C_{k}=\int dz\,g\,n_{C}(z)\,(\psi_{c}^{\ast}(z)\,\overline{u}_{k}(z)-\psi
_{c}(z)\,\overline{v}_{k}(z)) \label{Eq.AltExpC}%
\end{equation}
in terms of these modes. A generalised orthogonality condition $\int
dz\,(\psi_{c}^{\ast}(z)\,\overline{u}_{k}(z)-\psi_{c}(z)\,\overline{v}%
_{k}(z))=0$ follows from the standard BDG equations, but this does not mean
that the new modes $\overline{u}_{k},\overline{v}_{k}$ individually satisfy
the required orthogonality conditions (\ref{Eq.BogolCondModeOrthog}) with the
condensate mode function.

We can therefore obtain the true $u_{k},v_{k}$ by using the standard BDG
equations (\ref{Eq.StandardBogDeGennes}) to first determine the $\overline
{u}_{k},\overline{v}_{k}$ and then use (\ref{Eq.StandardModesB}) to determine
the $u_{k},v_{k}$. Since the $u_{k},v_{k}$ satisfy the orthogonality
conditions (\ref{Eq.BogolCondModeOrthog}), we can \ use this feature to
determine the quantity $C_{k}/\hslash\omega_{k}$. We find that
\begin{equation}
D_{k}=\frac{C_{k}}{\hslash\omega_{k}}=\int dz\,\psi_{c}^{\ast}(z)\,\overline
{u}_{k}(z)=\int dz\,\psi_{c}(z)\,\overline{v}_{k}(z) \label{Eq.Dk}%
\end{equation}
\smallskip

\subsubsection{Field Operators - Condensate and Bogoliubov Modes}

This gives the following expressions for the field operators after introducing
the $\overline{u}_{k},\overline{v}_{k}$ into the expressions
(\ref{Eq.FieldOpr1}) for the fluctuation field. We have%
\begin{align}
\hat{\Psi}(z)  &  =\widehat{c}_{0}\psi_{c}(z)+%
{\displaystyle\sum\limits_{k\neq0}}
\left[  \left(  \overline{u}_{k}(z)-D_{k}\psi_{c}(z)\right)  \,\widehat{b}%
_{k}-\left(  \overline{v}_{k}(z)-D_{k}\psi_{c}^{\ast}(z)\right)  ^{\ast
}\,\widehat{b}_{k}^{\dag}\right] \nonumber\\
\hat{\Psi}(z)^{\dag}  &  =\widehat{c}_{0}^{\dag}\psi_{c}^{\ast}(z)+%
{\displaystyle\sum\limits_{k\neq0}}
\left[  -\left(  \overline{v}_{k}(z)-D_{k}\psi_{c}^{\ast}(z)\right)
\,\widehat{b}_{k}+\left(  \overline{u}_{k}(z)-D_{k}\psi_{c}(z)\right)  ^{\ast
}\,\widehat{b}_{k}^{\dag}\right]  \label{Eq.FieldOprsBogolModes}%
\end{align}
Similar equations to (\ref{Eq.FieldOprsBogolModes}) are set out in Ref.
\cite{Blakie08}. We can use these expressions to relate the stochastic field
functions at the initial time. The Bogoliubov modes $\overline{u}%
_{k},\overline{v}_{k}$ involved are those obtained from the standard BDG
equations (\ref{Eq.StandardBogDeGennes}).\smallskip

\subsubsection{Field Functions - \ Condensate and Bogoliubov Modes}

\label{SubSubSection - Field Fns Cond and Bogol Modes}

As an alternative to Floquet or gravitational modes, the field functions
$\Psi(z)$, $\Psi^{+}(z)$\ and the stochastic field functions can also be
expanded in terms of Bogoliubov modes. The mode annihilation, creation
operators $\widehat{b}_{k},\widehat{b}_{k}^{\dag}$ are represented by
phase-space variables $\beta_{k},\beta_{k}^{+}$, which are replaced by
stochastic variables $\widetilde{\beta}_{k},\widetilde{\beta}_{k}^{+}$.
Similarly, the condensate mode annihilation, creation operators $\widehat{c}%
_{0},\widehat{c}_{0}^{\dag}$ are represented by phase-space variables
$\gamma_{0},\gamma_{0}^{+}$, which are replaced by stochastic variables
$\widetilde{\gamma}_{0},\widetilde{\gamma}_{0}^{+}$. Based on Eqs.
(\ref{Eq.FieldOprsBogolModes}) and\textbf{\ }re-introducing the original
Bogoliubov modes we have \ \
\begin{align}
\psi(z)  &  =\gamma_{0}\psi_{c}(z)+%
{\displaystyle\sum\limits_{k\neq0}}
\left[  u_{k}(z)\,\beta_{k}-v_{k}(z)^{\ast}\,\beta_{k}^{+}\right] \nonumber\\
\psi^{+}(z)  &  =\gamma_{0}^{+}\psi_{c}^{\ast}(z)+%
{\displaystyle\sum\limits_{k\neq0}}
\left[  -v_{k}(z)\,\beta_{k}+u_{k}(z)^{\ast}\,\beta_{k}^{+}\right]
\label{Eq.FieldsBogolModes}\\
\widetilde{\psi}(z)  &  =\widetilde{\gamma}_{0}\psi_{c}(z)+%
{\displaystyle\sum\limits_{k\neq0}}
\left[  u_{k}(z)\,\widetilde{\beta}_{k}-v_{k}(z)^{\ast}\,\widetilde{\beta}%
_{k}^{+}\right] \nonumber\\
\widetilde{\psi}^{+}(z)  &  =\widetilde{\gamma}_{0}^{+}\psi_{c}^{\ast}(z)+%
{\displaystyle\sum\limits_{k\neq0}}
\left[  -v_{k}(z)\,\widetilde{\beta}_{k}+u_{k}(z)^{\ast}\,\widetilde{\beta
}_{k}^{+}\right]  \label{Eq.StochFieldsBogolModesB}%
\end{align}
where $u_{k}(z),v_{k}(z)$ are given in Eqs, (\ref{Eq.StandardModesB}) and
(\ref{Eq.Dk}). Similar equations to (\ref{Eq.StochFieldsBogolModesB}) are set
out in Ref. \cite{Blakie08}. Note here that none of the stochastic quantities
are time dependent.\smallskip

\subsection{Stochastic Field Functions during Driven Evolution}

During the evolution of the system of interacting bosons driven by the
oscillating potential and bouncing under the influence of the gravitation
field, the stochastic field functions and their expansion coefficients (or
stochastic mode amplitudes) are time dependent. In the cases of Floquet modes
and gravitational modes the stochastic field functions are given in Eqs.
(\ref{Eq.StochFldFns}) and (\ref{Eq.StochFieldGravModes}), respectively.

However, during evolution the field operators, field functions and stochastic
field functions can also be expressed in terms of the time-dependent
condensate mode function $\psi_{c}(z,t)$ and any set of orthogonal
non-condensate mode functions $\psi_{i}(z,t)$ - as in Eq.
(\ref{Eq.FluctFieldOpr3}) for the fluctuation field operator. Thus, for the
stochastic field functions we have (see Eq. (\ref{Eq.StochFieldFnsCondNonCond}%
))%
\begin{align}
\widetilde{\psi}(z,t)  &  =\widetilde{\gamma}_{0}(t)\psi_{c}(z,t)+%
{\displaystyle\sum\limits_{i\neq0}}
\widetilde{\gamma}_{i}(t)\psi_{i}(z,t)\nonumber\\
\widetilde{\psi}^{+}(z,t)  &  =\widetilde{\gamma}_{0}^{+}(t)\psi_{c}^{\ast
}(z,t)+%
{\displaystyle\sum\limits_{i\neq0}}
\widetilde{\gamma}_{i}^{+}(t)\psi_{i}^{\ast}(z,t)
\label{Eq.StochFieldsCondNonCondModes}%
\end{align}
The stochastic mode amplitudes are $\widetilde{\gamma}_{0}%
(t),\widetilde{\gamma}_{0}^{+}(t)$\ for the condensate mode and
$\widetilde{\gamma}_{i}(t),\widetilde{\gamma}_{i}^{+}(t)$ for the
non-condensate modes. These expressions can be equated to those in Eqs.
(\ref{Eq.StochFieldsBogolModesB}), (\ref{Eq.StochFldFns}) and
(\ref{Eq.StochFieldGravModes}) at $t=0$. In regard to the last form we have
$\widetilde{\gamma}_{0}(0)=\widetilde{\gamma}_{0},$\ $\widetilde{\gamma}%
_{0}^{+}(0)=\widetilde{\gamma}_{0}^{+}$ in view of $\psi_{c}(z,0)=\psi_{c}(z)$
- hence the same notation.\smallskip

\subsubsection{Initial Conditions for Stochastic Amplitudes of Gravitational
Modes}

\label{SubSect Initial Cond Stoch Amp Grav Modes}

The initial conditions for the gravitational mode approach can be obtained
from the Bogoliubov mode approach via equating Eqs.
(\ref{Eq.StochFieldGravModes}) and (\ref{Eq.StochFieldsBogolModesB}) at $t=0$
and using the orthogonality properties of the gravitational modes. We find
that
\begin{align}
&  \widetilde{\eta}_{k}(0)=%
{\textstyle\int}
dz\,\xi_{k}^{\ast}(z)\,\left[  \widetilde{\gamma}_{0}\psi_{c}(z)+%
{\displaystyle\sum\limits_{k\neq0}}
\left[  u_{k}(z)\,\widetilde{\beta}_{k}-v_{k}(z)^{\ast}\,\widetilde{\beta}%
_{k}^{+}\right]  \right] \nonumber\\
&  \widetilde{\eta}_{k}^{+}(0)=%
{\textstyle\int}
dz\,\xi_{k}(z)\,\left[  \widetilde{\gamma}_{0}^{+}\psi_{c}^{\ast}(z)+%
{\displaystyle\sum\limits_{k\neq0}}
\left[  -v_{k}(z)\,\widetilde{\beta}_{k}+u_{k}(z)^{\ast}\,\widetilde{\beta
}_{k}^{+}\right]  \right] \nonumber\\
&  \label{Eq. StochFloqBogolRelns}%
\end{align}
where $u_{k}(z),v_{k}(z)$ are given in Eqs. (\ref{Eq.StandardModesB}) and
(\ref{Eq.Dk}). This shows how a particular choice of stochastic phase-space
variables $\widetilde{\gamma}_{0},\widetilde{\gamma}_{0}^{+},$
$\widetilde{\beta}_{q},\widetilde{\beta}_{q}^{+}$ for the Bogoliubov mode
treatment of the BEC preparation can be translated into the choice of
stochastic phase-space variables $\widetilde{\eta}_{k}(0),\widetilde{\eta}%
_{k}^{+}(0)$ for the gravitational mode treatment of the periodically driven
BEC. Essentially, the $\widetilde{\eta}_{k}(0),\widetilde{\eta}_{k}^{+}(0)$
for the gravitational mode treatment are linearly dependent on the
$\widetilde{\gamma}_{0},\widetilde{\gamma}_{0}^{+},\widetilde{\beta}%
_{q},\widetilde{\beta}_{q}^{+}$.

We then see that we can write the initial stochastic fields for gravitational
mode evolution as
\begin{equation}
\widetilde{\psi}(z,0)=\widetilde{\gamma}_{0}(0)\psi_{c}(z)+\delta
\widetilde{\psi}(z,0)\qquad\widetilde{\psi}^{+}(z,0)=\widetilde{\gamma}%
_{0}^{+}(0)\psi_{c}^{\ast}(z)+\delta\widetilde{\psi}^{+}(z,0)
\label{Eq.InitialStochFields}%
\end{equation}
with
\begin{align}
\delta\widetilde{\psi}(z,0)  &  =%
{\textstyle\sum\limits_{k}}
\xi_{k}(z)\,%
{\textstyle\int}
dz^{\#}\,\xi_{k}^{\ast}(z^{\#})\,\times%
{\displaystyle\sum\limits_{k\neq0}}
\left[  u_{k}(z^{\#})\,\widetilde{\beta}_{k}-v_{k}(z^{\#})^{\ast
}\,\widetilde{\beta}_{k}^{+}\right] \nonumber\\
\delta\widetilde{\psi}^{+}(z,0)  &  =%
{\textstyle\sum\limits_{k}}
\xi_{k}^{\ast}(z)\,%
{\textstyle\int}
dz^{\#}\,\xi_{k}(z^{\#})\,\times%
{\displaystyle\sum\limits_{k\neq0}}
\left[  -v_{k}(z^{\#})\,\widetilde{\beta}_{k}+u_{k}(z^{\#})^{\ast
}\,\widetilde{\beta}_{k}^{+}\right] \nonumber\\
&  \label{Eq.InitialStochFieldFluctns}%
\end{align}
where $u_{k}(z),v_{k}(z)$ are given in Eqs. (\ref{Eq.StandardModesB}) and
(\ref{Eq.Dk}). These expressions allow for quantum fluctuations in both the
condensate and non-condensate modes.

Analogous expressions to (\ref{Eq. StochFloqBogolRelns})\ apply for the
initial stochastic amplitudes $\widetilde{\alpha}_{k}(0)$, $\widetilde{\alpha
}_{k}^{+}(0)$ of Floquet modes with the replacement of the gravitational modes
$\,\xi_{k}(z)\,$by the Floquet modes $\phi_{k}(z,0)$, see Appendix
\ref{Appendix - Floquet Mode Treatment}.\smallskip

\subsection{Quantum State for Initial BEC}

\label{SubSection - Quantum State for Initial BEC}

The question now arises - Having described the prepared BEC via the condensate
mode and the Bogoliubov modes, what is the quantum density operator? This
density operator is needed to define the Wigner distribution function for this
state and hence the stochastic properties of the $\widetilde{\beta}%
_{q}(0),\widetilde{\beta}_{q}^{+}(0)$ and $\widetilde{\gamma}_{0}%
(0),\widetilde{\gamma}_{0}^{+}(0)$, which are equivalent to this initial
Wigner function. If the Bogoliubov modes are in thermal equilibrium at
temperature $T_{0}$ then the density operator for the non-condensate modes
should be given by

$\widehat{\rho}_{NC}=\exp(-%
{\textstyle\sum\limits_{k}}
\hslash\gamma_{k}\,\widehat{b}_{k}^{\dag}\widehat{b}_{k}/k_{B}T_{0})/Tr\left[
\exp(-%
{\textstyle\sum\limits_{k}}
\hslash\gamma_{k}\,\widehat{b}_{k}^{\dag}\widehat{b}_{k}/k_{B}T_{0})\right]
$. This of course ignores interactions between the condensate mode and the
non-condensate Bogoliubov modes.

At $T_{0}=0$ the non-condensate Bogoliubov modes are unoccupied and this would
be approximately the case at temperatures where the BEC exists. Consistent
with the GPE that has been used to describe the condensate bosons, all $N_{c}$
bosons are assumed to be in the condensate mode - whose mode function is
obtained from the time-independent Gross-Pitaevskii equation
(\ref{Eq.GPECondensPrepnA}) - and the state involved is a Fock state. The
combined state for both condensate and non-condensate modes will then be a
pure state whose density operator commutes with the number operator, and hence
complies with the requirement of being phase invariant. Descriptions in which
the condensate bosons are in a Glauber coherent state are often used for
mathematical convenience, but as such states are not phase invariant they are
unphysical. For completeness, this approach is set out in Appendix
\ref{Appendix - Alternative Initial States for Condensate}.

In this section the time value is $t=0$, which is left understood for
simplicity.\smallskip

\subsubsection{Density Operator for Initial State}

The density operator is given by
\begin{equation}
\widehat{\rho}(0)=\left\vert \Phi_{N0}\right\rangle \left\langle \Phi
_{N0}\right\vert \label{Eq.FockStateDensOpr}%
\end{equation}
where
\begin{align}
\left\vert \Phi_{N0}\right\rangle  &  =\left\vert N_{c}\right\rangle
_{0}\times\left\vert 0\right\rangle _{B}\nonumber\\
\left\vert N\right\rangle _{0}  &  =\frac{(\widehat{c}_{0}^{\dag})^{N}}%
{\sqrt{N!}}\left\vert 0\right\rangle _{c}\qquad\left\vert 0\right\rangle _{B}=%
{\textstyle\prod\limits_{k\neq0}}
\,\left\vert 0\right\rangle _{k} \label{Eq.FockState}%
\end{align}
This state is the product of vacuum states for the Bogoliubov modes and a Fock
state with $N$ bosons for the condensate mode.\smallskip

\subsubsection{First and Second-Order QCF for Initial State}

We then find that
\begin{align}
\hat{\Psi}(z)\,\left\vert \Phi_{N0}\right\rangle  &  =\sqrt{N}\,\psi
_{c}(z)\,\left\vert N-1\right\rangle _{0}\times\left\vert 0\right\rangle
_{B}+\left\vert N\right\rangle _{0}\times%
{\displaystyle\sum\limits_{k\neq0}}
(-v_{k}^{\ast}(z)\,\left\vert 0\right\rangle _{1}..\left\vert 1\right\rangle
_{k}...\left\vert 0\right\rangle _{n}\nonumber\\
\left\langle \hat{\Psi}(z)\right\rangle  &  =0\nonumber\\
\left\langle \hat{\Psi}(z)^{\dag}\,\hat{\Psi}(z)\right\rangle  &
=n_{C}(z)=N\,\psi_{c}(z)^{\ast}\psi_{c}(z)+%
{\displaystyle\sum\limits_{k\neq0}}
(v_{k}^{\ast}(z)\,v_{k}(z))\,\approx\Phi_{c}(z)^{\ast}\Phi_{c}(z)\nonumber\\
\left\langle \hat{\Psi}(z^{\#})^{\dag}\,\hat{\Psi}(z)\right\rangle  &
=P(z,z^{\#},0)=N\,\psi_{c}(z^{\#})^{\ast}\psi_{c}(z)+%
{\displaystyle\sum\limits_{k\neq0}}
(v_{k}^{\ast}(z)\,v_{k}(z^{\#}))\,\approx\Phi_{c}(z^{\#})^{\ast}\Phi
_{c}(z)\nonumber\\
&  \label{Eq.FockStateT0Props}%
\end{align}
where the terms involving $v_{k}(z)$ can be neglected since $N\gg1$. We see
that the mean value of the field operator is zero, as required for a
phase-invariant state. However, the mean value of the number density operator
is still obtained from the condensate wave-function. The Fock state is a good
description of the initial BEC state, and will therefore be used to determine
the initial conditions. Note that the last equation demonstrates long-range
spatial order in the BEC, since the QCF will be non-zero for $|z-z^{\#}|$
within the extent of the condensate wave function. The result $P(z,z^{\#}%
,0)=N\,\psi_{c}(z^{\#})^{\ast}\psi_{c}(z)$ expresses the QCF in terms of
natural orbitals - in this case the only natural orbital occupied is the
condensate mode function, where the occupancy is $N_{c}$.\smallskip

\subsubsection{Wigner Distribution Functions for Initial State}

For any given initial state, the stochastic averages of products of stochastic
field functions $\widetilde{\psi}(z),\widetilde{\psi}^{+}(z)$ based on
phase-space theory using the Wigner distribution functional can be obtained
from the quantum correlation functions for normally ordered products of field
operators $\hat{\Psi}(z),\hat{\Psi}(z)^{\dag}$ and then using Wick's theorem
\cite{Wick} to obtain the quantum correlation functions for symmetrically
ordered products of the same field operators. These QCF for the symmetrically
ordered products are then equal to the stochastic averages of the products of
stochastic field functions associated with the original field operators. From
these QCF the stochastic averages of products of stochastic phase-space
variables $\widetilde{\gamma}_{0}(0),\widetilde{\gamma}_{0}^{+}(0),$
$\widetilde{\beta}_{q}(0),\widetilde{\beta}_{q}^{+}(0)$ for the Bogoliubov
mode treatment of the BEC preparation could be obtained. This method would be
required when the condensate mode and the non-condensate modes are correlated
and the initial density operator does not factorise into separate density
operators for the condensate and non-condensate modes. For completeness, the
more general method is described in Appendix \ref{Appendix A - Gen Init State}.

However, for the initial state given by Eq. (\ref{Eq.FockStateDensOpr}) the
density operator factorises into a density operator $\widehat{\rho}_{C}(0)$
for the condensate mode and a density operator $\widehat{\rho}_{NC}(0)$ for
the non-condensate Bogoliubov modes, which are all in their vacuum states%
\begin{align}
\widehat{\rho}(0)  &  =\widehat{\rho}_{C}(0)\otimes\widehat{\rho}%
_{NC}(0)\nonumber\\
\widehat{\rho}_{C}(0)  &  =\left\vert \Phi_{N0}\right\rangle \left\langle
\Phi_{N0}\right\vert \qquad\widehat{\rho}_{NC}(0)=%
{\textstyle\prod\limits_{k\neq0}}
(\left\vert 0\right\rangle \left\langle 0\right\vert )_{k}
\label{Eq.NonCorrDensOpr}%
\end{align}
For this situation we can show that the Wigner distribution function
$W(\gamma_{0},\gamma_{0}^{+},\mathbf{\beta},\mathbf{\beta}^{+})$ factorises
into separate Wigner distribution functions $W_{C}(\gamma_{0},\gamma_{0}^{+})$
and $W_{NC}(\mathbf{\beta},\mathbf{\beta}^{+})$ for the condensate mode and
the non-condensate Bogoliubov modes, where $\mathbf{\beta}\equiv\{\beta
_{1},...\beta_{k},..)$ and $\mathbf{\beta}^{+}\equiv\{\beta_{1}^{+}%
,...\beta_{k}^{+},..)$ We can then calculate the stochastic averages of
products of stochastic phase-space variables by considering the condensate and
non-condensate modes separately.

The proof of the result
\begin{equation}
W(\gamma_{0},\gamma_{0}^{+},\mathbf{\beta},\mathbf{\beta}^{+})=W_{C}%
(\gamma_{0},\gamma_{0}^{+})\times W_{NC}(\mathbf{\beta},\mathbf{\beta}^{+})
\label{Eq.WignerFactorn}%
\end{equation}
where the Wigner functions are normalised to unity - $%
{\textstyle\int}
d^{2}\gamma_{0}d^{2}\gamma_{0}^{+}\,W_{C}(\gamma_{0},\gamma_{0}^{+})=1$ and $%
{\textstyle\int}
d^{2}\mathbf{\beta}\,d^{2}\mathbf{\beta}^{+}\,W_{NC}(\mathbf{\beta
},\mathbf{\beta}^{+})=1$ - is given in Appendix
\ref{Appendix B - Wigner Distn Bogol Modes}. This result means that
phase-space averages that give QCF for products of condensate and
non-condensate mode operators will factorise into a separate phase-space
average for the condensate mode and a phase-space average for the
non-condensate modes. This in turn means that the equivalent stochastic
average of products of stochastic phase-space variables for the condensate
mode and stochastic phase-space variables for the non-condensate modes will
factorise into separate stochastic averages for the condensate and
non-condensate stochastic variables. Thus, for example, $\overline
{\widetilde{\gamma}_{0}\widetilde{\beta}_{k}^{+}}=\overline{\widetilde{\gamma
}_{0}}$\thinspace$\times\overline{\widetilde{\beta}_{k}}$\thinspace.

The result uses Morgan's approach \cite{Morgan00} and involves first
establishing a relation between the standard non-condensate mode operators
$\widehat{c}_{i},\widehat{c}_{i}^{\dag}$ that were introduced in Eq.
(\ref{Eq.FluctFieldOpr3}) and the Bogoliubov mode operators $\widehat{b}%
_{k},\widehat{b}_{k}^{\dag}$ which are given in matrix form (the convention
differs from that in Ref \cite{Morgan00}) as
\begin{equation}
\left[
\begin{array}
[c]{c}%
\widehat{\mathbf{c}}\\
\widehat{\mathbf{c}}^{\dag}%
\end{array}
\right]  =\left[
\begin{array}
[c]{cc}%
U & -V\\
-V^{\ast} & U^{\ast}%
\end{array}
\right]  \times\left[
\begin{array}
[c]{c}%
\widehat{\mathbf{b}}\\
\widehat{\mathbf{b}}^{\dag}%
\end{array}
\right]
\end{equation}
where the matrices $U,V$ have elements given by
\begin{equation}
U_{i,k}=%
{\textstyle\int}
dz\,\psi_{i}(z)^{\ast}\,u_{k}(z)\qquad V_{i,k}=%
{\textstyle\int}
dz\,\psi_{i}(z)^{\ast}\,v_{k}(z)^{\ast}%
\end{equation}
This is a linear canonical transformation in which commutation rules are
preserved. For the matrices, $T$ denotes transverse, $\ast$ denotes complex
conjugation and $\dag$ denotes the Hermitian adjoint. The bosonic commutation
rules for both sets of non-condensate mode operators leads to the following
matrix equations.
\begin{align}
-U\,V^{T}+V\,U^{T}  &  =0\nonumber\\
U\,U^{\dag}-V\,V^{\dag}  &  =E
\end{align}
Further features of these matrices are set out in Appendix
\ref{Appendix B - Wigner Distn Bogol Modes}.\smallskip

\subsubsection{Stochastic Averages - Condensate Mode}

For the condensate mode the density operator is given by
(\ref{Eq.FockStateDensOpr}) and as $\widehat{c}_{0}\left\vert N\right\rangle
=\sqrt{N}\left\vert N-1\right\rangle $ it is easy to show that the first and
second-order QCF for normally ordered condensate mode operators are given by
\begin{align}
\left\langle \widehat{c}_{0}\right\rangle  &  =\left\langle \widehat{c}%
_{0}^{\dag}\right\rangle =0\label{Eq.FirstOrderQCFCondMode}\\
\left\langle \widehat{c}_{0}\widehat{c}_{0}\right\rangle  &  =\left\langle
\widehat{c}_{0}^{\dag}\widehat{c}_{0}^{\dag}\right\rangle =0\qquad\left\langle
\widehat{c}_{0}^{\dag}\widehat{c}_{0}\right\rangle =N
\label{Eq.SecondOrderQCFCondMode}%
\end{align}

Since $\left\{  \widehat{c}_{0}\widehat{c}_{0}\right\}  =\widehat{c}%
_{0}\widehat{c}_{0}$, $\left\{  \widehat{c}_{0}^{\dag}\widehat{c}_{0}^{\dag
}\right\}  =\widehat{c}_{0}^{\dag}\widehat{c}_{0}^{\dag}$ and $\left\{
\widehat{c}_{0}^{\dag}\widehat{c}_{0}\right\}  =\widehat{c}_{0}^{\dag
}\widehat{c}_{0}+1/2$ we see that the stochastic averages are
\begin{align}
\overline{\widetilde{\gamma}_{0}}  &  =\overline{\widetilde{\gamma}_{0}^{+}%
}=0\label{Eq.FirstOrderStochAverCond}\\
\overline{\widetilde{\gamma}_{0}\widetilde{\gamma}_{0}}  &  =\overline
{\widetilde{\gamma}_{0}^{+}\widetilde{\gamma}_{0}^{+}}=0\qquad\overline
{\widetilde{\gamma}_{0}^{+}\widetilde{\gamma}_{0}}=N+\frac{1}{2}
\label{Eq.SecondOrderStochAverCond}%
\end{align}

The distribution function generating the stochastic $\widetilde{\gamma}_{0}$
and $\widetilde{\gamma}_{0}^{+}$ thus must produce a mean value of zero for
the first order, and $N+\frac{1}{2}$ for the non-zero second-order case
$\overline{\widetilde{\gamma}_{0}^{+}\widetilde{\gamma}_{0}}$.

Higher order QCFs can also be considered. The third-order QCF for normally
ordered condensate mode operators are all zero. The only non-zero fourth-order
QCF for normally ordered condensate mode operators is $\left\langle
\widehat{c}_{0}^{\dag}\widehat{c}_{0}^{\dag}\widehat{c}_{0}\widehat{c}%
_{0}\right\rangle =N(N-1)$. Since $\{\widehat{c}_{0}^{\dag}\widehat{c}%
_{0}^{\dag}\widehat{c}_{0}\widehat{c}_{0}\}=(\widehat{c}_{0}^{\dag}%
\widehat{c}_{0}^{\dag}\widehat{c}_{0}\widehat{c}_{0}+\widehat{c}_{0}^{\dag
}\widehat{c}_{0}\widehat{c}_{0}^{\dag}\widehat{c}_{0}+\widehat{c}_{0}^{\dag
}\widehat{c}_{0}\widehat{c}_{0}\widehat{c}_{0}^{\dag}+\widehat{c}%
_{0}\widehat{c}_{0}^{\dag}\widehat{c}_{0}^{\dag}\widehat{c}_{0}+\widehat{c}%
_{0}\widehat{c}_{0}^{\dag}\widehat{c}_{0}\widehat{c}_{0}^{\dag}+\widehat{c}%
_{0}\widehat{c}_{0}\widehat{c}_{0}^{\dag}\widehat{c}_{0}^{\dag})/6$ we can use
Wick's theorem \cite{Wick} to show that $\{\widehat{c}_{0}^{\dag}%
\widehat{c}_{0}^{\dag}\widehat{c}_{0}\widehat{c}_{0}\}=\widehat{c}_{0}^{\dag
}\widehat{c}_{0}^{\dag}\widehat{c}_{0}\widehat{c}_{0}+2\widehat{c}_{0}^{\dag
}\widehat{c}_{0}+1/2$. We then see that the non-zero stochastic average is%
\begin{align}
\overline{(\widetilde{\gamma}_{0}^{+})^{2}(\widetilde{\gamma}_{0})^{2}}  &
=N(N-1)+2N+1/2\nonumber\\
&  =N^{2}+N+1/2 \label{Eq.FourthOrderStochAverCond}%
\end{align}
This result is given in Ref \cite{Olsen04a}. If we choose $\widetilde{\gamma
}_{0}=\sqrt{\widetilde{\Gamma}_{0}}\,\exp(i\widetilde{\phi}_{0})$\ and
$\widetilde{\gamma}_{0}^{+}=\sqrt{\widetilde{\Gamma}_{0}}\,\exp
(-i\widetilde{\phi}_{0})$, where $\widetilde{\Gamma}_{0}$\ is a normally
distributed Gaussian random variable with a mean equal to $N+\frac{1}{2}$\ and
a variance of $1/4$, and we choose $\widetilde{\phi}_{0}$\ to be a uniformly
distributed random phase over $0$\ to $2\pi$, then all the results in
Eqs.(\ref{Eq.FirstOrderStochAverCond}), (\ref{Eq.SecondOrderStochAverCond})
and (\ref{Eq.FourthOrderStochAverCond}) are obtained. The Wigner distribution
function for a Fock state is given in \cite{Olsen04a}.\smallskip

\subsubsection{Stochastic Averages - Non-Condensate Modes}

For the non-condensate mode the density operator is given by
(\ref{Eq.NonCorrDensOpr}) and as $\widehat{b}_{k}\left\vert 0\right\rangle
_{k}=0\times\left\vert 0\right\rangle _{k}$ it is easy to show that the first
and second-order QCF for normally ordered non-condensate mode operators are
given by%
\begin{align}
\left\langle \widehat{b}_{k}\right\rangle  &  =\left\langle \widehat{b}%
_{k}^{\dag}\right\rangle =0\label{Eq.FirstOrderQCFNonCondModes}\\
\left\langle \widehat{b}_{k}\widehat{b}_{l}\right\rangle  &  =\left\langle
\widehat{b}_{k}^{\dag}\widehat{b}_{l}^{\dag}\right\rangle =0\qquad\left\langle
\widehat{b}_{k}^{\dag}\widehat{b}_{l}\right\rangle =0
\label{Eq.SecondOrderQCFNonCondModes}%
\end{align}

Since $\left\{  \widehat{b}_{k}\widehat{b}_{l}\right\}  =\widehat{b}%
_{k}\widehat{b}_{l}$, $\left\{  \widehat{b}_{k}^{\dag}\widehat{b}_{l}^{\dag
}\right\}  =\widehat{b}_{k}^{\dag}\widehat{b}_{l}^{\dag}$ and $\left\{
\widehat{b}_{k}^{\dag}\widehat{b}_{l}\right\}  =\widehat{b}_{k}^{\dag
}\widehat{b}_{l}+(1/2)\,\delta_{k,l}$ we see that the stochastic averages are
\begin{align}
\overline{\widetilde{\beta}_{k}}  &  =\overline{\widetilde{\beta}_{k}^{+}%
}=0\label{Eq.FirstOrderStochAverNonCondModes}\\
\overline{\widetilde{\beta}_{k}\widetilde{\beta}_{l}}  &  =\overline
{\widetilde{\beta}_{k}^{+}\widetilde{\beta}_{l}^{+}}=0\qquad\overline
{\widetilde{\beta}_{k}^{+}\widetilde{\beta}_{l}}=\frac{1}{2}\,\delta_{k,l}
\label{Eq.SecondOrderStochAverNonCondModes}%
\end{align}
The distribution function generating the stochastic $\widetilde{\beta}_{k}$
and $\widetilde{\beta}_{k}^{+}$ thus must produce a mean value of zero for the
first order, and $\frac{1}{2}$ for the non-zero second order case
$\overline{\widetilde{\beta}_{k}^{+}\widetilde{\beta}_{k}}$.

It is of some interest to calculate the stochastic averages for the stochastic
phase-space variables associated with the standard modes $\widehat{c}%
_{i},\widehat{c}_{i}^{\dag}$ associated with Eq. (\ref{Eq.FluctFieldOpr3}). It
can be shown (see Appendix \ref{Appendix B - Wigner Distn Bogol Modes}) that
the vacuum state for Bogoliubov modes is equivalent to a squeezed vacuum state
for the standard non-condensate modes. \pagebreak

\section{Numerical Results}

\label{Section - Numerical Results}

The numerics in this paper are based on the truncated Wigner approximation
(TWA) and to a lesser extent on the mean-field Gross-Pitaevskii equation
(GPE). Our aim in the present paper is to study the DTC with
periodicity\textbf{ }$sT$\textbf{ }for the case of period doubling\textbf{
}$(s=2)$\textbf{, }where\textbf{ }$T$\textbf{ }is the periodicity of the
oscillating mirror.

The position probability density (PPD) $F(z,t)=Tr(\hat{\Psi}^{\dag}%
(z)\hat{\Psi}(z)\rho(t))$ versus $z$ plots for various times
(Eq.(\ref{Eq.PositProbStochasAver})) depend either on the time-dependent
stochastic field functions $\widetilde{\psi}(z,t)$ in the TWA case or on the
time-dependent condensate wave function $\Phi_{c}(z,t)$, both considered as a
function of time (see Eqs.(\ref{Eq.ItoSFEA}), (\ref{Eq.ItoSFAB}),
(\ref{Eq.GPE})). The same is true for the quantum correlation function (QCF)
$P(z,z^{\#},t)=Tr(\hat{\Psi}^{\dag}(z^{\#})\hat{\Psi}(z)\rho(t))$ (see
Eq.(\ref{Eq.QCFStochAver})) and the one-body projector (OBP) (both for the
condensate mode $M_{c}(t)=%
{\displaystyle\int}
{\displaystyle\int}
dz\,dz^{\#}\,\psi_{c}(z^{\#},0)\,\,\psi_{c}^{\ast}(z,0)\,P(z,z^{\#},t)$ and
for the Wannier modes $\Phi_{1}$ and $\Phi_{2}$ (see
Eq.(\ref{Eq.ApproxWannier}), with $s=2$) and its Fourier transform (FT) (see
Eqs.(\ref{Eq.ObservableStochResult}), (\ref{Eq.FTResult}), (\ref{Eq.OtherOBP}%
)). The natural orbitals and their occupation numbers are determined from the
QCF (see Eqs.(\ref{Eq.QCFNaturalOrbitals}), (\ref{Eq.NaturalOrb})). The mean
energy is determined from the stochastic field functions (see Eq.
(\ref{Eq.MeanEnergyStochastic})) with the most important contributions given
in Eq.(\ref{Eq.MeanEnergyGravModes}) using gravitational modes. The quantum
depletion from the condensate mode $N_{D}(t)=N-N_{C}(t)$ is determined from
Eq. (\ref{Eq.StochResultCondModeOccupn}).

In both the TWA and GPE numerics, the behaviour of these functions is
deterministic and depends on the initial conditions $\widetilde{\psi}(z,0)$
and $\Phi_{c}(z,0)$. The difference between the TWA and GPE is that in the
former a stochastic ensemble of $\widetilde{\psi}(z,0)$ is considered whereas
in the latter just a single $\Phi_{c}(z,t)$ is involved (see
Sect.\ref{SubSection - TWA}). In the TWA numerics we find it convenient to
choose stochastic field functions such that $\widetilde{\psi}^{+}%
(z,t)=\widetilde{\psi}(z,t)^{\ast}$\ and stochastic phase amplitudes such that
$\widetilde{\eta}_{k}^{+}(t)=\widetilde{\eta}_{k}(t)^{\ast}$,
etc.\textbf{\ \ }

For the initial conditions the BEC is assumed to be at zero temperature, with
all bosons in a single mode. For both the TWA and GPE cases the initial
condensate mode function is based on the time-independent GPE for a harmonic
trap with non-condensate modes being Bogoliubov modes. For the Wannier initial
condition a linear combination of two Wannier modes is taken as the initial
condensate mode function, with non-condensate modes being any orthogonal set
of modes. The linear combination is chosen to minimise the energy functional
as described in Ref. \cite{Sacha15a}. For the TWA case the initial condition
allows for unoccupied Bogoliubov modes in the harmonic trap case, and
unoccupied non-condensate modes in the Wannier case. There are no unoccupied
non-condensate modes in the GPE approach by definition. The loss of bosons
from the condensate mode is specified by the quantum depletion (see
Eqs.(\ref{Eq.QuantumDepletion}), (\ref{Eq.QuantumRetention}),
(\ref{Eq.StochAmpCondModes})). For the TWA case there is a stochastic ensemble
of phase-space amplitudes for the condensate and Bogoliubov modes (see Sect.
\ref{SubSubSection - Field Fns Cond and Bogol Modes}\ and Eq.
(\ref{Eq.StochFieldsBogolModesB})). Equivalent initial stochastic phase-space
amplitudes for gravitational modes are determined by equating expressions for
the initial stochastic field functions (see
Sect.\ref{SubSect Initial Cond Stoch Amp Grav Modes} and
Eq.(\ref{Eq. StochFloqBogolRelns})).

The TWA numerical calculations are based on expanding the stochastic field
functions in terms of time-independent gravitational modes, and during the
calculations $\widetilde{\psi}(z,t)$ is known at each time point (see Sect.
\ref{SubSubSect - Gravit Modes} and Eqs. (\ref{Eq.StochFieldGravModes}),
(\ref{Eq.StochPhaseGravA}), (\ref{Eq.StochPhaseGravB})). At each stage
$\widetilde{\psi}(z,t)$ and $\Phi_{c}(z,t)$ can also be re-expanded in terms
of Floquet modes (see Sect.\ref{SubSubSection - Floquet Modes} and
Eq.(\ref{Eq.StochFldFns})).

A suitable description of the behaviour for both TWA and GPE numerics in terms
of the theoretical approach in this paper would be to think of a
representative stochastic field function or the condensate wave-function as
moving wave-packets, starting from their initial space-localised forms
$\widetilde{\psi}(z,0)$ and $\Phi_{c}(z,0)$ and then changing in shape and
position as time evolves. If $\widetilde{\psi}(z,t)$ or $\Phi_{c}(z,t)$ are
expanded in terms of Wannier modes, it may be the case that only one or two
modes are important, and the evolution of each component can be considered
separately. This situation can be described as two or more wave-packets that
may move in opposite directions, then combine again to reform the initial
$\widetilde{\psi}(z,t)$ or $\Phi_{c}(z,t)$ after a characteristic period has elapsed.

As will be seen, there will be certain values for the boson-boson interaction
strength factor $gN$ that divide the observable behaviour between the presence
of a DTC and its absence, and also between where discrete time-translation
symmetry breaking occurs and where it does not. These changes occur in a
different way for Wannier and harmonic trap initial conditions; so to
distinguish the two cases, the values for $gN$ involved will be referred to as
the critical\ interaction strength or the threshold interaction strength,\ respectively.

The numerical calculations are performed for a quasi-one-dimensional
Bose-Einstein condensate of ultracold atoms released from a harmonic trap to
bounce resonantly under the influence of gravity on an atom mirror oscillating
with period $T$ and amplitude $\lambda$ (in the oscillating frame). In
carrying out these calculations we use (dimensionless) gravitational units:
length $l_{0}=(\hbar^{2}/m^{2}g_{E})^{1/3}$, time $t_{0}=(\hbar^{2}/mg_{E}%
^{2})^{1/3}$, energy $E_{0}=mg_{E}l_{0}$, where $\hbar$\ is the reduced
Planck's constant, $g_{E}$\ is the Earth's gravitational acceleration and $m$
is the mass of the atom. For $s=2$ (period-doubling) we set the drop height
$h_{0}$ so that the bounce period $t_{bounce}=(2h_{0})^{1/2}$ is equal to
$2T$. In gravitational units the potential in Eq. (\ref{Eq.PotentialTerm}) is
given by\textbf{ }%
\begin{equation}
\overline{V}(\overline{z},\overline{t})=\overline{z}(1-\lambda\cos
\overline{\omega}\overline{t})
\end{equation}
where $\overline{V}$ is the potential in units of $E_{0}$, and $\overline
{z},\overline{t}$ are position and time in units of $l_{0},t_{0}$
respectively, with $\overline{\omega}$ being the frequency in units of
$1/t_{0}$. Henceforth the over-bars will be left understood. \smallskip

\subsection{Choice of Parameters}

As we wish to compare our TWA calculations with the mean-field GPE and
time-dependent Bogoliubov calculations in Kuros et al. \cite{Kuros20a}, we
chose the parameters $\lambda$ and $\omega$ to be the same as theirs, namely
$\lambda=0.12$ and $\omega=1.4$ (in gravitational units). The choice made by
Kuros et al. \cite{Kuros20a} was informed by semi-classical treatments of the
DTC system \cite{Sacha15a}. These parameters allow a reasonably short atom
transfer time (about $2$ $s$ for $^{7}Li$) between the two wave-packets for
zero particle interaction compared with a typical BEC lifetime, a reasonably
small number of bounces ($<100$) during the atom transfer time, and an average
interaction energy per particle that is small compared with the energy gap
between the first and second bands in the single-band Bose-Hubbard model used
in Ref. \cite{Kuros20a}. The above values of $\lambda$\ and $\omega$\ result
in two Floquet modes $\phi_{1}$ and $\phi_{2}$ at $t=0.5T$ that are similar to
those in Fig. 2a of Ref. \cite{Sacha15a} for $t=0T$. The parameters for the
harmonic trap potential, $\widetilde{\omega}_{0}=0.68$ and drop height,
$h_{0}=9.82$, (in gravitational units) for our TWA calculations were chosen so
that at $t=0T$ the condensate mode function has a large overlap with the
Wannier function $\Phi_{2}$ that is obtained from the two Floquet modes
$\phi_{1}$ and $\phi_{2}$. The Floquet frequencies are chosen to create a
localised Wannier-like wave-packet which leads to $(\nu_{2}-\nu_{1})$ being
approximately $\omega/2$.

The various physical quantities for the simple case of a hard-wall mirror
potential $(V(z)=z$ \ for $z\geq0$ and $\infty$ for $z<0)$ are summarised in
Table 1, both in gravitational units and in SI units for the case of the
bosonic $^{7}Li$ atom, which is chosen as an example because its $s$-wave
scattering length can be tuned precisely via a broad Feshbach resonance
\cite{Giergiel20a}. The drop height required to satisfy the resonance
condition for $s=2$, i.e., $h_{0}l_{0}=20$\ $\mu m$\ for $^{7}Li$,\ is rather
challenging to work with in an experiment, but larger drop heights could be
used by operating with larger $s$\ resonances \cite{Giergiel20a}. The mirror
oscillation amplitude, $\lambda l_{0}/\omega^{2}=125$ $nm$,\ is for a
theoretical hard-wall potential mirror; in the case of a realistic soft
Gaussian potential mirror the oscillation amplitudes are typically an order of
magnitude larger in order to achieve the same driving effect
\cite{Giergiel20a}.

\begin{center}

\end{center}

\textbf{Table 1}

Physical quantities for $s=2$ resonance used in the calculations. All
quantities are in gravitational units except the expressions for $l_{0},t_{0}$
and $E_{0}.$%

\begin{tabular}
[c]{|l|l|l|l|}\hline
\textbf{Quantity} & \textbf{Symbol} & \textbf{Gravitational} & \textbf{MKS
units for }$^{7}Li$\\\hline
&  & \ \ \ \ \ \ \textbf{units} & \ \ \ \ \ \textbf{(lab frame)}\\\hline
Gravitational unit of length & $l_{0}=(\hbar^{2}/m^{2}g_{E})^{1/3}$ &  &
$2.03$ $\mu$m\\\hline
Gravitational unit of time & $t_{0}=(\hbar/mg_{E}^{2})^{1/3}$ &  & $0.455$
ms\\\hline
Gravitational unit of energy & $E_{0}=mg_{E}l_{0}$ &  & $E_{0}/k_{B}=16.8$
nK\\\hline
Mirror oscillation frequency & $\omega$ & $1.4$ & $0.49$ kHz\\\hline
Mirror oscillation period & $T=2\pi/\omega$ & $4.49$ & $2.04$ ms\\\hline
Mirror oscillation amplitude & $\lambda$ & $0.12$ & $\lambda l_{0}/\omega
^{2}=125$ nm\\\hline
Bounce period for $s=2$ & $2T$ & $1.43$ & $4.08$ ms\\\hline
Drop height for $s=2$ & $h_{0}=\frac{1}{2}T^{2}$ & $9.82$ & $20.0$ $\mu
$m\\\hline
Trap frequency & $\widetilde{\omega}_{0}$ & $0.68$ & $238$ Hz\\\hline
Boson transfer rate for $gN=0$ & $J$ & $7.14\times10^{-4}$ $^{\ast}$ & $1.6$
s$^{-1}$\\\hline
Boson transfer time for $gN=0$ & $t_{trans}=\pi/J$ & $440$ & $2.0$ s\\\hline
Number of bounces during $t_{trans}$ & $N_{b}=t_{trans}/T$ & $98$ &
$98$\\\hline
Longitudinal cloud width at $z=h_{0}$ & $\sigma_{z}=(\hbar/2m\omega_{0}%
)^{1/2}/l_{0}$ & $0.85$ & $1.7$ $\mu$m\\\hline
s-wave scattering length & $a_{S}/a_{0}$ & $9.0\times10^{3}|gN|/N$ &
$9.0\times10^{3}|gN|/N$\\\hline
\end{tabular}
\medskip

* Using mean -field calculations based on the overlap of two Wannier
wave-functions used as a basis for the condensate wave-function,\textbf{\ }%
similar to calculations in Ref. \cite{Kuros20a}. \smallskip

\subsection{Initial State - Linear Combination of Wannier-Like States}

\label{SubSection - Initial Wannier State}

We first carry out TWA calculations for an initial condensate mode function
given by the $t=0$ form of the stable periodic solution of the time-dependent
mean-field GPE (Eq. (\ref{Eq.GPE})). This solution is approximated as a linear
combination of two single-particle Wannier-like modes $\Phi_{1}$ and $\Phi
_{2}$ (Eq. (\ref{Eq.ApproxWannier})), as in Ref. \cite{Sacha15a}. These
Wannier-like modes each have period $2T$ and are delayed by time $T$ with
respect to each other, as shown in Figs. 1 (c), (d), with the related Floquet
modes shown in Figs. 1(a), (b). The linear combination coefficients are
obtained via the extrema of the energy functional given by Eq. (3) in Ref.
\cite{Sacha15a}. For$|gN|$ less than for the critical interaction strength
$-0.006$, the linear combination gives the Floquet mode $\phi_{2}(z,t)$, which
would have period $T$. When $|gN|\,>0.006$, a more general linear combination
of $\Phi_{1}(z,t)$ and $\Phi_{2}(z,t)$ results, which would have period $2T$.
Such a Wannier initial condition is useful for studying the dynamical
processes and demonstrating discrete time-translation symmetry breaking,
though it is not a realistic BEC state that can be prepared in the laboratory.
The position probability density (PPD) results are presented in Fig. 2 and the
one-body projector (OBP) and its Fourier transform (FT) are shown in Fig. 3.
The OBP for a many-body system, is similar to the autocorrelation function or
the fidelity used in Ref. \cite{Kuros20a} in a mean-field calculation (see
Sect. \ref{SubSubSection - One Body Projector} for details).

The PPD plots in Fig. 2 are TWA (blue) and mean-field GPE (red) calculations
for interaction strengths $gN=-0.005$, $-0.006$ and $-0.007$, with $N=600$
bosonic atoms, and evolution times out to $t/T=1998$ mirror oscillations.
Initially, at time $t=0T$, the two wave-packets move towards each other; the
$\Phi_{1}$ wave-packet is reflected coherently from the oscillating mirror
creating fringes due to interference between the incident and reflected parts
of the wave-packet, while the $\Phi_{2}$ wave-packet is at the classical
turning point $z=h_{0}$. At times $t=(k+0.5)T$ $(k=0,1,2\ldots),$ the two
wave-packets cross paths and interfere, again creating fringes.%

\begin{figure}[ptb]%
\centering
\includegraphics[
height=2.936in,
width=5.2762in
]%
{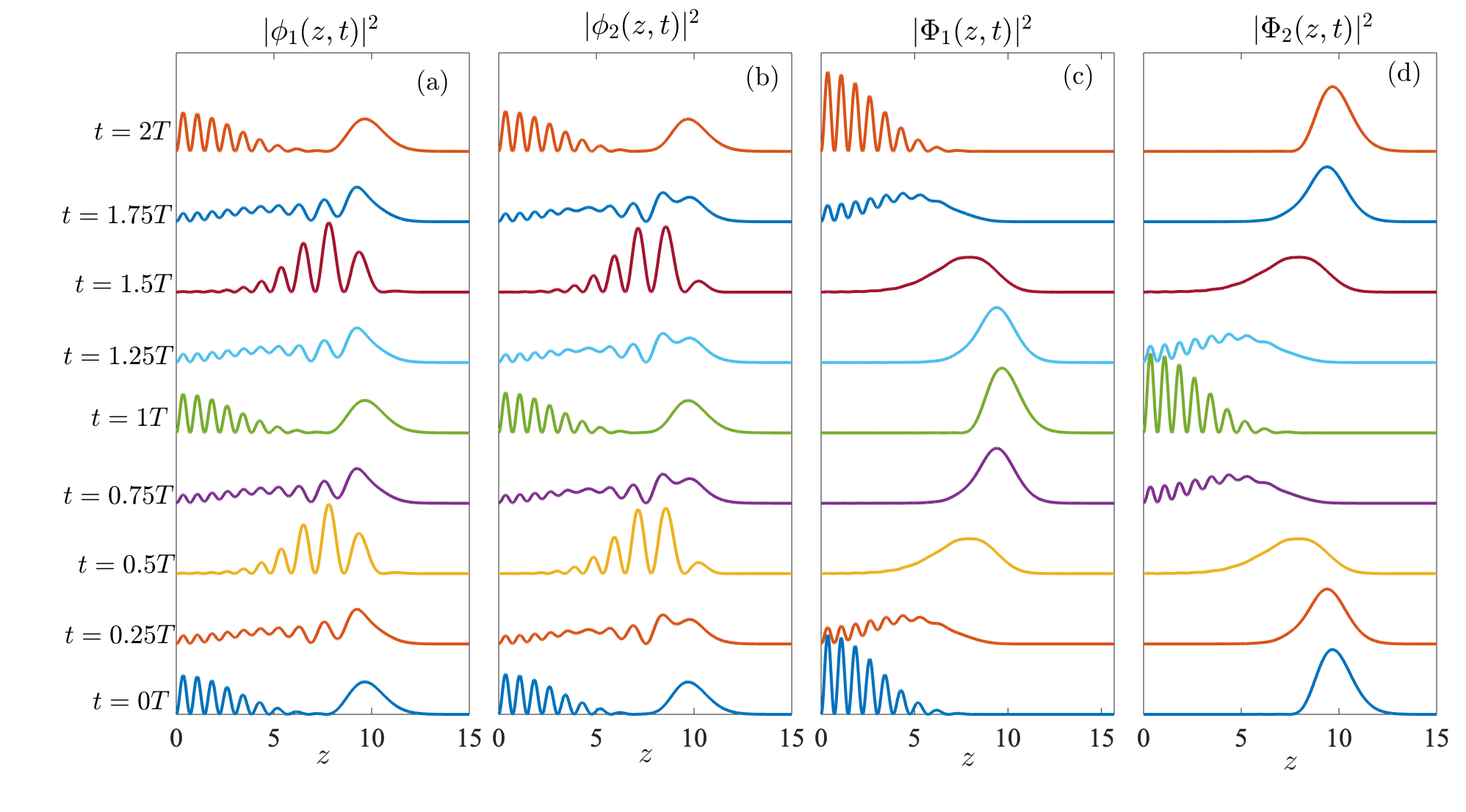}%
\caption{Single particle Floquet states and Wannier modes for $s=2$, initial
position $h_{0}=9.82$, trap frequency $\protect\widetilde{\omega}_{0}=0.68$,
oscillation amplitude $\lambda=-0.12$ and drive frequency $\omega=1.4$. (a)
and (b) show the two Floquet states $\phi_{1}(z,t)$ and $\phi_{2}(z,t)$ with
the corresponding Floquet quasi-energies $\nu_{1}\approx0.410$ and $\nu
_{2}\approx1.109$. (c) and (d) show the two Wannier modes $\Phi_{1}(z,t)$ and
$\Phi_{2}(z,t)$ constructed from the two Floquet states via Eq.
(\ref{Eq.ApproxWannier})}%
\end{figure}
%

\begin{figure}[ptb]%
\centering
\includegraphics[
height=2.9369in,
width=6.9081in
]%
{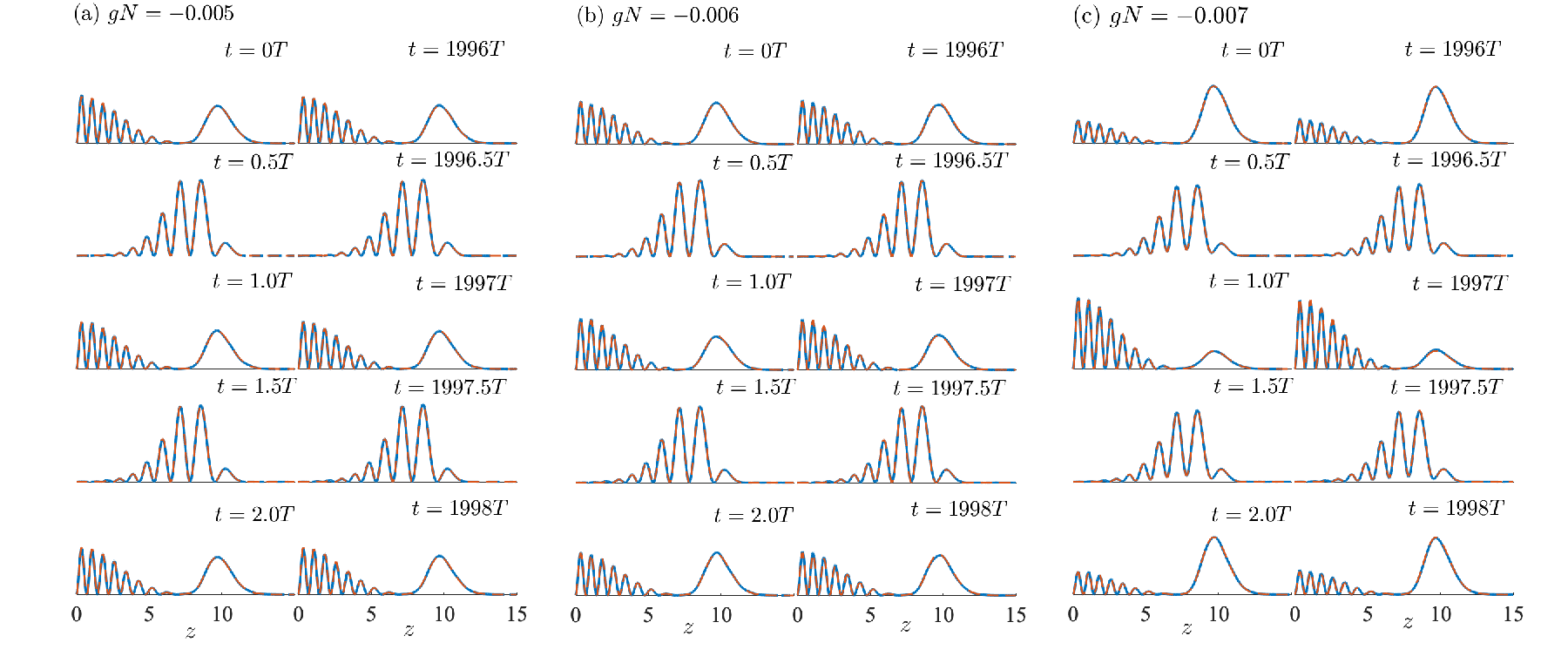}%
\caption{Position probability density (PPD) TWA (blue) and mean-field GPE
(red) calculations and for Wannier initial conditions as a function of $z$ and
$t$ for $s=2$ and for different interaction strengths $gN=-0.005$, $-0.006$
and $-0.007$ in (a), (b) and (c), respectively. The calculations are carried
out for $N=600$, $h_{0}=9.82$, $\protect\widetilde{\omega}_{0}=0.68$,
$\lambda=0.12$ and $\omega=1.4$.}%
\end{figure}
\begin{figure}[ptb]%
\centering
\includegraphics[
height=2.936in,
width=3.0381in
]%
{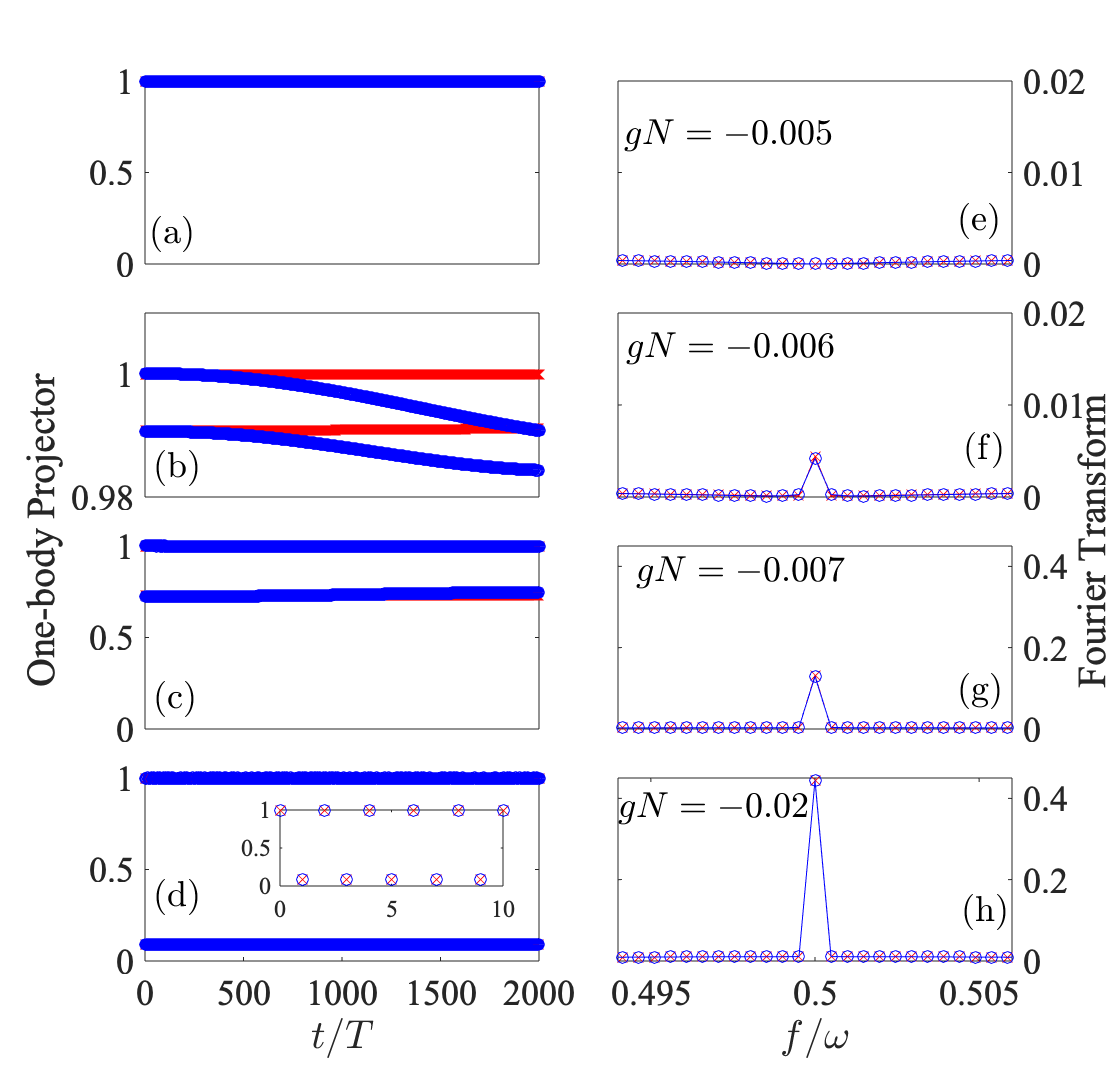}%
\caption{One-body projector (OBP) and the corresponding Fourier transform (FT)
for Wannier initial conditions. (a), (b), (c) and (d) show the OBP for
$gN=-0.005$, $-0.006$, $-0.007$ and $-0.02$, respectively; (e), (f), (g) and
(h) shows the corresponding FT. The inset of (d) shows a zoom-in of the OBP at
a shorter time interval. Note that the pairs of OBP versus $t$ plots are not
continuous; the OBP oscillates between the two curves, such as every $T$ as
the inset for $gN=-0.02$ shows. The blue circle (red cross) symbol show the
TWA (GPE) results. Note the different $y$-axis scale of (b) from (a), (c) and
(d), as well as the different $y$-axis scale of (e) and (f) from (g) and (h).
The other parameters are the same as Figure 2.}%
\end{figure}

For $gN=-0.005$ (Fig. 2 (a)), the PPDs are almost perfectly $T$-periodic and
the two Wannier wave-packets have about the same magnitude, while for
$gN=-0.007$ (Fig. 2 (c)), the PPDs are $2T$-periodic and the two wave-packets
have different magnitude. The change in response period of the BEC is clearly
illustrated in plots of the one-body projector versus evolution time $t/T$ and
its Fourier transform (Fig. 3). For $gN=-0.005,$ the OBP versus $t/T$ is
essentially a single oscillating horizontal line (Fig. 3 (a)), indicating a
period $T$ equal to the drive period, while for $gN=-0.006$, the OBP exhibits
a closely spaced double-curve structure (Fig. 3 (b)) and a small sub-harmonic
peak appears in the FT at half the driving frequency $\omega/2$ (Fig. 3 (f)).
Thus, discrete time-translation symmetry is broken at a critical interaction
strength near $gN\approx-0.006$ to form a DTC. For $gN=-0.007$ and
$\ gN=-0.02$, the OBP exhibits a clear double-curve structure (Figs. 3(c),
(d)) and a larger sub-harmonic peak in the FT at $\omega/2$ (Figs.3 (g), (h)),
indicating a single stable wave-packet for evolution times out to a least
$2000$ mirror oscillations.

For $gN=-0.006$ and $gN=-0.007$ (Figs. 2 (b), (c)), there is no noticeable
change in the PPDs between the set for $t=0$ to $2T$ and the set for $t=1996T$
to $1998T$, indicating the DTC shows no sign of decay for times out to at
least $t/T=1998$ mirror oscillations. Furthermore, there is no noticeable
difference between the TWA (red) and the mean-field GPE (blue) PPDs for both
$gN=-0.005$ and $gN=-0.007$. However, at the\textbf{ }critical interaction
strength, $gN\approx-0.006$, there is a small difference in the OBP (up to
$\approx1\%$) between the TWA and the mean-field GPE calculations (Fig. 3 (b))
for times longer than $500$ mirror oscillations, indicating a quantum
depletion of this order. This peak in the quantum depletion at the
\textbf{critical} interaction strength for discrete time-translation symmetry
breaking is further illustrated in Fig. 11.

In summary, in the case of the Wannier initial condition, for small $|gN|$ the
special linear combination of Wannier-like modes as the initial condition
evolves with a period $T$ as for a Floquet mode, while for $|gN|$ larger than
for the critical value of the interaction strength $gN\approx-0.006$, discrete
time-translation symmetry breaking occurs, and the more general linear
combination of Wannier-like modes now evolves with periodicity $2T$, to form a
DTC.\textbf{ }\smallskip

\subsection{ Initial State - Gaussian-like State Prepared in a Harmonic Trap}

\label{SubSection - Initial Gaussian State}

We now consider an initial Gaussian-like state that is prepared in a harmonic
trap and matches the Wannier wave-packet at the classical turning point. Such
an initial state can be realistically prepared in the laboratory. Our
calculations for the harmonic trap case are mainly for negative $gN.$

In Figs. 4-8, we present TWA (blue) and mean-field GPE (red) calculations of
the PPD for a range of interaction strengths $gN$, with $N=600$ bosonic atoms,
and evolution times out to $t/T=1998$ mirror oscillations. In Fig. 9 we show
the corresponding OBP and FT for the original condensate mode function and
Fig. 10 shows the OBP for the two Wannier modes $\Phi_{1}$, $\Phi_{2}$. Note
that the OBP for the original condensate mode involves a time-independent
function, whereas the OBP for the Wannier modes involve functions that are
time dependent.\textbf{ }

For $gN=0$ (Fig. 4, based on the GPE\textbf{)}, the PPD commences with
$2T$-periodicity - reflecting the bounce period - but after times
$t=498T-500T$, about half the bosonic atoms have transferred from the initial
wave-packet to the second wave-packet, while after times $t=998T-1000T$
essentially all of the atoms have transferred to the second wave-packet, and
then after times $t=1996T-1998T$, all of the atoms have transferred back to
the original wave-packet, which is now a slightly asymmetric Wannier-like
wave-packet (see Fig. 2). The transfer of bosonic atoms back and forth between
the two wave-packets appears as a periodic modulation of the OBP (Fig. 9 (a))
and a splitting of the peak in the FT (Fig. 9 (e)). The magnitude of the
splitting of the FT peak matches a mean-field calculation of the coupling
constant, $J=7.14\times10^{-4}$, based on the overlap of the two Wannier
wave-functions used as a basis for the condensate wave-function, similar to
calculations presented in Ref. \cite{Kuros20a}. Confirmation that the bosonic
atoms transfer back and forth between the two wave-packets is provided by
calculations of the number of atoms occupying the first ($N_{1}$) and second
wave-packets ($N_{2}$) versus time (Fig. 10). Complete transfer of bosonic
atoms between Wannier states $\Phi_{1}$, $\Phi_{2}$\ occurs at $gN=-0.006$%
\ but the transfer starts to cease at about $gN=-0.012$. Note that the PPD
plot for $gN=-0.006$ (Fig. 5) is almost identical to that for $gN=0$ (Fig. 4);
so in Fig. 10 the $gN=-0.006$\ plot also applies to $gN=0$. Also, in Fig. 5
for $gN=-0.006$ the TWA and GPE plots of the PPD are indistinguishable, as
they also would be for $gN=0$.

For $gN=-0.02$ (Fig. 7 - blue solid curve), the interaction is sufficiently
strong to suppress the transfer of bosonic atoms between the two wave-packets,
so that the initial wave-packet propagates as a single stable\ localised
wave-packet for times out to at least $t/T=1998$ mirror oscillations,
indicating the creation of a DTC. As may be seen in Fig. 10, this wave-packet
is the Wannier mode $\Phi_{2}$, which has a periodicity $2T$. The OBP (Fig. 9
(c)) further indicates little or no transfer of atoms between the two
wave-packets and the FT (Fig. 9 (g)) exhibits a single sub-harmonic peak at
half the driving frequency $\omega/2.$ Further evidence that there is little
or no transfer of atoms between the two wave-packets for $gN=-0.02$ is
provided by calculations of the atom numbers in the two wave-packets versus
time (Fig. 10). For an even larger interaction strength $gN=-0.1$, the PPDs
(Fig. 8) and OBPs (Fig. 9 (d)) are similar to those for $gN=-0.02$ but now
there is essentially no transfer of atoms between the two wave-packets for
times out to at least $t/T=1998$ mirror oscillations. In addition, there is no
significant broadening of the wave-packet for times out to $t/T=1998$,
indicating no significant heating of the atom cloud by the periodic driving or
by the quantum many-body fluctuations. The creation of a single stable
localised wave-packet at such a large interaction strength indicates that the
TWA calculations are valid at interaction strengths up to at least $gN=-0.1$.
This is in a regime where for models based on a single-band description of the
Bose-Hubbard model \cite{Sacha15a}, \cite{Giergel18a}, \cite{Giergiel20a} the
interaction energy per particle becomes comparable with the energy gap between
the first and second bands and the model is no longer valid. For
$gN=0,-0.006,-0.02$ and $-0.1$, the TWA (blue) and the mean-field GPE (red)
PPDs are indistinguishable for times out to at least $1998T$.

For $gN=-0.012$ (Fig. 6), the PPDs show only a weak transfer of bosonic atoms
between the two wave-packets, which is illustrated by the OBP and its FT
(Figs. 9 (b),(f)). Thus, $gN=-0.012$ represents the threshold interaction
strength for the creation of a single stable localised wave-packet (see Fig.
10). This represents the onset of DTC creation, since the atoms never
completely transfer out of the wave-packet $\Phi_{2}$, which is $2T$-periodic.
Furthermore, for times $t/T=1498$ $-1998$ mirror oscillations the wave-packet
for the TWA calculation is significantly smaller than for the mean-field GPE
calculation, indicating significant quantum depletion near this threshold
value. The difference (up to $\approx30\%$) between the TWA and the mean-field
GPE calculations due to quantum depletion near the threshold interaction
strength is clearly illustrated in the OBP and its FT for times
$t/T=1498-1998$ mirror oscillations (Figs. 9 (b),(f)).

The red dashed PPD curves in Fig. 7 represent TWA calculations for a repulsive
interaction $gN=+0.02$. Interestingly, these are very similar to the PPD
curves for $gN=-0.02$\ (blue solid curves), apart from a small shift in phase
of the interference fringes. In particular, there is no sign of decay or
broadening at times out to at least $1998$ mirror oscillations.

In summary, for harmonic trap initial conditions, we have a transition from a
wave-packet evolving with period $2T$ coupled (with coupling constant $J$ ) to
a second wave-packet with period $2T$, to a single stable wave-packet evolving
with period $2T$, and hence a DTC, at a threshold interaction strength
$gN\approx-0.012$. Below the threshold $gN$, the wave-packet has periodicity
$2T$, modulated at frequency $J$.

For the non-driving case with $gN=-0.10$ the PPD, OBP and its FT are presented
in Appendix \ref{Appendix - No Driving} in the Supplementary Material (see
Figs. 14, 15). No well-defined periodicity is apparent in either the PPD or
OBP results.\smallskip%

\begin{figure}[ptb]%
\centering
\includegraphics[
height=2.936in,
width=4.3803in
]%
{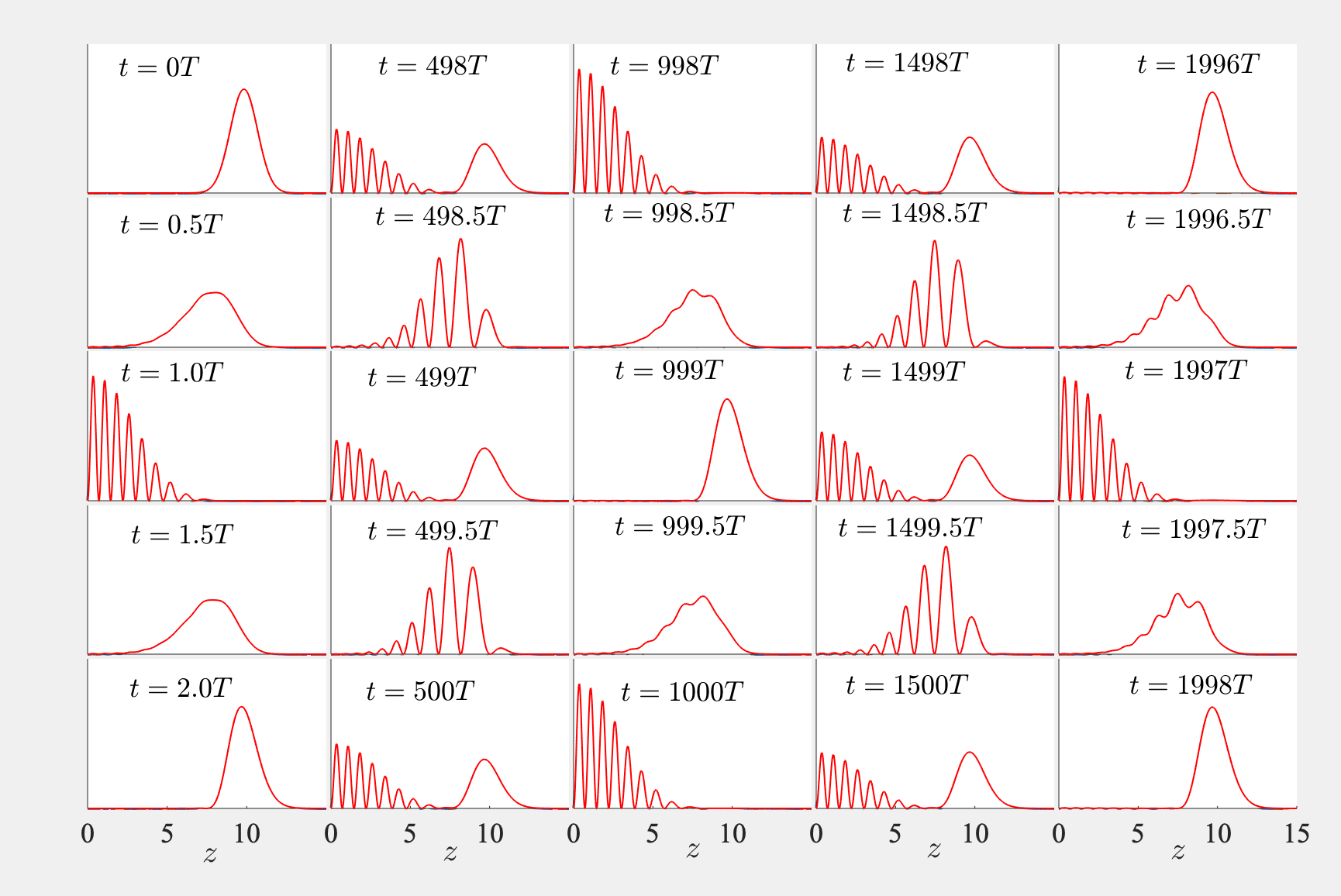}%
\caption{PPD for harmonic trap initial conditions as a function of $z$ and $t$
for $s=2$ and non-interacting case ($gN=0$) for $h_{0}=9.82$,
$\protect\widetilde{\omega}_{0}=0.68$, $N=600$, $\lambda=0.12$ and
$\omega=1.4$ by solving the time-dependent Schr\"{o}dinger equation.}%
\end{figure}
\begin{figure}[ptb]%
\centering
\includegraphics[
height=2.936in,
width=4.3803in
]%
{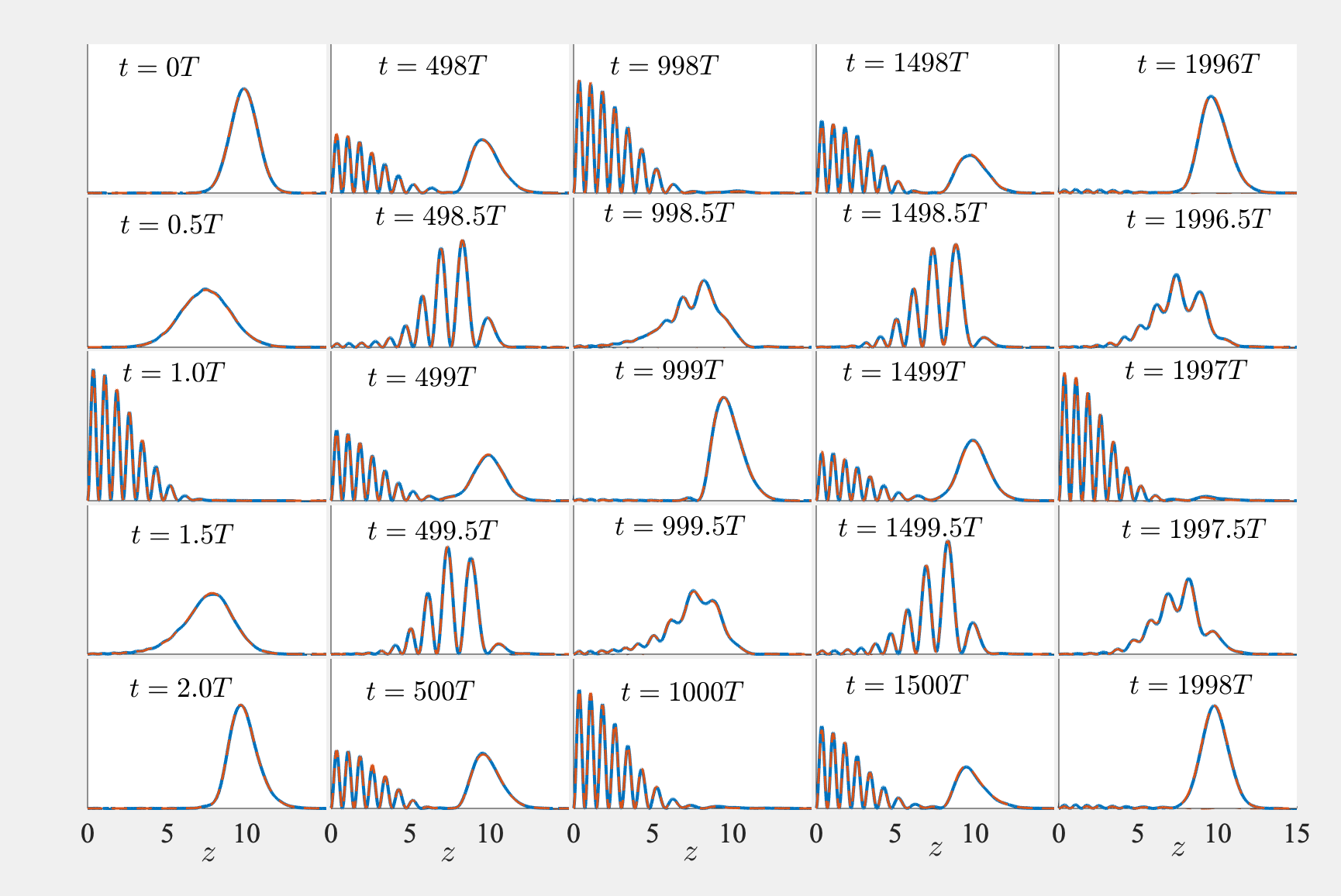}%
\caption{PPD for harmonic trap initial conditions as a function of $z$ and $t$
for interaction $gN=-0.006$ using the same parameters as Figure 2. The blue
solid (red dashed) curves are calculated using the TWA (GPE) approach.}%
\end{figure}
\begin{figure}[ptb]%
\centering
\includegraphics[
height=2.936in,
width=4.3803in
]%
{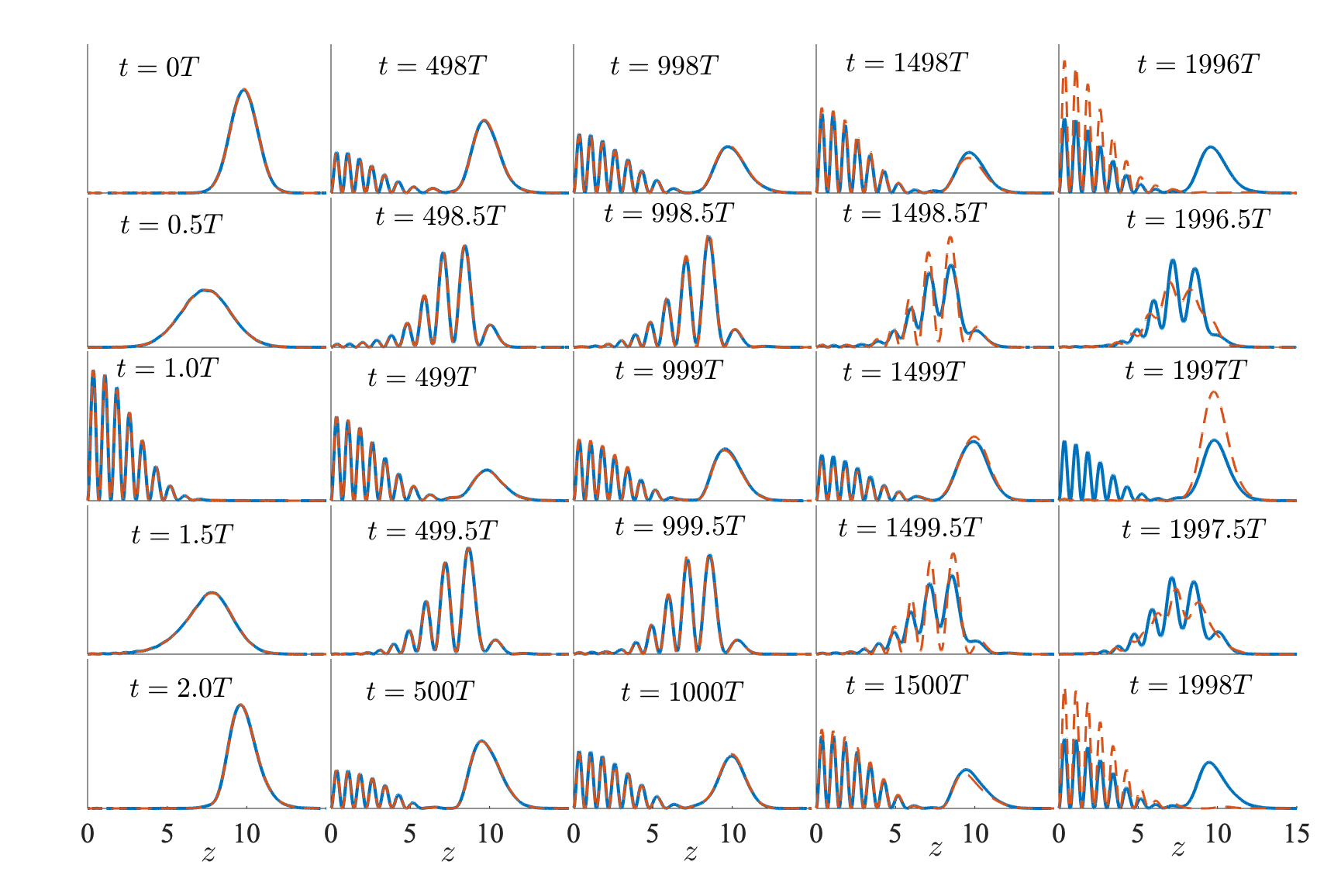}%
\caption{{}The same as Figure 5 but for $gN=-0.012$. The blue solid (red
dashed) curves are calculated using the TWA (GPE) approach.}%
\end{figure}
\begin{figure}[ptb]%
\centering
\includegraphics[
height=2.936in,
width=4.3803in
]%
{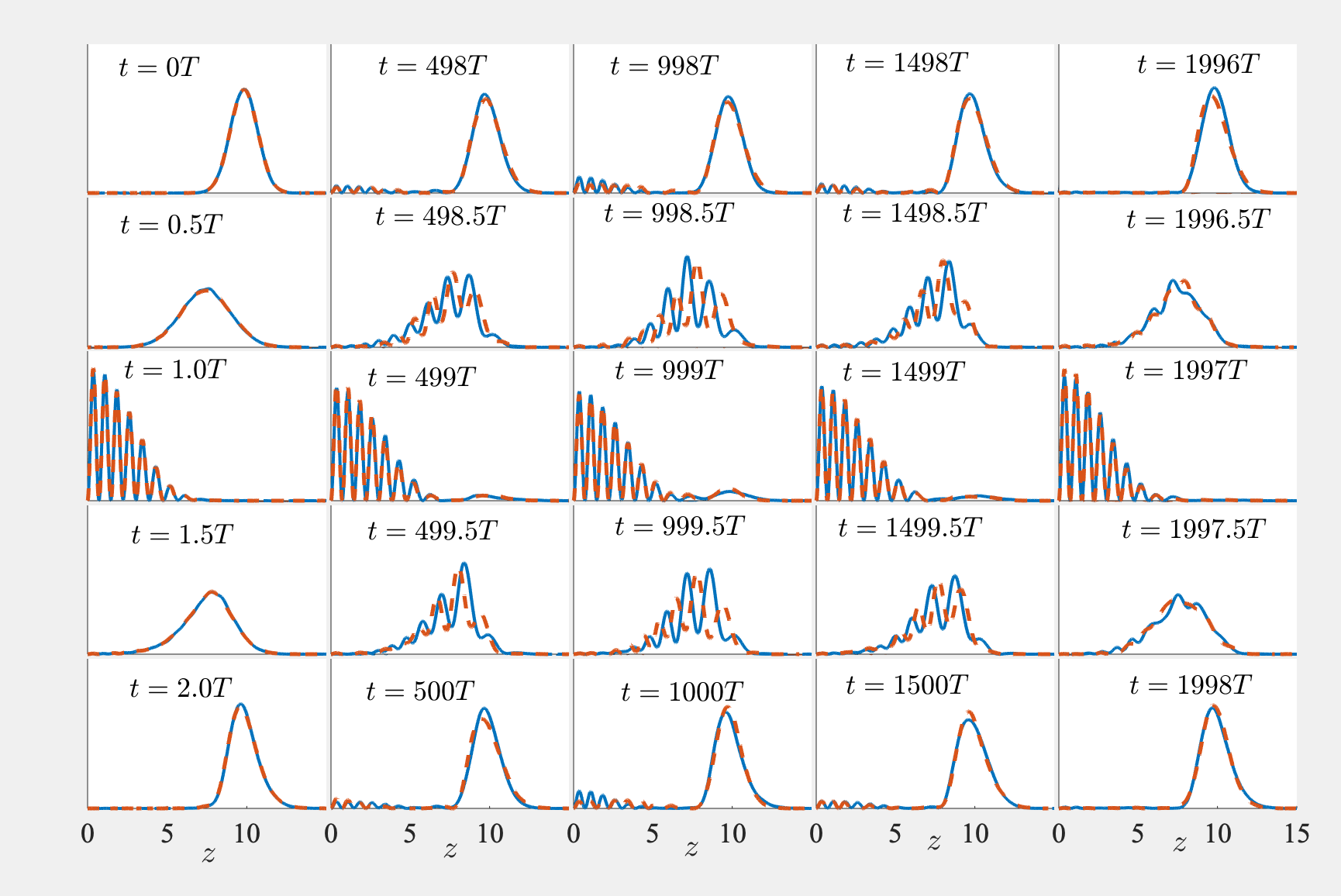}%
\caption{PPD for harmonic trap initial conditions as a function of $z$ and $t$
for interaction $|gN|=0.02$ using the same parameters as Figure 2. The blue
solid (red dashed) curves correspond to the attractive (repulsive) interaction
case.}%
\end{figure}
\begin{figure}[ptb]%
\centering
\includegraphics[
height=2.936in,
width=4.2739in
]%
{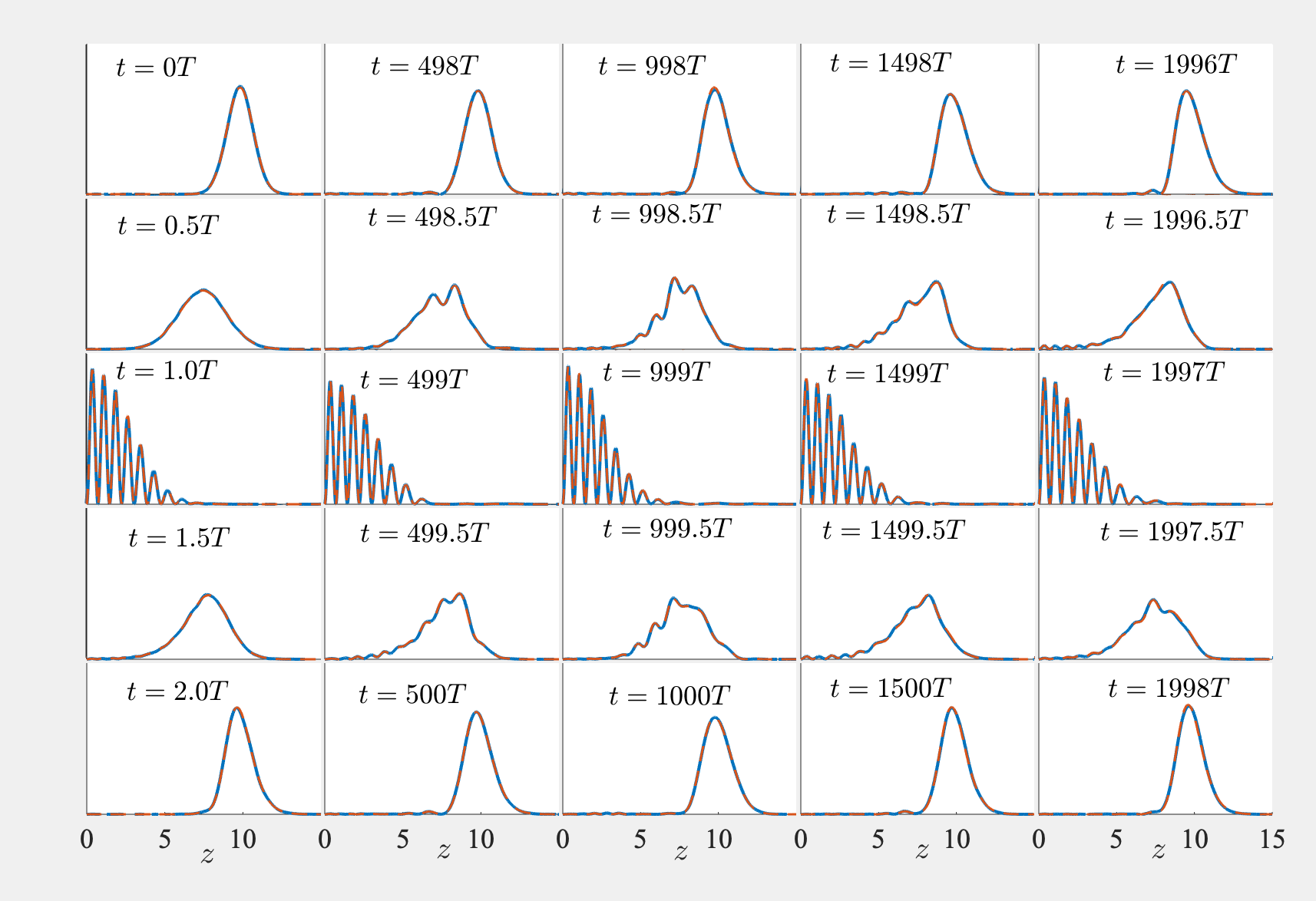}%
\caption{The same as Figure 5 but for $gN=-0.1$. The blue solid (red dashed)
curves are calculated using the TWA (GPE) approach.}%
\end{figure}
\begin{figure}[ptb]%
\centering
\includegraphics[
height=2.936in,
width=3.0597in
]%
{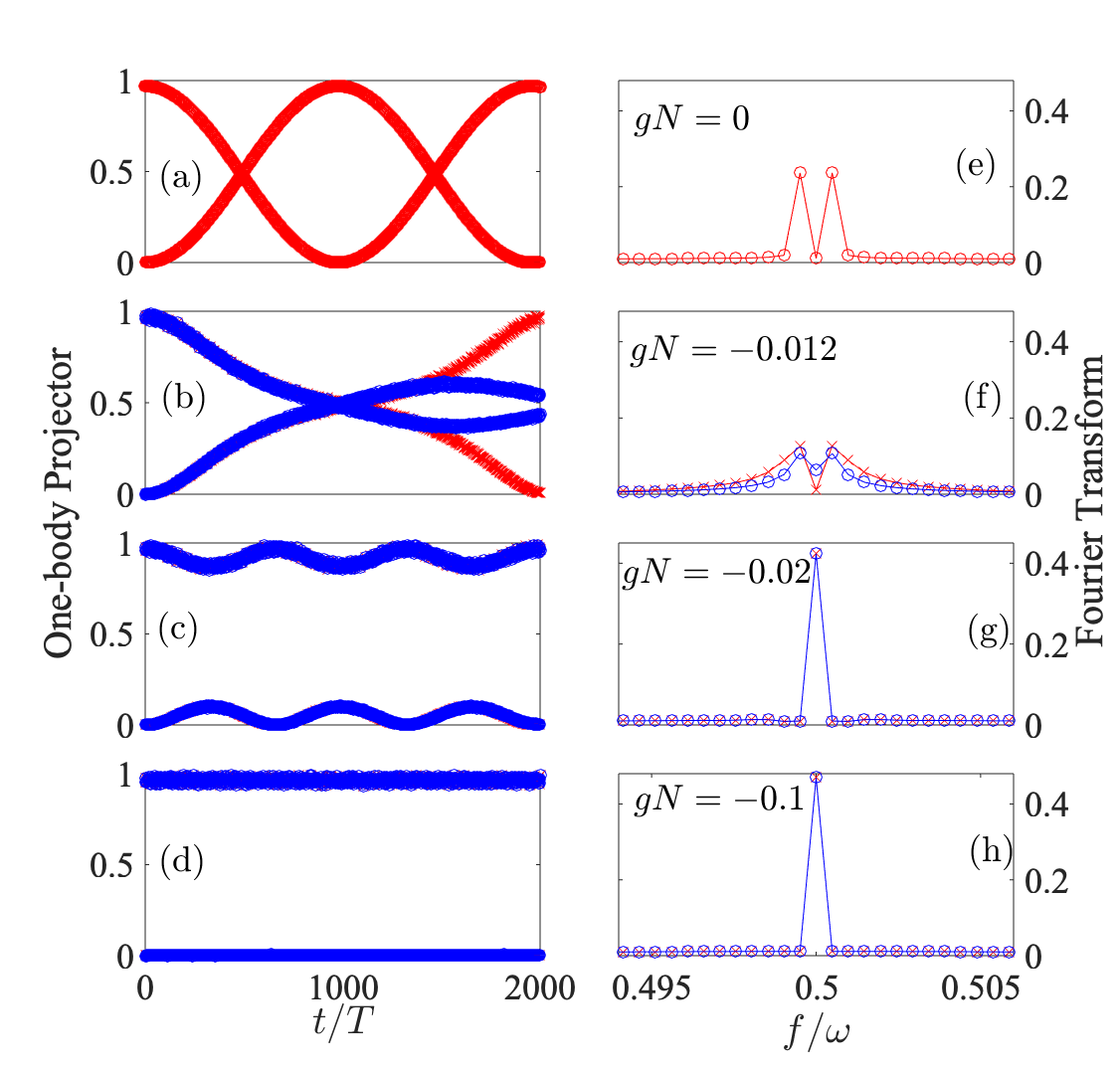}%
\caption{.OBP and the corresponding FT for harmonic trap initial conditions
(a), (b), (c), and (d) show the OBP for $gN=0$, $-0.012$, $-0.02$ and $-0.1$,
respectively; (e), (f), (g) and (h) show the corresponding FT. The red circle
symbols in (a) and (e) show a Schr\"{o}dinger equation calculation for the
non-interacting case. In other cases, the blue circle (red cross) symbols show
the TWA (GPE) results. The parameters are the same as in Figure 2.}%
\end{figure}
\begin{figure}[ptb]%
\centering
\includegraphics[
height=2.9369in,
width=3.9055in
]%
{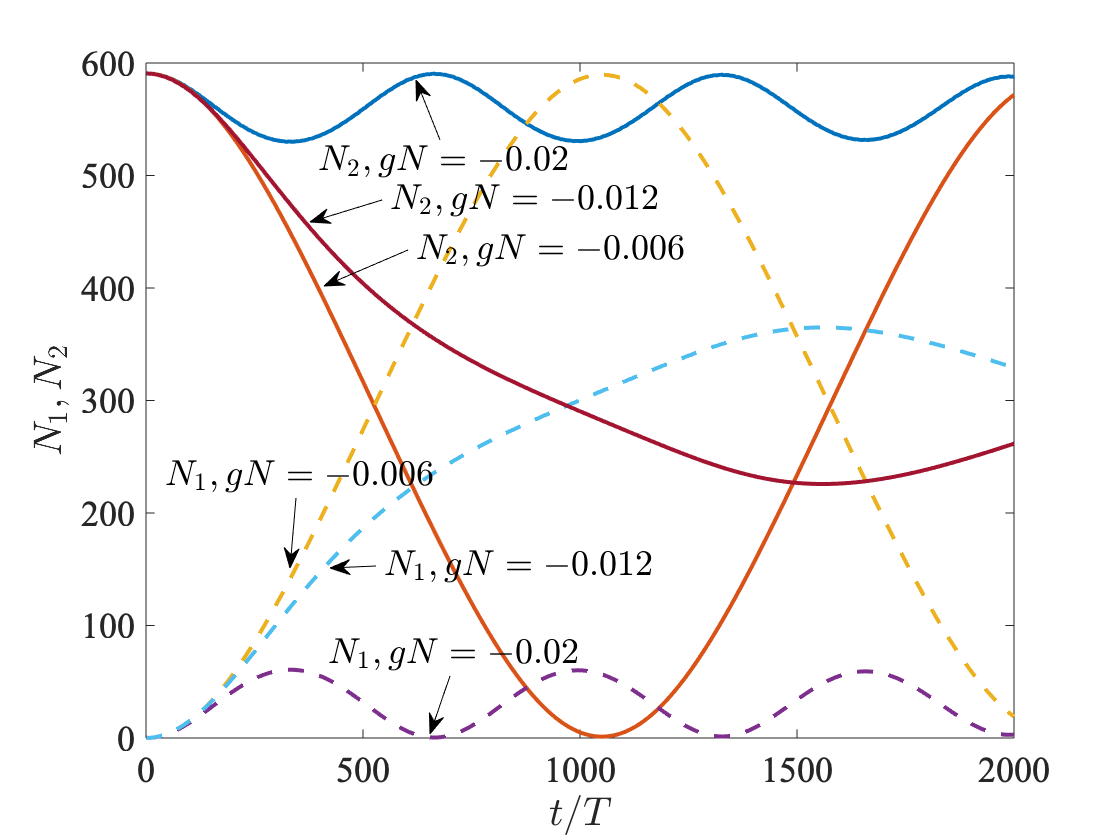}%
\caption{Expectation value of particle numbers $N_{1}$ and $N_{2}$ in single
particle Wannier modes $\Phi_{1}$ and $\Phi_{2}$ as a function of time for
different interaction strengths $gN$ shown in the figure. The parameters are
the same as in Figure 2.}%
\end{figure}

\subsection{Quantum Depletion}

Figure 11 shows a logarithmic plot of the maximum quantum depletion due to
quantum many-body fluctuations obtained for the TWA calculations as a function
of both attractive and repulsive interaction strengths for a BEC of $N=600$
atoms and evolution times $t/T\leq2000$ mirror oscillations. The calculations
were performed for both a Wannier-like initial state and a harmonic-trap
initial state with and without periodic driving of the mirror. Interestingly,
the plots are symmetrical about $gN=0$, as a result of the PPDs for an
attractive interaction being almost the same as those for a repulsive
interaction of the same magnitude (see Fig. 7).

For the case of a Wannier-like initial state, the quantum depletion in Fig. 11
is essentially zero for evolution times out to at least $t/T=2000$ mirror
oscillations, except close to the \textbf{critical} interaction strength for
breaking discrete time-translation symmetry, $|gN|=0.006$, where there are
small peaks in the quantum depletion corresponding to about $6$ atoms out of a
total of $N=600.$

For the case of a harmonic-trap initial state with driving, the quantum
depletion is less than $2$ atoms out of $N=600$ for evolution times out to at
least $t/T=2000$, except close to the threshold interaction strength for
creating a single localised wave-packet, $|gN|=0.012$, where the depletion is
as high as $260$ atoms out of $N=600$. To confirm the reliability of the TWA
calculations near the threshold interaction strength, we implemented an
analytical two-mode model for comparison and found excellent agreement within
the investigated time window. The two-mode model will be detailed elsewhere.
We note that our TWA calculations can handle a situation where the quantum
depletion is very large and where the time-dependent Bogoliubov theory
treatment of depletion \cite{Kuros20a} would break down.

For the case of a static mirror ($\lambda=0$), the TWA calculations indicate a
relatively large quantum depletion, i.e., $40-220$ atoms out of $N=600$ for
$|gN|\leq0.1$. A comparison with the case of a resonantly driven mirror
indicates that the DTC created by resonant periodic driving together with a
sufficiently strong particle interaction significantly suppresses the quantum
depletion due to quantum many-body fluctuations.

Figure 12 (a) shows the quantum depletion and the occupation number of atoms
in the condensate mode ($p_{0}$) and the second occupied mode ($p_{1}$) versus
time out to $t/T=2000$ mirror oscillations for $gN=-0.015,N=600$, while Fig.
12 (b) shows the natural orbitals for the condensate mode and the second
occupied mode at$\;t/T=2000$. The total quantum depletion closely matches the
difference $N-p_{0}$ over the range $t/T=2000$, indicating that the depletion
is essentially due to the escape of atoms from the condensate mode 1 to the
second occupied mode 2. For $gN=-0.015$, the difference $N-p_{0}$ is very
close to the occupation number of the second occupied mode $p_{1}$, indicating
that the occupation number for all other modes is essentially zero ($<$
$1.6\%$) for times out to at least $t/T=2000$. This implies that models based
on just two modes should work well. For $gN=-0.015$, i.e., fairly close to the
threshold interaction strength for creating a single localised wave-packet,
the quantum depletion is about $5$ atoms out of $N=600$ for times out to
$t/T=1500$ and about $15$ atoms out of $N=600$ for times out to $t/T=2000$
mirror oscillations.

The quantum depletion versus $t/T$\ curve in Fig. 12 suggests the quantum
depletion may be still increasing at $t/T=2000$. In Ref. \cite{Kuros20a} it is
found that in a two-mode theory treatment the quantum depletion for $gN=-0.02$
saturates at times of about $t/T\approx500\sqrt{N}$, or at $t/T\approx
12,000$\ mirror oscillations for $N=600$.\smallskip%
\begin{figure}[ptb]%
\centering
\includegraphics[
height=2.9369in,
width=4.8871in
]%
{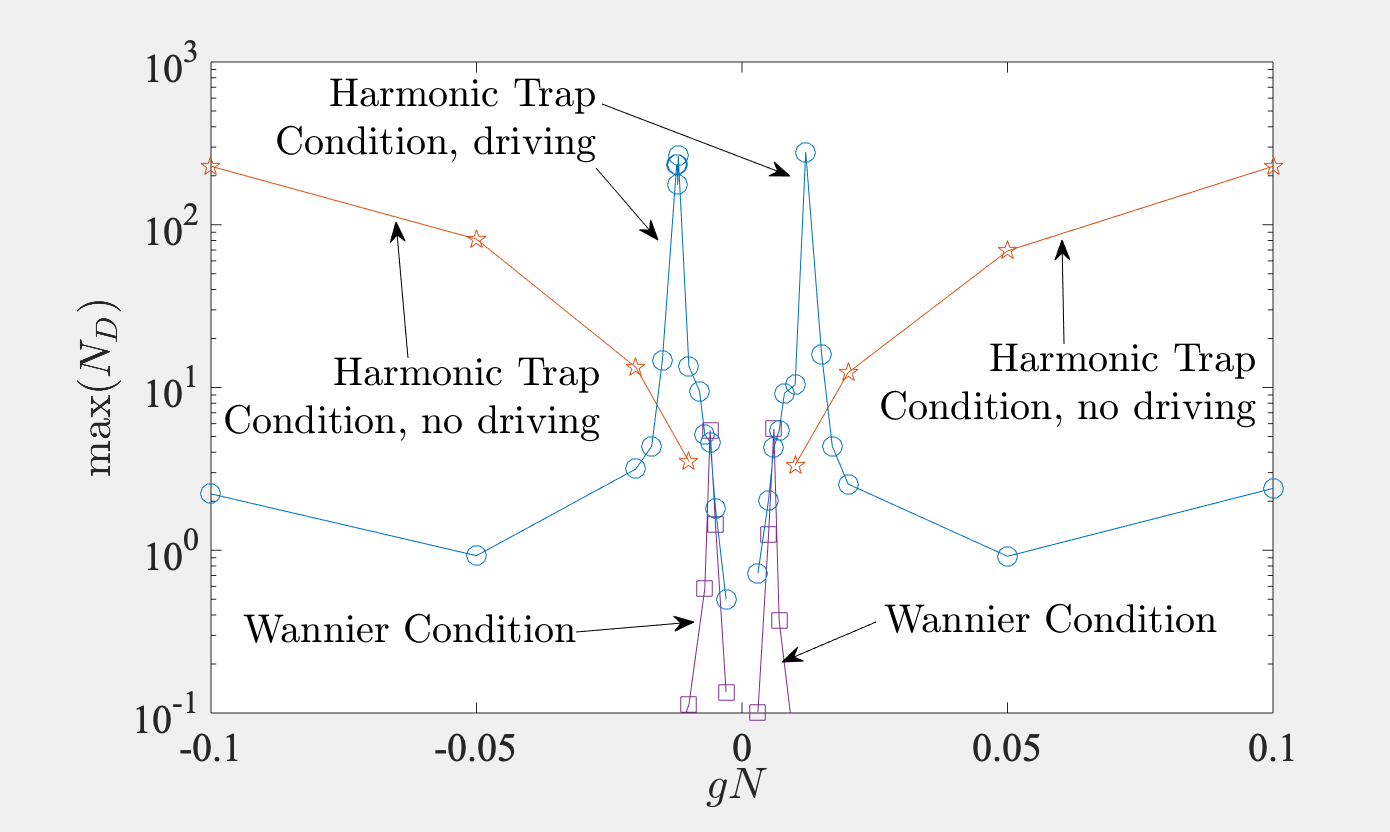}%
\caption{Maximum quantum depletion for $t<2000T$ as a function of $gN$. The
blue circles and purple squares correspond to the harmonic trap and Wannier
initial condition, respectively, where the calculation is carried out for
$N=600$ and other parameters are the same as Figure 2. We also show the
quantum depletion in red pentagrams for the harmonic trap initial condition
and a static mirror (no driving) for comparison.}%
\end{figure}
\begin{figure}[ptb]%
\centering
\includegraphics[
height=2.9369in,
width=3.9055in
]%
{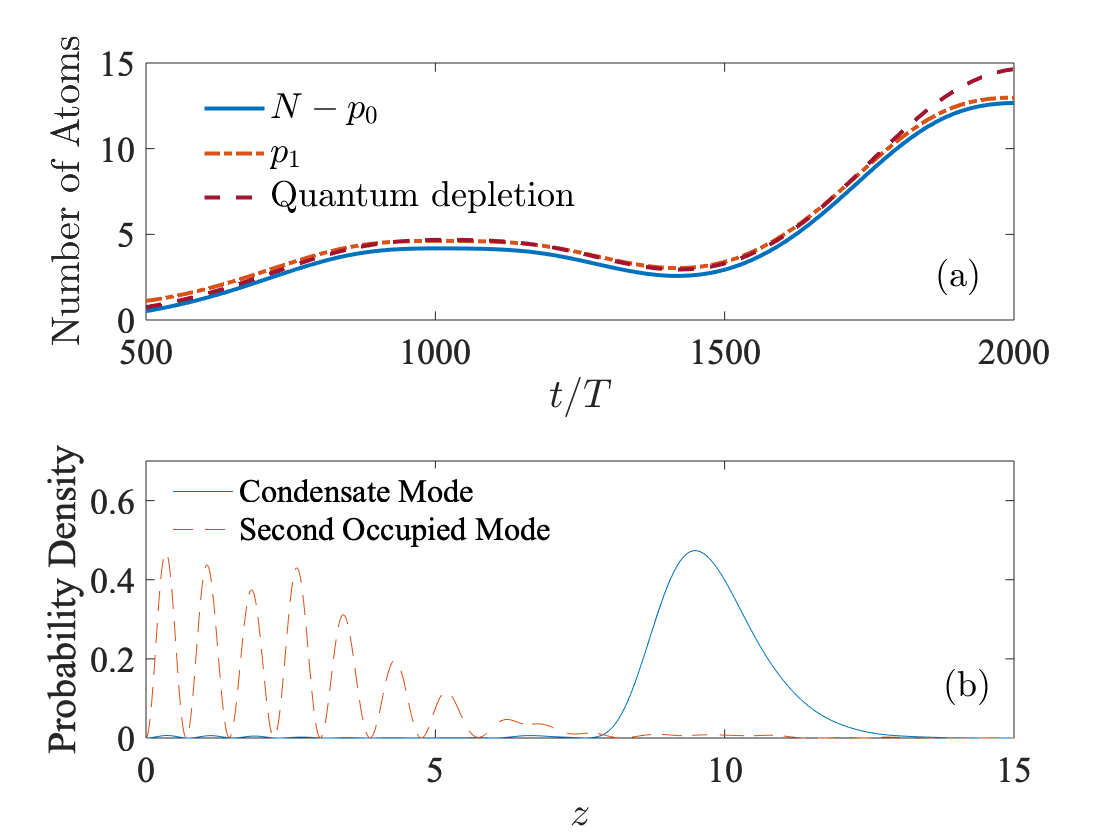}%
\caption{(a) Occupation number as a function of time for harmonic trap initial
condition with $gN=-0.015$, $N=600$, and other parameters are the same as
Figure 2. The blue solid (orange dash-dotted) curve shows $N-p_{0}$ ($p_{1}$),
respectively. We also show the quantum depletion as red dashed curve. (b) The
two mode functions for the condensed mode (blue solid curve) and second
occupied mode (orange dashed curve) correspond to $p_{0}$ and $p_{1}$,
respectively.}%
\end{figure}

\subsection{Mean Energy of Atoms at Long Times}

Figure 13 presents TWA calculations of the mean energy $\left\langle
H\right\rangle $ for the case of a harmonic-trap initial state for times out
to $t/T=2000$ mirror oscillations and different interaction strengths $gN$. We
observe that the mean energy does not significantly increase and typically
oscillates around an average value close to its initial value. The evolution
of the mean energy indicates that the system reaches a steady state with no
net energy pumped from the drive and which is consistent with DTC behaviour.

Since our TWA method is a fully multi-mode approach that allows thermalisation
and occupations of many modes in the non-driving case with the same
interaction strength, we conclude that the absence of thermalisation in a DTC
is due to quantum effects of the resonant driving.\smallskip\
\begin{figure}[ptb]%
\centering
\includegraphics[
height=2.9369in,
width=3.9055in
]%
{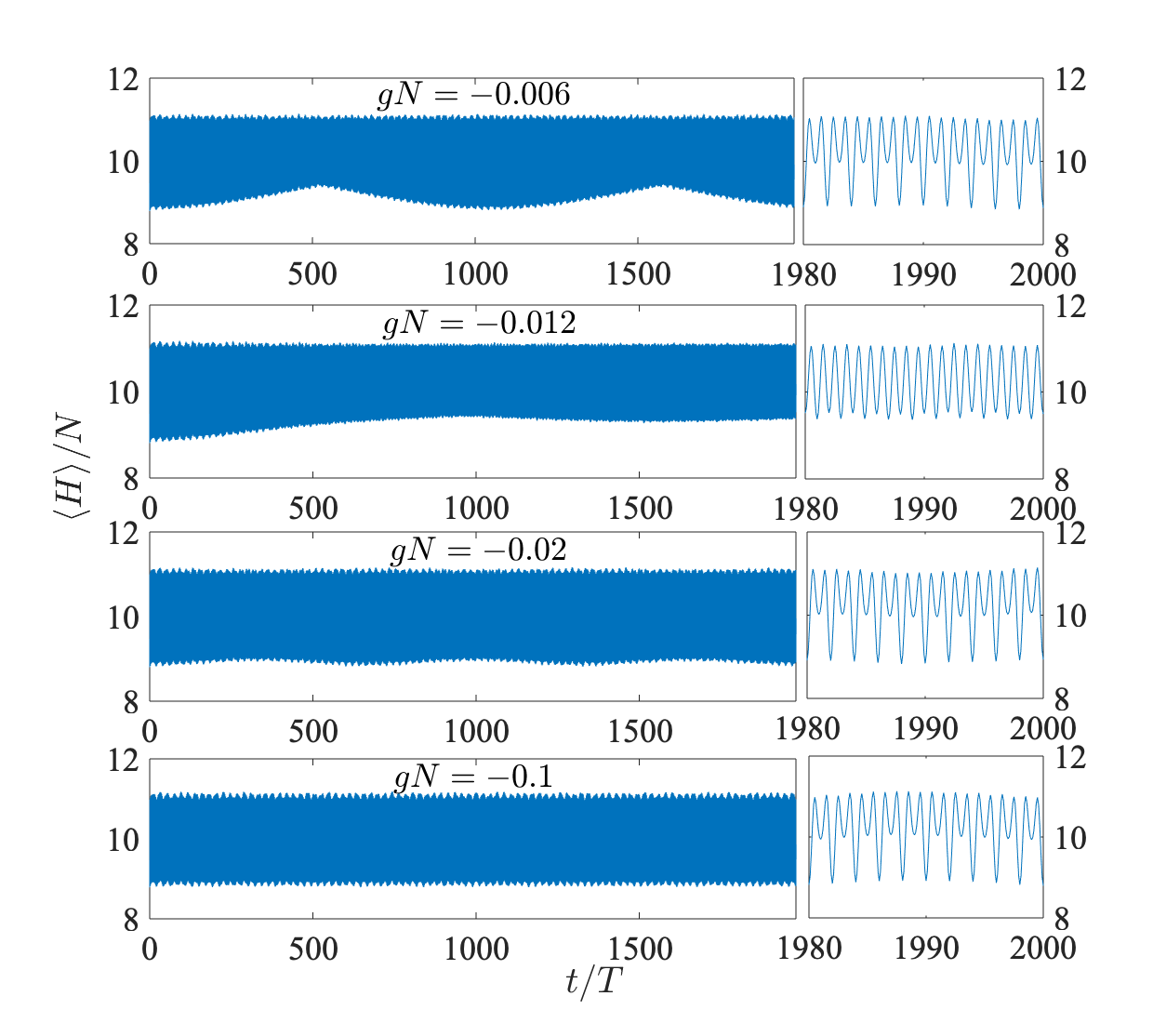}%
\caption{Mean energy as a function of time for harmonic trap initial condition
and different interaction strengths $gN$ shown in the figure. In these
calculations, $N=600$ and other parameters are the same as Figure 2.}%
\end{figure}
\bigskip\pagebreak

\section{Discussion}

\label{Section - Discussion}

For a realistic initial condition corresponding to a harmonic trap condensate
mode function, our many-body TWA calculations for $s=2$ agree closely with the
mean-field GPE\ calculations for times out to at least $2000$ mirror
oscillations, except at interaction strengths $gN$ very close to the threshold
value for transition to a single stable wave-packet and DTC formation, where
the PPDs differ significantly from those determined from the mean-field theory
GPE approach. On the other hand, for a hypothetical initial condition in which
the BEC wavefunction is treated as a linear combination of Wannier-like
states, our TWA calculations agree broadly with the mean-field
GPE\ calculations, with only a small difference at interactions close to the
critical interaction for discrete time-translation symmetry breaking and DTC
formation\textbf{.}

For typical attractive interaction strengths greater in magnitude than the
threshold value for DTC formation and for the chosen parameters, the TWA
calculations indicate a quantum depletion less than about two atoms out of a
total of $600$ atoms at times corresponding to $2000$ mirror oscillations.
Here, our TWA calculations are also found to be in close agreement with recent
many-body calculations based on a time-dependent Bogoliubov approach
\cite{Kuros20a}. Together with our results, this indicates that the occupation
of additional non-condensate modes is insignificant for most of the range of
parameters investigated. However, using the TWA approach we find that for
interaction strengths very close to the threshold value for DTC formation, the
quantum depletion due to the quantum fluctuations is as high as about $260$
atoms out of a total of $600$ atoms, at times corresponding to $2000$ mirror
oscillations. Since Bogoliubov theory assumes that the non-condensate field is
relatively small, the calculations based on a time-dependent Bogoliubov
approach \cite{Kuros20a} would break down here, as would the PPDs calculations
based on mean-field theory \cite{Sacha15a}, since the atoms are no longer in a
single mode. The PPD calculations based on the GPE differ from those obtained
via the TWA approach, as we have seen. For the range of conditions studied,
the quantum depletion in the vicinity of the threshold for creating a DTC
results mainly from the escape of atoms from the condensate mode to the second
occupied mode and the occupation of other modes is essentially zero
($<1.6\%$). This suggests that many-body models based on just the two dominant
modes should work well for certain applications.

The general agreement we have found for our TWA calculations with those based
on the mean-field GPE and time-dependent Bogoliubov theory (TDBT) approaches
in Refs. \cite{Sacha15a}, \cite{Giergel18a}, \cite{Kuros20a},
\cite{Giergiel20a} (except near the threshold value for $gN$) is not by
chance. In Appendix
\ref{Appendix - Mean Field Theory and Time-Dependent Bogoliubov Theory} we
show that the mean-field GPE and TDBT equations can be derived from the
phase-space TWA\ theory via an approximation in situations where quantum
depletion is small.

We also observe that the mean energy does not significantly increase for times
out to at least $2000$ mirror oscillations, and typically oscillates around an
average value close to its initial value. The evolution of the mean energy
indicates that the system reaches a steady state with no net energy pumped
from the drive and is consistent with a DTC behaviour. Thus, our TWA approach
predicts that thermalisation does not occur, at least out to $2000$ mirror
oscillations. Thermalisation cannot be treated using mean-field theory approaches.

When the driving field is turned off and for a relatively strong interaction
($0.05\leq|gN|\leq0.1$), the interaction is strong enough to couple many modes
and we find using the TWA that the quantum depletion can be more than $220$
atoms out of a total of $600$ atoms at times corresponding to $2000$ mirror
oscillations. However, with driving, the quantum depletion associated with
quantum fluctuations is strongly suppressed, and thermalisation quenched,
implying that the absence of our system's thermalisation is a genuine
many-body effect, and along with DTC formation, is a direct consequence of
driving in the presence of a sufficiently strong interaction.

Thus, in certain circumstances, namely for realistic harmonic trap initial
conditions and where the interaction parameter $gN$ is close to the threshold
value for DTC formation ($gN=-0.012$ for our parameters), previous theories
based on mean-field theory (\cite{Sacha15a}, \cite{Kuros20a}) or
time-dependent Bogoliubov theory \cite{Kuros20a} break down, and the
phase-space theory TWA approach is required in order to provide a
comprehensive treatment of DTC behaviour covering all regimes.

We also find that the dynamical behaviour of our system is largely independent
of whether the boson-boson interaction is attractive or repulsive, and that
for sufficiently large repulsive interactions it is possible to create a
stable DTC based on repulsive interactions. The ability to create a stable
time crystal with repulsive interactions should allow more flexibility; for
example, the use of repulsive interactions may allow us to create a DTC at
still larger interaction strengths where there is no limitation imposed by
bosenova collapse \cite{Roberts01a} or the formation of a bright soliton as
there is with attractive interactions.

The explanation of the behaviour being essentially the same for negative and
positive $g$ can be provided from the Ito stochastic field equations
(\ref{Eq.ItoSFEA}), (\ref{Eq.ItoSFAB}), noting that all the observable
quantities are determined from $\widetilde{\psi}(z,t)$ and $\widetilde{\psi
}^{+}(z,t)$. For large $N$ the $\delta_{C}(z,z)$ term can be ignored, and the
solutions for $-g$ can be obtained from those for $+g$ via the substitution
$\widetilde{\psi}(z,t)=i\,\widetilde{\psi}(z,t)^{\#},\widetilde{\psi}%
^{+}(z,t)=i\,\widetilde{\psi}^{+}(z,t)^{\#}$ - which is possible for the
double phase-space Wigner distribution $W^{+}$ where the field functions and
stochastic field functions $\widetilde{\psi}(z,t)$ and $\widetilde{\psi}%
^{+}(z,t)$ are allowed to be unrelated. The PPD, etc., calculated from the new
fields would be for $-g$, and would have the same $z,t$ behaviour as for
$+g$.\smallskip

\section{Summary and Conclusions}

\label{Section - Conclusions and Summary}

We have presented a full theoretical many-body quantum study based on
physically realisable initial conditions for creating discrete time crystals
using a Bose-Einstein condensate which is allowed to bounce resonantly on an
oscillating mirror in a gravitational field, as proposed in \cite{Sacha15a}.
Our theory of DTC creation allows for the effects of there being many modes
that the bosons could occupy; it allows for the mutually important effects of
both interactions and driving; it determines whether or not thermalisation
occurs; and it enables the sensitivity of DTC behaviour to changes in the
initial conditions to be studied - including the effect of the initial BEC temperature.

The significance of the theory presented here is that it is not based on
restricted assumptions that have applied to previous work on DTC formation in
such systems, such as a mean-field theory (\cite{Sacha15a} and \cite{Kuros20a}%
), time-dependent Bogoliubov theory (\cite{Kuros20a}) or a two-mode model. In
a mean-field theory, all the bosons are assumed to remain in a single mode,
and a time-dependent Bogoliubov theory is restricted to situations where the
depletion from the condensate mode is small. In contrast, our TWA theory
allows for the large depletion and quantum fluctuations that occur near the
threshold $gN$ value for a DTC to form. Thermalisation can also be treated,
unlike in mean-field theory.

The ability to create a DTC over a broad range of attractive and repulsive
interaction strengths also provides flexibility for applications of DTCs; for
example, to the study of a wide range of non-trivial condensed matter
phenomena in the time domain. Condensed matter phenomena that have been
proposed based on a driven bouncing BEC platform include Anderson localization
\cite{Sacha15c} and many-body localization \cite{Mierzejewski17a} in the time
domain due to temporal disorder; Mott insulator-like phases in the time domain
\cite{Sacha15c}; quasi-crystalline structures in time \cite{Giergiel19a};
topological time crystals \cite{Giergiel19b}; time crystals with exotic
long-range interactions \cite{Giergiel18b}; and dynamical quantum phase
transitions in time crystals \cite{Kosier18a}.

Possible future directions of this work include applying the TWA approach for
times much longer than $2000T$. We propose to investigate whether long-time
calculations could be based on using relatively small numbers of Wannier
modes, rather than the large number of gravitational modes used in the present
study. We also propose to apply the TWA approach to treat higher order
$sT$-periodicity, such as in the range $s=5-10$. This type of DTC has been
predicted using mean-field theory (\cite{Giergel18a}, \cite{Giergiel20a}), but
the conditions for creating a DTC could be investigated using a theoretical
approach that does not rely on the mean-field assumption. The TWA theory is
already formulated for treating finite temperature effects, and could be used
to investigate the conditions under which a DTC can be created at non-zero
temperatures. Finally, we wish to explore more fully the conditions for
$sT$-periodicity than is possible in this initial paper, such as the range of
the boson-boson interaction strength, different initial conditions and the
stability of the DTC. The upper bound of useable attractive interaction
strengths is limited by bosenova collapse \cite{Roberts01a}, and it would be
interesting to also study this regime.\smallskip

\subsection{Acknowledgements}

We thank Krzysztof Sacha for many fruitful discussions and suggestions, and
for a critical reading of this article. Also, Peter Drummond, King Lun Ng and
Run Yan Teh are acknowledged for discussions on phase-space theory. Support of
the Australian Research Council (DP190100815) is gratefully acknowledged. JW
acknowledges an ARC Discovery Early Career Researcher Award (DE180100592).
\smallskip

\subsection{Orcid ID}

These are:

Jia Wang 0000-0002-9064-5245

Peter Hannaford 0000-0001-9896-7284

Bryan Dalton 0000-0002-1176-7528\pagebreak

\section{Appendix A\ - Mean-Field Theory and Time-Dependent Bogoliubov Theory}

\label{Appendix - Mean Field Theory and Time-Dependent Bogoliubov Theory}

In this appendix we show how the mean-field Floquet theory equation set out as
Eq. (2) in Ref \cite{Sacha15a}, and which is the key starting equation in
subsequent papers \cite{Giergel18a}, \cite{Kuros20a}, \cite{Giergiel20a}, can
be derived as an approximation from the phase-space truncated Wigner
approximation (TWA)\ theory. As we will see, Eq (2) is the equation that a
condensate field function contribution in the stochastic field function would
satisfy. We also show how the time-dependent Bogoiliubov theory, which is a
further development in Ref. \cite{Kuros20a} can also be seen as an
approximation to the TWA approach. The starting point is the Ito stochastic
field equation for the stochastic field function, but now based on the grand
canonical Hamiltonian $\widehat{H}_{G}=\widehat{H}-\mu\widehat{N}$, where
$\widehat{H}$ is the Hamiltonian, $\widehat{N}$ is the number operator and
$\mu$ is the chemical potential. We only consider number conserving states. As
usual we consider a regime where the boson number $N$ is large, $N\gg1$. The
advantage of this approach is that Eq (2) is now based on a standard many-body
theory approach, whereas at present its justification rests on it being a
physically reasonable combination of the Gross-Pitaeskii equation (GPE) for a
single-mode condensate and the Floquet equation for a single-particle system
in a periodic potential.\smallskip

\subsection{FFPE and Ito SFE - Extra Term}

The density operator $\widehat{\rho}$ satisfies the Liouville-von Neumann
equation
\begin{equation}
\frac{\partial}{\partial t}\widehat{\rho}=\frac{1}{i\hbar}[\widehat{H}%
,\widehat{\rho}]
\end{equation}
but for number conserving states $[\widehat{N},\widehat{\rho}]=0$, so that we
also have
\begin{equation}
\frac{\partial}{\partial t}\widehat{\rho}=\frac{1}{i\hbar}[\widehat{H}%
_{G},\widehat{\rho}] \label{Eq.LVNGrandCanonH}%
\end{equation}
involving the grand canonical Hamitonian.

The functional Fokker-Planck equation (see Eqs.
(\ref{Eq.FFPEKineticFermiFldModel}), (\ref{Eq.FFPEPotentialFermiFldModel}) and
(\ref{Eq.FFPEInteractFermiFldModel})) for the Wigner distribution functional
$W$ then has an additional term
\begin{equation}
\left(  \frac{\partial}{\partial t}W\right)  _{\mu}=-\frac{\mu}{i\hbar}\left(
-\frac{\delta}{\delta\psi}(\psi W)+\frac{\delta}{\delta\psi^{+}}(\psi
^{+}W)\right)  \label{Eq.FFPEMuTerm}%
\end{equation}
and this results in an extra term in the Ito stochastic field equations on the
right hand side
\begin{equation}
\left(  \frac{\partial\widetilde{\psi}}{\partial t}\right)  _{\mu}=\frac
{1}{i\hbar}(-\mu\,\widetilde{\psi})\qquad\qquad\left(  \frac{\partial
\widetilde{\psi}^{+}}{\partial t}\right)  _{\mu}=\frac{1}{i\hbar}%
(+\mu\,\widetilde{\psi}^{+}) \label{Eq.ItoSDEMuTerm}%
\end{equation}

Consequently, from Eqs. (\ref{Eq.ItoSFEA}) and (\ref{Eq.ItoSFAB}), the full
Ito stochastic field equations are now
\begin{align}
&  \frac{\partial}{\partial t}\widetilde{\psi}(z,t)\nonumber\\
&  =-\frac{i}{\hslash}\left[  -\frac{\hslash^{2}}{2m}\frac{\partial^{2}%
}{\partial z^{2}}\widetilde{\psi}(z,t)+V(z,t)\widetilde{\psi}%
(z,t)+g\{\widetilde{\psi}^{+}(z,t)\widetilde{\psi}(z,t)-\delta_{C}%
(z,z)\}\widetilde{\psi}(z,t)-\mu\widetilde{\psi}(z,t)\right] \nonumber\\
&  \label{Eq.ItoSFEAMu}%
\end{align}
and
\begin{align}
&  \frac{\partial}{\partial t}\widetilde{\psi}^{+}(z,t)\nonumber\\
&  =+\frac{i}{\hslash}\left[  -\frac{\hslash^{2}}{2m}\frac{\partial^{2}%
}{\partial z^{2}}\widetilde{\psi}^{+}(z,t)+V(z,t)\widetilde{\psi}%
^{+}(z,t)+g\{\widetilde{\psi}^{+}(z,t)\widetilde{\psi}(z,t)-\delta
_{C}(z,z)\}\widetilde{\psi}^{+}(z,t)-\mu\widetilde{\psi}^{+}(z,t)\right]
\nonumber\\
&  \label{Eq.ItoSFABMu}%
\end{align}
\smallskip

\subsection{Condensate and Non-Condensate Components}

We first make the large $N$ approximation of neglecting the terms $\delta
_{C}(z,z)$ in comparison to $\widetilde{\psi}^{+}(z,t)\widetilde{\psi}(z,t) $
since the former term is of order $1$ whereas the latter is of order $N$.

We then write the stochastic field functions as the sum of a deterministic
term $\Phi_{0}(z,t)$ (or its conjugate) corresponding to the condensate mode
and a stochastic term $\delta\widetilde{\psi}(z,t)$ (or $\delta\widetilde{\psi
}^{+}(z,t)$) corresponding to the non-condensate modes. The stochastic terms
$\delta\widetilde{\psi}(z,t)$ or $\delta\widetilde{\psi}^{+}(z,t)$ are assumed
to be small. A similar approach is made in time-dependent Bogoliubov theory in
regard to the field operators.

Thus, we have
\begin{equation}
\widetilde{\psi}(z,t)=\Phi_{0}(z,t)+\delta\widetilde{\psi}(z,t)\qquad
\widetilde{\psi}^{+}(z,t)=\Phi_{0}^{\ast}(z,t)+\delta\widetilde{\psi}^{+}(z,t)
\label{Eq.CondNonCondFields}%
\end{equation}

If we substitute into Eqs. (\ref{Eq.ItoSFEAMu}) and (\ref{Eq.ItoSFABMu}) and
retain only the lowest order terms we find that
\begin{align}
&  \frac{\partial}{\partial t}\Phi_{0}(z,t)\nonumber\\
&  =-\frac{i}{\hslash}\left[  -\frac{\hslash^{2}}{2m}\frac{\partial^{2}%
}{\partial z^{2}}\Phi_{0}(z,t)+V(z,t)\Phi_{0}(z,t)+g\{\Phi_{0}^{\ast}%
(z,t)\Phi_{0}(z,t)\}\Phi_{0}(z,t)-\mu\Phi_{0}(z,t)\right] \nonumber\\
&  \label{Eq.GPEFloquet}%
\end{align}
This is the same as Eq.(2) in Ref. \cite{Sacha15a}.

The stochastic non-condensate terms satisfy the equations
\begin{align}
&  \frac{\partial}{\partial t}\delta\widetilde{\psi}(z,t)=\nonumber\\
&  -\frac{i}{\hslash}\left[  -\frac{\hslash^{2}}{2m}\frac{\partial^{2}%
}{\partial z^{2}}+V(z,t)-\mu+2g\,\Phi_{0}^{\ast}(z,t)\Phi_{0}(z,t)\right]
\;\delta\widetilde{\psi}(z,t)\nonumber\\
&  -\frac{i}{\hslash}\left[  g\,\Phi_{0}(z,t)^{2}\right]  \;\delta
\widetilde{\psi}^{+}(z,t) \label{Eq.NonCondStochA}%
\end{align}
and
\begin{align}
&  \frac{\partial}{\partial t}\delta\widetilde{\psi}^{+}(z,t)=\nonumber\\
&  +\frac{i}{\hslash}\left[  -\frac{\hslash^{2}}{2m}\frac{\partial^{2}%
}{\partial z^{2}}+V(z,t)-\mu+2g\,\Phi_{0}^{\ast}(z,t)\Phi_{0}(z,t)\right]
\;\delta\widetilde{\psi}^{+}(z,t)\nonumber\\
&  +\frac{i}{\hslash}\left[  g\,\Phi_{0}^{\ast}(z,t)^{2}\right]
\;\delta\widetilde{\psi}(z,t) \label{Eq.NonCondStochB}%
\end{align}
correct to the lowest order in $\delta\widetilde{\psi}(z,t)$ and
$\delta\widetilde{\psi}^{+}(z,t)$. Here, we see that the stochastic
fluctuation fields $\delta\widetilde{\psi}(z,t)$ and $\delta\widetilde{\psi
}^{+}(z,t)$ are coupled. The last two equations are analogous to
time-dependent Bogoliubov-de Gennes equations (compare the time-independent
Eqs. (\ref{Eq.StandardBogDeGennes})). Note the well-established $2g$ factor
that appears in Bogoliubov theory. The difference is that within the TWA
approach the non-condensate fields $\delta\widetilde{\psi}(z,t)$ and
$\delta\widetilde{\psi}^{+}(z,t)$ are treated as stochastic quantities,
whereas in time-dependent Bogoliubov theory, the equivalent quantities would
be time-dependent Bogoliubov mode functions $\overline{u}_{k}(z,t)$,
$\overline{v}_{k}(z,t)$.

The conclusion then is that the approach to mean-field theory and
time-dependent Bogoliubov theory in Refs. \cite{Sacha15a}, \cite{Giergel18a},
\cite{Kuros20a}, \cite{Giergiel20a} are now shown to be an approximation to
the TWA theory, valid in regimes where almost all the bosons are in one
condensate mode and the quantum depletion into non-condensate modes is
relatively small. There will of course be regimes where this approximate
solution to the TWA equations will break down, and in future work we hope to
explore some of these. \smallskip

\subsection{Energy Functional}

In Ref. \cite{Sacha15a} Eq.(2) can be derived by minimising the energy
functional set out in Eq.(3) of Ref. \cite{Sacha15a}, subject to the
normalisation constraint that $%
{\displaystyle\int}
dz\;\Phi_{0}^{\ast}(z,t)\Phi_{0}(z,t)=N.$ The energy functional involves an
integral over time $t$, which in Ref. \cite{Sacha15a} is taken over the
interval $0$ to $2T$, where $T$ is the period of the driving potential. The
energy functional could be taken as an average over interval $sT$ in general
\begin{equation}
E[\Phi_{0}]=\frac{1}{sT}%
{\displaystyle\int}
dz%
{\displaystyle\int\limits_{0}^{sT}}
dt\;\Phi_{0}^{\ast}\left[  -\frac{\hslash^{2}}{2m}\frac{\partial^{2}}{\partial
z^{2}}+V(z,t)-i\hbar\frac{\partial}{\partial t}+\frac{g}{2}\Phi_{0}^{\ast}%
\Phi_{0}\right]  \Phi_{0} \label{Eq.EnergyFnal}%
\end{equation}
so minimising this with respect to changes in $\Phi_{0}$ subject to the
contraint leads to Eq. (\ref{Eq.GPEFloquet}) above for $\Phi_{0}$. The
chemical potential appears via a Lagrange undetermined multiplier. The choice
of the time interval for a different $s$ does not result in a different
equation for $\Phi_{0}$.

The advantage of this linkage is that if the condensate wave-function is
expanded in terms of Wannier functions $\Phi_{i}(z,t)$ which have periodicity
$sT$%
\begin{equation}
\Phi_{0}(z,t)=%
{\displaystyle\sum\limits_{i}}
a_{i}(t)\Phi_{i}(z,t) \label{Eq.WannierExpn}%
\end{equation}
the energy functional becomes a function $E(a_{i})$ of the coefficients
involving a fourth-order polynomial. This can then be minimised to give the
condensate wave-function. The energy function will then involve coefficients
that are space-time integrals involving the Wannier functions up to the fourth
power. However, there are no extra exponential factors of the form
$\exp(-in\omega t)$ that occur in many-body formulations of Floquet theory
based on Shirley's paper \cite{Shirley65a}, and which can be eliminated via a
fast-rotating approximation. \pagebreak

\section{Appendix B - Details regarding TWA Validity}

\label{Appendix - Details re TWA Validity}

Some specific questions regarding the TWA validity and reliability are as follows:

(a) Does the accuracy of the TWA calculations change when the interactions
change from weak to strong? Does its accuracy depend on the extent of
correlations between different modes of the system?

At present we have no reason to believe that our TWA calculations would not be
accurate for the weak interaction situation we studied. Even if (as according
to Ref. \cite{Blakie08}) the TWA is not suitable to describe systems with very
strong interaction and strong correlations, in our system we focus on the weak
interaction situation. We strongly believe though that the TWA can treat the
strong interaction regime, as the derivation of the final Ito stochastic field
equations does not depend on the size of the coupling constant $g$. In regard
to correlations, our multi-mode TWA approach allows for correlations between
modes to develop, so the TWA should be able to treat strong correlations. Many
gravitational modes become correlated during the evolution, but for our weak
interaction case the main correlation is between only two Wannier modes.

(b) Do the TWA predictions hold for long evolution times or will build up of
errors become uncontrollable? Is the time over which the TWA remains accurate
long enough to confirm time crystal behaviour?

The evolution error is a numerics issue rather than one involving the theory's
formalism. In principle, the TWA method can accumulate error in our adopted
Runge-Kutta numerical method as the evolution time increases. However, in our
chosen time-window $(t<2000T)$, convergent tests with respect to time steps
have been performed and show very good convergence. Stable time-crystal
behaviour has been seen within this time window for $|gN|>0.012$ with the
harmonic trap initial condition. The time evolution for even longer times is
currently limited by computational resources. As indicated in the last
paragraph of the Summary, we plan a further investigation using more advanced
computers regarding whether the time crystal will remain stable after much
longer evolution times.

(c) What will the effect of the TWA approximation be on the results? Which
features will display differences with respect to exact behaviour?

Estimating the error coming from the approximation that neglects the
third-order functional derivative term is very difficult. As far as we know,
no exact calculation exists for the solution of the full FFPE for the sort of
many-boson systems we are studying and no equivalent Ito stochastic field
equations have even been derived when third-order derivative terms are
included. Therefore, this important question will remain unanswered until
researchers find a way to obtain exact solutions for the full FFPE. This is
beyond the scope of the present paper. \pagebreak

\pagebreak

\begin{center}
\bigskip

\bigskip\bigskip{\Huge SUPPLEMENTARY\bigskip\ MATERIAL}\pagebreak
\end{center}

\section{Appendix S1 - Mean Energy}

\label{Appendix - Mean Energy}

\subsection{General Expression}

To derive the TWA stochastic expression (\ref{Eq.MeanEnergyStochastic}) for
the mean energy Eq. (\ref{Eq.MeanEnergy}) for the Hamiltonian in Eq.
(\ref{Eq.HamiltonianFieldModel}) we must first replace the \emph{normally
ordered} products involving the field operators $\hat{\Psi}(z)$, $\hat{\Psi
}(z)^{\dag}$ in each of the kinetic energy $\widehat{K}$, potential energy
$\widehat{V}$ and interaction energy $\widehat{U}$ terms by their
\emph{symmetrically ordered} forms (see for example Sect 7.1.2 in Ref
\cite{Dalton15a}). This can be done using Wick's theorem \cite{Wick}.
Expansions of the field operators in terms of a suitable set of orthonormal
mode functions $\phi_{k}(z)$ are also used.

We find that in the kinetic energy term
\begin{align}
&  \frac{\partial}{\partial z}\hat{\Psi}(z)^{\dag}%
\mbox{\rule{-0.5mm}{0mm}}\,\mbox{\rule{-0.5mm}{0mm}}\frac{\partial}{\partial
z}\hat{\Psi}(\ z)\nonumber\\
&  =\left\{  \frac{\partial}{\partial z}\hat{\Psi}(z)^{\dag}%
\mbox{\rule{-0.5mm}{0mm}}\,\mbox{\rule{-0.5mm}{0mm}}\frac{\partial}{\partial
z}\hat{\Psi}(\ z)\right\}  -\frac{1}{2}\triangle\delta_{C}(z,z)
\label{Eq.ResK}%
\end{align}
where
\begin{equation}
\triangle\delta_{c}(z,z)=%
{\displaystyle\sum\limits_{k}}
\left(  \frac{\partial}{\partial z}\phi_{k}(z)\right)  \left(  \frac{\partial
}{\partial z}\phi_{k}(z)\right)  ^{\ast}%
\end{equation}

In the potential energy term we have
\begin{equation}
\hat{\Psi}(z)^{\dag}\hat{\Psi}(z)=\left\{  \hat{\Psi}(z)^{\dag}\hat{\Psi
}(z)\right\}  -\frac{1}{2}\delta_{C}(z,z) \label{Eq.ResV}%
\end{equation}

The interaction energy term is more difficult to deal with, but in this term
we find that
\begin{align}
&  \hat{\Psi}(z)^{\dag}\hat{\Psi}(z)^{\dag}\hat{\Psi}(z)\hat{\Psi
}(z)\nonumber\\
&  =\left\{  \hat{\Psi}(z)^{\dag}\hat{\Psi}(z)^{\dag}\hat{\Psi}(z)\hat{\Psi
}(z)\right\} \nonumber\\
&  -2\,\delta_{C}(z,z)\,\left\{  \hat{\Psi}(z)^{\dag}\hat{\Psi}(z)\right\}
\nonumber\\
&  +\frac{1}{2}(\delta_{C}(z,z))^{2} \label{Eq.ResU}%
\end{align}

Hence, overall we have for the Hamiltonian
\begin{align}
\widehat{K}  &  =\int\mbox{\rule{-1mm}{0mm}}dz\,\frac{\hbar^{2}}{2m}\left(
\left\{  \frac{\partial}{\partial z}\hat{\Psi}(z)^{\dag}%
\mbox{\rule{-0.5mm}{0mm}}\,\mbox{\rule{-0.5mm}{0mm}}\frac{\partial}{\partial
z}\hat{\Psi}(\ z)\right\}  -\frac{1}{2}\triangle\delta_{C}(z,z)\right)
\nonumber\\
\widehat{V}  &  =\int\mbox{\rule{-1mm}{0mm}}dz\,V(z,t)\left(  \left\{
\hat{\Psi}(z)^{\dag}\hat{\Psi}(z\right\}  -\frac{1}{2}\delta_{C}(z,z)\right)
\nonumber\\
\widehat{U}  &  =\int\mbox{\rule{-1mm}{0mm}}dz\,\frac{g}{2}\left(  \left\{
\hat{\Psi}(z)^{\dag}\hat{\Psi}(z)^{\dag}\hat{\Psi}(z)\hat{\Psi}(z)\right\}
-2\,\delta_{C}(z,z)\,\left\{  \hat{\Psi}(z)^{\dag}\hat{\Psi}(z)\right\}
+\frac{1}{2}(\delta_{C}(z,z))^{2}\right) \nonumber\\
&  \label{Eq,HamSymmetric}%
\end{align}

The final step is to use the results for the Wigner distribution functional
giving the mean value for a symmetrically ordered forms involving the field
operators
\begin{align}
\left\langle \left\{  \hat{\Psi}(z^{\#})^{\dag}\hat{\Psi}(z)\right\}
\right\rangle  &  =%
{\displaystyle\int}
D^{2}\psi\,D^{2}\psi^{+}\;\psi(z)\,\psi^{+}(z^{\#})\,W[\psi,\psi
^{+}]\nonumber\\
&  =\overline{\widetilde{\psi}(z,t)\,\widetilde{\psi}^{+}(z^{\#}%
,t)}\,\nonumber\\
\left\langle \left\{  \partial_{z}\hat{\Psi}(z^{\#})^{\dag}\,\partial_{z}%
\hat{\Psi}(z)\right\}  \right\rangle  &  =%
{\displaystyle\int}
D^{2}\psi\,D^{2}\psi^{+}\;\partial_{z}\psi(z)\,\partial_{z}\psi^{+}%
(z^{\#})\,W[\psi,\psi^{+}]\nonumber\\
&  =\overline{\partial_{z}\widetilde{\psi}(z,t)\,\partial_{z}\widetilde{\psi
}^{+}(z^{\#},t)}\nonumber\\
\left\langle \left\{  \hat{\Psi}(z_{1}^{\#})^{\dag}\hat{\Psi}(z_{2}%
^{\#})^{\dag}\hat{\Psi}(z_{1})\hat{\Psi}(z_{2})\right\}  \right\rangle  &  =%
{\displaystyle\int}
D^{2}\psi\,D^{2}\psi^{+}\;\psi(z_{1})\,\psi(z_{2})\,\psi^{+}(z_{1}^{\#}%
)\,\psi^{+}(z_{2}^{\#})\,W[\psi,\psi^{+}]\nonumber\\
&  =\overline{\widetilde{\psi}(z_{1},t)\,\widetilde{\psi}(z_{2}%
,t)\,\widetilde{\psi}^{+}(z_{1}^{\#},t)\,\widetilde{\psi}^{+}(z_{2}^{\#},t)}
\label{Eq.MeanOfSymmetOrderedForms}%
\end{align}
where we have expressed the mean value of symmetrically ordered forms
involving the field operators first as phase space functional integrals
involvinng the field functions $\psi(z),\psi^{+}(z)$ then as stochastic
averages involving the stochastic field functions $\widetilde{\psi
}(z,t),\widetilde{\psi}^{+}(z,t)$ (see Sects 12.3.3 and 15.1.9 in Ref
\cite{Dalton15a}). The final stochastic average expression for $\left\langle
\widehat{H}\right\rangle $ is given in Eq (\ref{Eq.MeanEnergyStochastic}%
).\smallskip

\subsection{Gravitational Modes}

The mean energy expression (\ref{Eq.MeanEnergyStochastic}) can be simplified
via expanding the stochastic field functions $\widetilde{\psi}%
(z,t),\widetilde{\psi}^{+}(z,t)$ in some terms via gravitational modes.

The potential energy is $V(z,t)=mg_{E}z-mg_{E}\lambda z\cos\omega t$ so the
mean value of $\widehat{T}+\widehat{V}$ is%
\begin{align}
\left\langle \widehat{T}+\widehat{V}\right\rangle  &  =-mg_{E}\lambda
\cos\omega t\,%
{\displaystyle\int}
dz\,z\left(  \overline{\widetilde{\psi}^{+}(z,t)\,\widetilde{\psi}(z,t)}%
-\frac{1}{2}\delta_{C}(z,z)\right) \nonumber\\
&  +%
{\displaystyle\int}
dz\,\overline{\widetilde{\psi}^{+}(z,t)\left(  -\frac{\hbar^{2}}{2m}%
\frac{\partial^{2}}{\partial z^{2}}+mg_{E}z\right)  \,\widetilde{\psi}%
(z,t)}\nonumber\\
&  -\frac{1}{2}%
{\displaystyle\int}
dz\,\left(  \frac{\hbar^{2}}{2m}\Delta\delta_{C}(z,z)-mg_{E}z\delta
_{C}(z,z)\right)  \label{Eq.TPlusV}%
\end{align}
In the second and third lines we expand $\widetilde{\psi}(z,t),\widetilde{\psi
}^{+}(z,t)$ using Eq.(\ref{Eq.StochFieldGravModes}) and $\Delta\delta
_{C}(z,z),\delta_{C}(z,z)$ via (\ref{Eq.DefnDelta}) and (\ref{Eq.DefnDzDelta})
using gravitational modes. This gives after spatial integration by parts
\begin{align}
&
{\displaystyle\int}
dz\,\overline{\widetilde{\psi}^{+}(z,t)\left(  -\frac{\hbar^{2}}{2m}%
\frac{\partial^{2}}{\partial z^{2}}+mg_{E}z\right)  \,\widetilde{\psi}%
(z,t)}\nonumber\\
&  -\frac{1}{2}%
{\displaystyle\int}
dz\,\left(  \frac{\hbar^{2}}{2m}\Delta\delta_{C}(z,z)-mg_{E}z\delta
_{C}(z,z)\right) \nonumber\\
&  =%
{\displaystyle\sum\limits_{k,l}}
\overline{\widetilde{\eta}_{k}^{+}(t)\widetilde{\eta}_{l}(t)}\,%
{\displaystyle\int}
dz\,\xi_{k}^{\ast}(z)\,\left(  -\frac{\hbar^{2}}{2m}\frac{\partial^{2}%
}{\partial z^{2}}+mg_{E}z\right)  \xi_{l}(z)\nonumber\\
&  -\frac{1}{2}%
{\displaystyle\int}
dz\,%
{\displaystyle\sum\limits_{k}}
\xi_{k}^{\ast}(z)\,\left(  -\frac{\hbar^{2}}{2m}\frac{\partial^{2}}{\partial
z^{2}}-mg_{E}z\right)  \xi_{k}(z)\nonumber\\
&  =%
{\displaystyle\sum\limits_{k,}}
\overline{\widetilde{\eta}_{k}^{+}(t)\widetilde{\eta}_{k}(t)}\,\hbar
\epsilon_{k}-\frac{1}{2}%
{\displaystyle\sum\limits_{k,}}
\hbar\epsilon_{k} \label{Eq.TermsResult}%
\end{align}
where the defining equation (\ref{Eq.GravModes}) for the gravitational modes
has been used, along with their orthogonality. Hence
\begin{align}
\left\langle \widehat{T}+\widehat{V}\right\rangle  &  =-mg_{E}\lambda
\cos\omega t\,%
{\displaystyle\int}
dz\,z\left(  \overline{\widetilde{\psi}^{+}(z,t)\,\widetilde{\psi}(z,t)}%
-\frac{1}{2}\delta_{C}(z,z)\right) \nonumber\\
&  +%
{\displaystyle\sum\limits_{k,}}
\hbar\epsilon_{k}\,\left(  \overline{\widetilde{\eta}_{k}^{+}%
(t)\widetilde{\eta}_{k}(t)}\,-\frac{1}{2}\right)  \label{Eq.FinalTplusV}%
\end{align}

The mean value for the interaction energy $\widehat{U}$ is left unchanged.
\pagebreak

\section{Appendix S2 - Floquet Mode Treatment}

\label{Appendix - Floquet Mode Treatment}

\subsection{Position Probability Density and QCF - Floquet Modes}

The position probability density in Eq. (\ref{Eq.PositProbStochasAver}) can be
expressed in terms of Floquet mode functions as
\begin{equation}
F(z,t)=%
{\displaystyle\sum\limits_{k,l}}
\phi_{k}(z,t)\,\phi_{l}^{\ast}(z,t)\,\left[  \overline{\widetilde{\alpha}%
_{k}(t)\,\widetilde{\alpha}_{l}^{+}(t)}\,-\frac{1}{2}\delta_{k,l}\right]
\label{Eq.PositProbFloquetModes}%
\end{equation}
and involves the stochastic average of products of stochastic phase space variables.

Similarly, the QCF in Eq. (\ref{Eq.QCFStochAver}) \ can also be expressed in
terms of Floquet mode functions as%
\begin{equation}
P(z,z^{\#},t)=%
{\displaystyle\sum\limits_{k,l}}
\phi_{k}(z,t)\,\phi_{l}^{\ast}(z^{\#},t)\,\left[  \overline{\widetilde{\alpha
}_{k}(t)\,\widetilde{\alpha}_{l}^{+}(t)}\,-\frac{1}{2}\delta_{k,l}\right]
\label{Eq.QCFFloquetModes}%
\end{equation}
The quantity $\left[  \overline{\widetilde{\alpha}_{k}(t)\,\widetilde{\alpha
}_{l}^{+}(t)}\,-\frac{1}{2}\delta_{k,l}\right]  $ is the $k,l$ element of a
Hermitian matrix $H$, since $\overline{\widetilde{\alpha}_{k}%
(t)\,\widetilde{\alpha}_{l}^{+}(t)}=(<\widehat{a}_{l}^{\dag}$ $\widehat{a}%
_{k}\,>+<\widehat{a}_{k}\,\widehat{a}_{l}^{\dag}>)/2$.\smallskip

\subsection{Evolution of Stochastic Phase Space Variables for Floquet Modes}

By substituting the expressions in Eq (\ref{Eq.StochFldFns}) for
$\widetilde{\psi}(z,t),\widetilde{\psi}^{+}(z,t)$ into the Ito SFE in Eqs.
(\ref{Eq.ItoSFEA}), (\ref{Eq.ItoSFAB}) and using Eq (\ref{Eq.FloquetModes})
for the Floquet modes together with their orthonormality property, we obtain
sets of non-linear coupled equations for the stochastic phase space variables
(SPSV)
\begin{align}
\frac{\partial}{\partial t}\widetilde{\alpha}_{k}  &  =-i\nu_{k}%
\widetilde{\alpha}_{k}-i\frac{g}{\hbar}\,%
{\displaystyle\sum\limits_{l,m,n}}
\,L_{k,l,m,n}\,\widetilde{\alpha}_{l}^{+}\,\widetilde{\alpha}_{m}%
\,\widetilde{\alpha}_{n}+i\frac{g}{\hbar}\,%
{\displaystyle\sum\limits_{l,n}}
\,L_{k,l,l,n}\,\widetilde{\alpha}_{n}\label{Eq.StochPhaseA}\\
\frac{\partial}{\partial t}\widetilde{\alpha}_{k}^{+}  &  =+i\nu
_{k}\widetilde{\alpha}_{k}^{+}+i\frac{g}{\hbar}\,%
{\displaystyle\sum\limits_{l,m,n}}
\,L_{k,l,m,n}^{\ast}\,\widetilde{\alpha}_{l}\,\widetilde{\alpha}_{m}%
^{+}\,\widetilde{\alpha}_{n}^{+}-i\frac{g}{\hbar}\,%
{\displaystyle\sum\limits_{l,n}}
\,L_{k,l,l,n}^{\ast}\,\widetilde{\alpha}_{n}^{+} \label{Eq.StochPhaseB}%
\end{align}
where the matrix $L$ involves integrals of products of the Floquet mode
functions, defined as%
\begin{equation}
L_{k,l,m,n}=%
{\displaystyle\int}
dz\,\phi_{k}^{\ast}(z,t)\,\phi_{l}^{\ast}(z,t)\,\phi_{m}(z,t)\,\phi_{n}(z,t)
\label{Eq. LMatrix}%
\end{equation}
This matrix is time dependent and periodic with period $T$.

Though non-linear, the equations for the SPSV are deterministic and can be
solved numerically if the initial values $\widetilde{\alpha}_{k}%
(0),\widetilde{\alpha}_{k}^{+}(0)$ are known. These initial values are of
course stochastic and lead to the field functions given by Eq
(\ref{Eq.StochFldFns}) being stochastic. The distribution function for the
initial values is chosen to represent the known properties of the initial
quantum state. Note that the $\widetilde{\alpha}_{k}$ and $\widetilde{\alpha
}_{k}^{+}$ SPSV\ do not evolve independently. The coupled equations for the
SPSV no longer involve the periodic potential directly - this has been taken
into account via the introduction of the Floquet mode functions $\phi
_{k}(z,t)$ and the Floquet frequencies $\nu_{k}$, resulting in equations that
are now focused on the many-body quantum effects.

An alternative way of writing the non-linear coupled eqations is
\begin{align}
\frac{\partial}{\partial t}\widetilde{\alpha}_{k}  &  =-i\nu_{k}%
\widetilde{\alpha}_{k}-i\frac{g}{\hbar}%
{\displaystyle\sum\limits_{n}}
D_{k,n}\,\widetilde{\alpha}_{n}\label{Eq,StochPhaseC}\\
\frac{\partial}{\partial t}\widetilde{\alpha}_{k}^{+}  &  =+i\nu
_{k}\widetilde{\alpha}_{k}^{+}+i\frac{g}{\hbar}%
{\displaystyle\sum\limits_{n}}
D_{k,n}^{+}\,\widetilde{\alpha}_{n}^{+} \label{Eq.StochPhaseD}%
\end{align}
where
\begin{align}
D_{k,n}  &  =%
{\displaystyle\int}
dz\,\phi_{k}^{\ast}(z,t)\left(  \widetilde{\psi}^{+}(z,t)\,\widetilde{\psi
}(z,t)\,-\delta_{C}(z,z)\right)  \,\phi_{n}(z,t)\nonumber\\
D_{k,n}^{+}  &  =%
{\displaystyle\int}
dz\,\phi_{k}(z,t)\left(  \widetilde{\psi}^{+}(z,t)\,\widetilde{\psi
}(z,t)\,-\delta_{C}(z,z)\right)  \,\phi_{n}^{\ast}(z,t) \label{Eq.DMatrix}%
\end{align}
This alternative form is more convenient for numerical calculations. The
non-linearity is embodied in the matrices $D,D^{+\text{ }}.$ In this method
the stochastic fields are determined at each time point from Eq.
(\ref{Eq.StochFldFns}), which then can also be used to determine the quantum
depletion (see Eq. (\ref{Eq.StochAmpCondModes})).\smallskip

\subsection{Initial Conditions - Floquet Modes}

These will be specified via the initial stochastic field
\begin{equation}
\widetilde{\psi}(z,0)=%
{\displaystyle\sum\limits_{k}}
\widetilde{\alpha}_{k}(0)\,\phi_{k}(z,0)\qquad\widetilde{\psi}^{+}(z,0)=%
{\displaystyle\sum\limits_{k}}
\widetilde{\alpha}_{k}^{+}(0)\,\phi_{k}^{\ast}(z,0)
\label{Eq.InitialCondFloquetModes}%
\end{equation}
which require knowing the Floquet mode functions at $t=0$ and choosing a
stochastic distribution of the $\widetilde{\alpha}_{k}(0),\widetilde{\alpha
}_{k}^{+}(0)$ to match the quantum state that has been prepared in the
trapping potential. \pagebreak

\section{Appendix S3 - Bogoliubov Theory}

\label{Appendix - Bogoliubov Theory}

As foreshadowed in Sect. \ref{SubSection - Preparation of BEC}, in this
Appendix we outline the derivation of the Bogoliubov theory form of the Grand
Canonical Hamiltonian, showing that it is the sum of Hamiltonians for
independent quantum harmonic oscilators for each Bogoliubov mode, plus a term
for the energy of the condensate mode and some unimportant constant terms.
\smallskip

\subsection{Grand Canonical Hamiltonian and Bogoliubov Approximation}

The Hamiltonian $\widehat{H}$ describing the evolution of the quantum state is
given by Eq. (\ref{Eq.HamiltonianFieldModel}), but with $V(z,t)$ replaced by
$V_{trap}(z)$. If the quantum state $\widehat{\rho}$ is invariant under the
$U(1)$ symmetry group of phase changing unitary operators $\widehat{U}%
(\theta)=\exp(-i\widehat{N}\,\theta)$, it follows that $[\widehat{N}%
,\widehat{\rho}]=0$. Hence the evolution for $\widehat{\rho}$ can be described
by the grand canonical Hamiltonian $\widehat{K}=\widehat{H}-\mu\widehat{N}$,
with $\mu$ chosen so that $\left\langle \widehat{N}\right\rangle =N_{c}$.

In Bogoliubov theory the grand canonical Hamiltonian $\widehat{K}$ is expanded
correct to the second order in the fluctuation field and by applying the
Bogoliubov approximation in which quantum fluctuations of the condensate mode
are ignored by replacing $\widehat{c}_{0}$ with $\sqrt{N_{c}} $ giving%

\begin{align}
\widehat{K}  &  =%
{\displaystyle\int}
dz\,\Phi_{c}(z)^{\ast}\left[  -\frac{\hslash^{2}}{2m}\frac{\partial^{2}%
}{\partial z^{2}}+V_{trap}(z)+\frac{1}{2}g\,n_{C}(z)-\mu\right]  \,\Phi
_{c}(z)\nonumber\\
&  +%
{\displaystyle\int}
dz\,\delta\widehat{\Psi}(z)\,\left[  -\frac{\hslash^{2}}{2m}\frac{\partial
^{2}}{\partial z^{2}}+V_{trap}(z)+g\,n_{C}(z)-\mu\right]  \,\Phi_{c}(z)^{\ast
}\nonumber\\
&  +%
{\displaystyle\int}
dz\,\delta\widehat{\Psi}(z)^{\dag}\,\left[  -\frac{\hslash^{2}}{2m}%
\frac{\partial^{2}}{\partial z^{2}}+V_{trap}(z)+g\,n_{C}(z)-\mu\right]
\,\Phi_{c}(z)\,\nonumber\\
&  +%
{\displaystyle\int}
dz\,\delta\widehat{\Psi}(z)^{\dag}\left[  -\frac{\hslash^{2}}{2m}%
\frac{\partial^{2}}{\partial z^{2}}+V_{trap}(z)+2g\,n_{C}(z)-\mu\right]
\,\delta\widehat{\Psi}(z)\nonumber\\
&  +%
{\displaystyle\int}
dz\,\frac{1}{2}g\,\left[  \left\{  \Phi_{c}(z)^{\ast}\delta\widehat{\Psi
}(z)\right\}  ^{2}+\left\{  \Phi_{c}(z)\,\delta\widehat{\Psi}(z)^{\dag
}\right\}  ^{2}\right]  \label{Eq.GrandCanHamSecondOrder}%
\end{align}
This is equivalent to writing the field operator as $\hat{\Psi}(z)=\Phi
_{c}(z)+\delta\widehat{\Psi}(z)$, so the condensate field term is replaced by
a non-operator field $\sqrt{N_{c}}\psi_{c}(z)=\Phi_{c}(z)$. Note however that
we still will require the condensate mode to be orthogonal to the modes
associated with the fluctuation field, and the commutation rules in Eq.
(\ref{Eq.CommRule}) to still apply. In the expression for $\widehat{K} $ the
first line gives the zero order contribution as a constant term, the next two
lines the first order contribution and the last two lines the second order
contribution. The boson number density associated with the condensate is given
by $n_{C}(z)=\Phi_{c}(z)^{\ast}\Phi_{c}(z)$. Note that a further approximation
has been made - there are terms involving $\,(\delta\widehat{\Psi}(z)^{\dag
})^{2}\,\delta\widehat{\Psi}(z)^{2},\,(\delta\widehat{\Psi}(z)^{\dag}%
)^{2}\,\delta\widehat{\Psi}(z),\,\delta\widehat{\Psi}(z)^{\dag}\,\delta
\widehat{\Psi}(z)^{2}$ that have been discarded.

Since $\Phi_{c}(z)$ satisfies Eq. (\ref{Eq.GPECondensPrepnA}), the linear
terms in the grand canonical Hamiltonian are zero. The constant term has no
dynamical effect. The second order term that remains may be diagonalised via
the Bogoliubov transformation, which is an example of a linear canonical
transformation (see Ref. \cite{Dalton15a}, Sections 6.1, 6.4.3) in which
commutation rules for the mode operators are preserved.\smallskip

\subsection{Bogoliubov Hamiltonian}

On substituting for the fluctuation field in terms of the Bogoliubov mode
operators using Eq. (\ref{Eq.FluctFieldOpr}) and with $u_{k}(z),v_{k}(z)$
satisfying the generalised BDG equations (\ref{Eq. BogolDeGennes}) , the grand
canonical Hamiltonian given by Eq. (\ref{Eq.GrandCanHamSecondOrder}) can be
expressed as the sum of Hamiltonians for independent quantum harmonic
oscilators for each Bogoliubov mode, plus a term for the energy of the
condensate mode and some unimportant constant terms \cite{Fetter72}. Thus
\begin{align}
\widehat{K}  &  =%
{\displaystyle\int}
dz\,\Phi_{c}(z)^{\ast}\left[  -\frac{\hslash^{2}}{2m}\frac{\partial^{2}%
}{\partial z^{2}}+V_{trap}(z)+\frac{1}{2}g\,n_{C}(z)-\mu\right]  \,\Phi
_{c}(z)\nonumber\\
&  +%
{\displaystyle\sum\limits_{k\neq0}}
\left[  -%
{\displaystyle\int}
dz\,|\,V_{k}(z)|^{2}\right] \nonumber\\
&  +%
{\displaystyle\sum\limits_{k\neq0}}
\hslash\omega_{k}\,\widehat{b}_{k}^{\dag}\,\widehat{b}_{k}
\label{Eq.BogolGrandCanHamilt}%
\end{align}
In deriving Eq (\ref{Eq.BogolGrandCanHamilt}) the Hermitiancy properties of $%
\mathcal{L}%
$, the reality of the $\omega_{k}$ , the commutation rules for the
$\widehat{b}_{k},\widehat{b}_{k}^{\dag}$, the biorthogonality conditions
(\ref{Eq.BiorthogCond}) and the orthogonality conditions
(\ref{Eq.BogolCondModeOrthog}) are all used, in addition to the generalised
BDG equations (\ref{Eq. BogolDeGennes}).\pagebreak

\section{Appendix S4 - Alternative Initial States for Condensate}

\label{Appendix - Alternative Initial States for Condensate}

One simple idea would be to assume the quantum state is a pure state given by
a single Glauber coherent state $\left\vert \gamma0\right\rangle _{c}$ for the
condensate mode with amplitude $\gamma_{0}$, and with each Bogoliubov mode $k$
in its vacuum state $\left\vert 0\right\rangle _{k}$. Thus
\begin{equation}
\widehat{\rho}(0)=\left\vert \Phi_{\gamma0}\right\rangle \left\langle
\Phi_{\gamma0}\right\vert \label{Eq.PureCoherStateT0}%
\end{equation}
with
\begin{align}
\left\vert \Phi_{\gamma0}\right\rangle  &  =\left\vert \gamma0\right\rangle
_{c}\left\vert 0\right\rangle _{1}\left\vert 0\right\rangle _{2}....\left\vert
0\right\rangle _{k}....\nonumber\\
&  =\left\vert \gamma0\right\rangle _{c}\times%
{\textstyle\prod\limits_{k\neq0}}
\,\left\vert 0\right\rangle _{k} \label{Eq.PureCoherStateT0Defn}%
\end{align}
For such a state we would have, if $\gamma0\,=\sqrt{N_{c}}$
\begin{align}
\hat{\Psi}(z)\,\left\vert \Phi_{\gamma0}\right\rangle  &  =\sqrt{N_{c}}%
\,\psi_{c}(z)\,\left\vert \Phi_{\gamma0}\right\rangle =\Phi_{c}(z)\,\left\vert
\Phi_{\gamma0}\right\rangle \nonumber\\
\left\langle \hat{\Psi}(z)\right\rangle  &  =Tr\,\hat{\Psi}(z)\,\widehat{\rho
}(0)=\Phi_{c}(z)\,\nonumber\\
\left\langle \hat{\Psi}(z)^{\dag}\,\hat{\Psi}(z)\right\rangle  &
=Tr\,\hat{\Psi}(z)^{\dag}\,\hat{\Psi}(z)\,\widehat{\rho}(0)=\Phi_{c}(z)^{\ast
}\Phi_{c}(z)=n_{C}(z)\nonumber\\
\left\langle \widehat{N}\right\rangle  &  =\int dz\,\left\langle \hat{\Psi
}(z)^{\dag}\,\hat{\Psi}(z)\right\rangle =N_{c}
\label{Eq. PureCoherStateT0Props}%
\end{align}
using $\widehat{c}_{0}\left\vert \gamma0\right\rangle _{c}=\gamma0\left\vert
\gamma0\right\rangle _{c}$ for Glauber coherent states. This state does
obviously resemble a BEC with all $N_{c}$ bosons having the same wave function
$\Phi_{c}(z)$. However, it is not invariant under phase change
transformations, so this is inconsistent with the requirement of
$\widehat{\rho}$ being phase invariant in order to treat evolution via the
grand canonical Hamiltonian $\widehat{K}$.

Nevertheless, a simple quantum state that is phase invariant can easily be
constructed as a mixed state based on the $\left\vert \Phi_{\gamma
0}\right\rangle $. Writing $\gamma0\,=\sqrt{N_{c}}\exp(i\phi_{c})$ we now
consider the mixed state given by%
\begin{equation}
\widehat{\rho}(0)=\int_{0}^{2\pi}\frac{d\phi_{c}}{2\pi}\left\vert \Phi
_{\gamma0}\right\rangle \left\langle \Phi_{\gamma0}\right\vert
\label{Eq.MixedCoherStates}%
\end{equation}
We then find that
\begin{align}
\hat{\Psi}(z)\,\left\vert \Phi_{\gamma0}\right\rangle  &  =\sqrt{N_{c}}%
\exp(i\phi_{c})\,\psi_{c}(z)\,\left\vert \Phi_{\gamma0}\right\rangle
=\exp(i\phi_{c})\,\Phi_{c}(z)\,\left\vert \Phi_{\gamma0}\right\rangle
\nonumber\\
\left\langle \hat{\Psi}(z)\right\rangle  &  =0\nonumber\\
\left\langle \hat{\Psi}(z)^{\dag}\,\hat{\Psi}(z)\right\rangle  &
=n_{C}(z)=\Phi_{c}(z)^{\ast}\Phi_{c}(z) \label{Eq.MixedCoherStateT0Props}%
\end{align}
We see that the mean value of the field operator is zero, as required for a
phase invariant state. However, the mean value of the number density operator
is still obtained from the condensate wave function. This mixed state is
therefore one good description of the initial BEC.\pagebreak

\section{Appendix S5 - General Initial States}

\label{Appendix A - Gen Init State}

For the state given by Eq (\ref{Eq.MixedCoherStates}) the QCF for normally
ordered products of the field operators are given by%
\begin{align}
\left\langle \hat{\Psi}(z)\right\rangle  &  =\left\langle \hat{\Psi}%
(z^{\#})^{\dag}\right\rangle =0\label{Eq.FirstOrderNormQCF}\\
\left\langle \hat{\Psi}(z^{\#})^{\dag}\,\hat{\Psi}(z)\right\rangle  &
=\Phi_{c}(z^{\#})^{\ast}\Phi_{c}(z)\nonumber\\
\left\langle \hat{\Psi}(z_{1})\,\hat{\Psi}(z_{2})\right\rangle  &
=\left\langle \hat{\Psi}(z_{1}^{\#})^{\dag}\,\hat{\Psi}(z_{2}^{\#})^{\dag
}\right\rangle =0\label{Eq.SecondOrderNormQCF}\\
&  .....\nonumber\\
\left\langle \hat{\Psi}(z_{n}^{\#})^{\dag}\,.....\hat{\Psi}(z_{1}^{\#})^{\dag
}\,\hat{\Psi}(z_{1}).....\hat{\Psi}(z_{n})\right\rangle  &  =\Phi_{c}%
(z_{n}^{\#})^{\ast}.....\Phi_{c}(z_{1}^{\#})^{\ast}\Phi_{c}(z_{1}%
).....\Phi_{c}(z_{n})\label{Eq.2NthOrderNormQCF}\\
\left\langle \hat{\Psi}(z_{n}^{\#})^{\dag}\,.....\hat{\Psi}(z_{1}^{\#})^{\dag
}\,\hat{\Psi}(z_{1})......\hat{\Psi}(z_{m})\right\rangle  &  =0\qquad n\neq m
\label{Eq.NPlusMOrderNormQCF}%
\end{align}
Note that the normally ordered QCF is zero unless there are the same numbers
of $\hat{\Psi}(z)$ and $\hat{\Psi}(z)^{\dag}$. When the numbers are the same,
the QCF is determined from the condensate wave function.

Normally ordered QCF are related to symmetrically ordered QCF via expressions
such as $\{\hat{\Psi}^{\dag}(\ z^{\#})\hat{\Psi}(\ z)\}=\hat{\Psi}^{\dag
}(\ z^{\#})\hat{\Psi}(\ z)+\frac{1}{2}\delta_{C}(z,z^{\#})$. This is a
particular case of Wick's theorem which states that a product of field
operators such as $\hat{\Psi}(z_{n}^{\#})^{\dag}\,.....\hat{\Psi}(z_{1}%
^{\#})^{\dag}\,\hat{\Psi}(z_{1})......\hat{\Psi}(z_{m})$ may be written as
\begin{align}
\widehat{A}_{1}\widehat{A}_{2}.....\widehat{A}_{n}  &  =N[\widehat{A}%
_{1}\widehat{A}_{2}....\widehat{A}_{n}]+%
{\textstyle\sum\limits_{(i,j)}}
N[\widehat{A}_{1}\widehat{A}_{2}..\overbrace{\widehat{A}_{i}...\widehat{A}%
_{j}}..\widehat{A}_{n}]+%
{\textstyle\sum\limits_{(i,j)(k,l)}}
N[\widehat{A}_{1}\widehat{A}_{2}..\overbrace{\widehat{A}_{i}...\widehat{A}%
_{j}}...\overbrace{\widehat{A}_{k}...\widehat{A}_{l}}....\widehat{A}%
_{n}]+...\nonumber\\
&  \label{Eq.Wick2}%
\end{align}
where $N[...]$ is the normally ordered form of the quantity in square
brackets, and $\overbrace{\widehat{A}_{i}\widehat{A}_{j}}$ is the so-called
contraction of the pair of operators $\widehat{A}_{i}\widehat{A}_{j}$ which is
defined by $\overbrace{\widehat{A}_{i}\widehat{A}_{j}}=\widehat{A}%
_{i}\widehat{A}_{j}-N[\widehat{A}_{i}\widehat{A}_{j}]$. In the second and
subsequent terms on the right side, the contractions (which are c-numbers )
are removed as multiplying factors to the remaining normalised ordered
expression. In our case the relevant contractions are
\begin{align}
\overbrace{\hat{\Psi}(z_{1}^{\#})^{\dag}\hat{\Psi}(z_{2}^{\#})^{\dag}}  &
=0\qquad\qquad\overbrace{\hat{\Psi}(z_{1})\hat{\Psi}(z_{2})}%
=0\label{Eq.Contractions}\\
\overbrace{\hat{\Psi}(z_{1}^{\#})^{\dag}\hat{\Psi}(z_{1})}  &  =0\qquad
\qquad\overbrace{\hat{\Psi}(z_{1})\hat{\Psi}(z_{1}^{\#})^{\dag}}=\delta
(z_{1}-z_{1}^{\#})\nonumber
\end{align}

Hence we have
\begin{align}
\overline{\widetilde{\psi}(z)}  &  =\left\langle \left\{  \hat{\Psi
}(z)\right\}  \right\rangle =\left\langle \hat{\Psi}(z)\right\rangle
=0\nonumber\\
\overline{\widetilde{\psi}^{+}(z)}  &  =\left\langle \left\{  \hat{\Psi
}(z)^{\dag}\right\}  \right\rangle =\left\langle \hat{\Psi}(z)^{\dag
}\right\rangle =0 \label{Eq.FirstOrderQCF}%
\end{align}
for first order QCF, and
\begin{align}
\overline{\widetilde{\psi}(z_{1})\widetilde{\psi}(z_{2})}  &  =\left\langle
\left\{  \hat{\Psi}(z_{1})\hat{\Psi}(z_{2})\right\}  \right\rangle
=\left\langle \hat{\Psi}(z_{1})\hat{\Psi}(z_{2})\right\rangle =0\nonumber\\
\overline{\widetilde{\psi}^{+}(z_{1}^{\#})\widetilde{\psi}^{+}(z_{2}^{\#})}
&  =\left\langle \left\{  \hat{\Psi}(z_{1}^{\#})^{\dag}\hat{\Psi}(z_{2}%
^{\#})^{\dag}\right\}  \right\rangle =\left\langle \hat{\Psi}(z_{1}%
^{\#})^{\dag}\hat{\Psi}(z_{2}^{\#})^{\dag}\right\rangle =0\nonumber\\
\overline{\widetilde{\psi}^{+}(z_{1}^{\#})\widetilde{\psi}^{+}(z_{1})}  &
=\left\langle \left\{  \hat{\Psi}(z_{1}^{\#})^{\dag}\hat{\Psi}(z_{1})\right\}
\right\rangle =\left\langle \hat{\Psi}(z_{1}^{\#})^{\dag}\hat{\Psi}%
(z_{1})\right\rangle +\delta(z_{1}-z_{1}^{\#})=\Phi_{c}(z_{1}^{\#})^{\ast}%
\Phi_{c}(z_{1})+\delta(z_{1}-z_{1}^{\#})\nonumber\\
&  \label{Eq.SecondOrderQCF}%
\end{align}
for second order QCF.

By substituting for the stochastic fields from Eq
(\ref{Eq.StochFieldsBogolModesB}) (reverting to the $u_{k}$ and $v_{k}$
notation for the Bogoliubov modes) and making use of the orthogonality
properties (\ref{Eq.BogolCondModeOrthog}) of the condensate and Bogoliubov
modes, we can show from the first order QCF results (\ref{Eq.FirstOrderQCF})
that%
\begin{align}
\overline{\widetilde{\gamma}_{0}(0)}  &  =\overline{\widetilde{\gamma}_{0}%
^{+}(0)}=0\nonumber\\
\overline{\,\widetilde{\beta}_{k}(0)}  &  =\overline{\,\widetilde{\beta}%
_{k}^{+}(0)}=0 \label{Eq.StochasticAverInitialConBogolModes}%
\end{align}
showing that the stochastic averages of the initial stochastic amplitudes for
the condensate mode and the Bogoliubov modes are all zero.

Similarly, by making use of both the orthogonality properties
(\ref{Eq.BogolCondModeOrthog}) of the condensate and Bogoliubov modes along
with the biorthogonality conditions (\ref{Eq.BiorthogCond}) for the Bogoliubov
modes, we can show from the second order QCF$\ $results
(\ref{Eq.SecondOrderQCF}) that
\begin{align}
\overline{\widetilde{\gamma}_{0}^{+}(0)\widetilde{\gamma}_{0}(0)}  &
=N_{c}+\frac{1}{2}\qquad\overline{\widetilde{\gamma}_{0}(0)\widetilde{\gamma
}_{0}(0)}=0\qquad\overline{\widetilde{\gamma}_{0}^{+}(0)\widetilde{\gamma}%
_{0}^{+}(0)}=0\label{Eq.StochAverSecondOrderCondMode}\\
\overline{\widetilde{\gamma}_{0}(0)\widetilde{\beta}_{k}(0)}  &
=\overline{\widetilde{\gamma}_{0}(0)\widetilde{\beta}_{k}^{+}(0)}%
=\overline{\widetilde{\gamma}_{0}^{+}(0)\widetilde{\beta}_{k}(0)}%
=\overline{\widetilde{\gamma}_{0}^{+}(0)\widetilde{\beta}_{k}^{+}(0)}=0
\label{Eq.SecondOrderCondBogolModes}%
\end{align}
This shows that the quantity $\overline{\widetilde{\gamma}_{0}^{+}%
(0)\widetilde{\gamma}_{0}(0)}$ is never less than one half - reflecting the
quantum nature of the condensate mode, and increases linearly with the number
of bosons in this mode. The second set of results reflects the lack of initial
correlation between the condensate and Bogoliubov modes.

The second order QCF results (\ref{Eq.SecondOrderQCF}) also could be used to
find expressions for stochastic averages involving pairs of Bogoliubov modes
such as $\,\overline{\widetilde{\beta}_{k}^{+}(0)\widetilde{\beta}_{m}(0)}$.
However, this is more easily accomplished by considering the modes
separately.\pagebreak

\section{Appendix S6 - Wigner Distribution Function}

\label{Appendix B - Wigner Distn Bogol Modes}

\subsection{Factorisation of Wigner Distribution Function}

We first show that the Wigner distribution function (see Sections 7.2, 7.3 in
Ref \cite{Dalton15a} for basic definiions of Wigner distribution functions)
for condensate and non-condensate modes factorises into a Wigner distribution
function $W_{C}(\gamma_{0},\gamma_{0}^{+})$ for the condensate mode and a
Wigner distribution function $W_{NC}(\mathbf{\gamma},\mathbf{\gamma}^{+})$ for
the non-condensate modes, where $\gamma_{0},\gamma_{0}^{+}$ are the phase
space variables for the condensate mode representing $\widehat{c}%
_{0},\widehat{c}_{0}^{\dag}$ and where $\mathbf{\gamma\equiv\{\gamma}%
_{1},...,\gamma_{i},...\}$ and $\mathbf{\gamma}^{+}\mathbf{\equiv\{\gamma}%
_{1}^{+},...,\gamma_{i}^{+},...\}$ are the phase space variables for the
standard non-condensate modes $\widehat{c}_{i},\widehat{c}_{i}^{\dag}$
$(i\neq0)$. We begin with the characteristic function $\chi_{NC}(\xi_{0,}%
\xi_{0}^{+},\mathbf{\xi,\xi}^{+})$ for the condensate mode $\widehat{c}%
_{0},\widehat{c}_{0}^{\dag}$ and standard non-condensate modes $\widehat{c}%
_{i},\widehat{c}_{i}^{\dag}$ $(i\neq0)$, which is defined by%
\begin{equation}
\chi_{W}(\xi_{0,}\xi_{0}^{+},\mathbf{\xi,\xi}^{+})=Tr\,\exp((i\{\widehat{c}%
_{0}\times\xi_{0}^{+}+\xi_{0}\times\widehat{c}_{0}^{\dag}+\widehat{\mathbf{c}%
}\cdot\mathbf{\xi}^{+}+\mathbf{\xi\cdot}\widehat{\mathbf{c}}^{\dag
}\})\,\widehat{\rho}) \label{Eq.CharFn}%
\end{equation}
where $\widehat{\mathbf{c}}\equiv\{\widehat{c}_{1},...,\widehat{c}_{i}%
,....\}$, $\widehat{\mathbf{c}}^{\dag}\equiv\{\widehat{c}_{1}^{\dag
},...,\widehat{c}_{i}^{\dag},....\}$, $\mathbf{\xi\equiv\{\xi}_{1},...,\xi
_{i},...\}$ and $\mathbf{\xi}^{+}\mathbf{\equiv\{\xi}_{1}^{+},...,\xi_{i}%
^{+},...\}$. This is related to the Wigner distribution function by%
\begin{equation}
\chi_{W}(\xi_{0,}\xi_{0}^{+},\mathbf{\xi,\xi}^{+})=%
{\textstyle\int}
d^{2}\gamma_{0}\,d^{2}\gamma_{0}^{+}\,d^{2}\mathbf{\gamma\,}d^{2}%
\mathbf{\gamma}^{+}\mathbf{\,}\exp(i\{\gamma_{0}\,\xi_{0}^{+}+\xi_{0}%
\,\gamma_{0}^{+}+\mathbf{\gamma}\cdot\mathbf{\xi}^{+}+\mathbf{\xi\cdot\gamma
}^{+}\})\,W(\gamma_{0},\gamma_{0}^{+},\mathbf{\gamma},\mathbf{\gamma}^{+})
\label{Eq.WignerDistnFn}%
\end{equation}
Now since $\widehat{\rho}=\widehat{\rho}_{C}\otimes\widehat{\rho}_{NC}$ we see
that the characteristic function factorises as
\begin{align}
\chi_{W}(\xi_{0,}\xi_{0}^{+},\mathbf{\xi,\xi}^{+})  &  =Tr_{C}\,\exp
((i\{\widehat{c}_{0}\times\xi_{0}^{+}+\xi_{0}\times\widehat{c}_{0}^{\dag
}\})\,\widehat{\rho}_{C})\times Tr_{NC}\,\exp((i\{\widehat{\mathbf{c}}%
\cdot\mathbf{\xi}^{+}+\mathbf{\xi\cdot}\widehat{\mathbf{c}}^{\dag
}\})\,\widehat{\rho}_{NC})\nonumber\\
&  =\chi_{C}(\xi_{0,}\xi_{0}^{+})\times\chi_{NC}(\mathbf{\xi,\xi}^{+})
\label{Eq.CharFnFactn}%
\end{align}
Hence we can write
\begin{align}
\chi_{C}(\xi_{0,}\xi_{0}^{+})  &  =%
{\textstyle\int}
d^{2}\gamma_{0}\,d^{2}\gamma_{0}^{+}\,\exp(i\{\gamma_{0}\,\xi_{0}^{+}+\xi
_{0}\,\gamma_{0}^{+}\})\,W_{C}(\gamma_{0},\gamma_{0}^{+}%
)\label{Eq.WignerCharRelnCondMode}\\
\chi_{NC}(\mathbf{\xi,\xi}^{+})  &  =%
{\textstyle\int}
d^{2}\mathbf{\gamma\,}d^{2}\mathbf{\gamma}^{+}\mathbf{\,}\exp
(i\{\mathbf{\gamma}\cdot\mathbf{\xi}^{+}+\mathbf{\xi\cdot\gamma}%
^{+}\})\,W_{NC}(\mathbf{\gamma},\mathbf{\gamma}^{+}
\label{Eq,WignerCharRelnNonCondModes}%
\end{align}
This then defines the Wigner distribution functions for the condensate mode
and for the non-condensate modes.\smallskip

\subsection{Relation between Standard and Bogoliubov Non-Condensate Modes}

By equating expressions (\ref{Eq.FluctFieldOpr3}) and (\ref{Eq.FluctFieldOpr})
for the fluctuation field operator in terms of standard modes $\psi_{i}(z)$
and Bogoliubov modes $u_{k}(z),v_{k}(z)$ and using the orthogonality of the
standard modes we follow the approach of Morgan \cite{Morgan00} and express
the standard non-condensate mode annihilation, creation operators
$\widehat{c}_{i},\widehat{c}_{i}^{\dag}$ in terms of the Bogoliubov operators
$\widehat{b}_{k},\widehat{b}_{k}^{\dag}$. In column matrix form this is
\begin{equation}
\left[
\begin{array}
[c]{c}%
\widehat{\mathbf{c}}\\
\widehat{\mathbf{c}}^{\dag}%
\end{array}
\right]  =\left[
\begin{array}
[c]{cc}%
U & -V\\
-V^{\ast} & U^{\ast}%
\end{array}
\right]  \times\left[
\begin{array}
[c]{c}%
\widehat{\mathbf{b}}\\
\widehat{\mathbf{b}}^{\dag}%
\end{array}
\right]  \label{Eq.StandModeBogolModeRln}%
\end{equation}
where the matrices $U,V$ have elements given by
\begin{equation}
U_{i,k}=%
{\textstyle\int}
dz\,\psi_{i}(z)^{\ast}\,u_{k}(z)\qquad V_{i,k}=%
{\textstyle\int}
dz\,\psi_{i}(z)^{\ast}\,v_{k}(z)^{\ast} \label{Eq.MatrixElemUandV}%
\end{equation}
This is a linear canonical transformation in which commutation rules are
preserved. For the matrices $T$ denotes transverse, $\ast$ denotes complex
conjugation and $\dag$ denotes the Hermitian adjoint.

Using the bosonic commutation rules for $\widehat{c}_{i},\widehat{c}_{i}%
^{\dag}$ and for $\widehat{b}_{k},\widehat{b}_{k}^{\dag}$ leads to the
following matrix equations
\begin{align}
-U\,V^{T}+V\,U^{T}  &  =0\nonumber\\
U\,U^{\dag}-V\,V^{\dag}  &  =E \label{Eq.MatrixEqnsUV}%
\end{align}
It then follows from the last equations that
\begin{equation}
\left[
\begin{array}
[c]{cc}%
U & -V\\
-V^{\ast} & U^{\ast}%
\end{array}
\right]  \times\left[
\begin{array}
[c]{cc}%
U^{\dag} & V^{T}\\
V^{\dag} & U^{T}%
\end{array}
\right]  =\left[
\begin{array}
[c]{cc}%
E & 0\\
0 & E
\end{array}
\right]  \label{Eq.MatrixEqnsUVMk2}%
\end{equation}

We can then use the last equation to express the Bogoliubov operators
$\widehat{b}_{k},\widehat{b}_{k}^{\dag}$ in terms of the standard mode
annihilation, creation operators $\widehat{c}_{i},\widehat{c}_{i}^{\dag}$.
This gives
\begin{equation}
\left[
\begin{array}
[c]{c}%
\widehat{\mathbf{b}}\\
\widehat{\mathbf{b}}^{\dag}%
\end{array}
\right]  =\left[
\begin{array}
[c]{cc}%
U^{\dag} & V^{T}\\
V^{\dag} & U^{T}%
\end{array}
\right]  \times\left[
\begin{array}
[c]{c}%
\widehat{\mathbf{c}}\\
\widehat{\mathbf{c}}^{\dag}%
\end{array}
\right]  \label{Eq.BogolStandModeReln}%
\end{equation}
and finally we can also derive a further result involving the $U,V$ matrices%
\begin{equation}
\left[
\begin{array}
[c]{cc}%
U & -V\\
-V^{\ast} & U^{\ast}%
\end{array}
\right]  ^{(\dag)}\times\left[
\begin{array}
[c]{cc}%
E & 0\\
0 & -E
\end{array}
\right]  \times\left[
\begin{array}
[c]{cc}%
U & -V\\
-V^{\ast} & U^{\ast}%
\end{array}
\right]  =\left[
\begin{array}
[c]{cc}%
E & 0\\
0 & -E
\end{array}
\right]  \label{Eq.MatrixEqnsUVMk3}%
\end{equation}
The last result involves evaluating the matrix elements for $(U^{\dag}%
U-V^{T}V^{\ast})$ and $(-V^{T}U+U^{T}V^{\ast})$, using Eqs.
(\ref{Eq.MatrixElemUandV}) for the matrix elements $U_{i,k}$ and $V_{i,k}$,
together with the completeness relation $%
{\displaystyle\sum\limits_{i}}
\psi_{i}(z)\,\psi_{i}^{\ast}(z^{\#})=\delta(z-z^{\#})$ for standard mode
functions as well as the biorthogonality conditions (\ref{Eq.BiorthogCond})
and the condensate mode orthogonality conditions (\ref{Eq.BogolCondModeOrthog}%
).\smallskip

\subsection{Wigner Distribution Function for Bogoliubov Modes}

To show that a Wigner distribution for the non-condensate modes exists in
terms of phase variables $\mathbf{\beta}\equiv\{\beta_{1},...\beta_{k},..)$
and $\mathbf{\beta}^{+}\equiv\{\beta_{1}^{+},...\beta_{k}^{+},..)$ for the
Bogoliubov modes, we begin with the characteristic function $\chi
_{NC}(\mathbf{\xi,\xi}^{+})$ for the standard modes $\widehat{c}%
_{i},\widehat{c}_{i}^{\dag}$ introduced previously
\begin{equation}
\chi_{NC}(\mathbf{\xi,\xi}^{+})=Tr_{NC}(\exp(i\{\widehat{\mathbf{c}}%
\cdot\mathbf{\xi}^{+}+\mathbf{\xi\cdot}\widehat{\mathbf{c}}^{\dag
}\})\,\widehat{\rho}_{NC}) \label{Eq.CharFnNonCond}%
\end{equation}
where $\mathbf{\xi\equiv\{\xi}_{1},...,\xi_{i},...\}$ and $\mathbf{\xi}%
^{+}\mathbf{\equiv\{\xi}_{1}^{+},...,\xi_{i}^{+},...\}$. This is related to
the Wigner distribution function by%
\begin{equation}
\chi_{NC}(\mathbf{\xi,\xi}^{+})=%
{\textstyle\int}
d^{2}\mathbf{\gamma\,}d^{2}\mathbf{\gamma}^{+}\mathbf{\,}\exp
(i\{\mathbf{\gamma}\cdot\mathbf{\xi}^{+}+\mathbf{\xi\cdot\gamma}%
^{+}\})\,W_{NC}(\mathbf{\gamma},\mathbf{\gamma}^{+}) \label{Eq.WignerCharReln}%
\end{equation}
where $\mathbf{\gamma\equiv\{\gamma}_{1},...,\gamma_{i},...\}$ and
$\mathbf{\gamma}^{+}\mathbf{\equiv\{\gamma}_{1}^{+},...,\gamma_{i}^{+},...\}$
are the phase space variables for the standard modes.

Now the phase space variables $\mathbf{\gamma},\mathbf{\gamma}^{+}$ associated
with the standard modes will be related to the phase space variables
$\mathbf{\beta},\mathbf{\beta}^{+}$ associated with the Bogoliubov modes via
the same form (\ref{Eq.StandModeBogolModeRln}) that applies for the mode
annihilation, creation operators. Thus
\begin{equation}
\left[
\begin{array}
[c]{c}%
\mathbf{\gamma}\\
\mathbf{\gamma}^{+}%
\end{array}
\right]  =\left[
\begin{array}
[c]{cc}%
U & -V\\
-V^{\ast} & U^{\ast}%
\end{array}
\right]  \times\left[
\begin{array}
[c]{c}%
\mathbf{\beta}\\
\mathbf{\beta}^{+}%
\end{array}
\right]  \label{Eq.RelnPSVStandardToBogol}%
\end{equation}
The same applies to the stochastic phase space variables. Hence we have%
\begin{align}
\left[
\begin{array}
[c]{c}%
\widetilde{\mathbf{\gamma}}\\
\widetilde{\mathbf{\gamma}}^{+}%
\end{array}
\right]   &  =\left[
\begin{array}
[c]{cc}%
U & -V\\
-V^{\ast} & U^{\ast}%
\end{array}
\right]  \times\left[
\begin{array}
[c]{c}%
\widetilde{\mathbf{\beta}}\\
\widetilde{\mathbf{\beta}}^{+}%
\end{array}
\right] \label{Eq.RelnStochPSVStandardToBogolMk2}\\
\left[
\begin{array}
[c]{c}%
\widetilde{\mathbf{\beta}}\\
\widetilde{\mathbf{\beta}}^{+}%
\end{array}
\right]   &  =\left[
\begin{array}
[c]{cc}%
U^{\dag} & V^{T}\\
V^{\dag} & U^{T}%
\end{array}
\right]  \times\left[
\begin{array}
[c]{c}%
\widetilde{\mathbf{\gamma}}\\
\widetilde{\mathbf{\gamma}}^{+}%
\end{array}
\right]  \label{Eq.RelnStochPSVBogolToStandard}%
\end{align}
where we have also written down the inverse relation, based on
(\ref{Eq.BogolStandModeReln}).

Using the result (\ref{Eq.StandModeBogolModeRln}) we can write $\exp
(i\{\mathbf{\gamma}\cdot\mathbf{\xi}^{+}+\mathbf{\xi\cdot\gamma}^{+}%
\})=\exp(i\{\mathbf{\beta}\cdot\mathbf{\eta}^{+}+\mathbf{\eta\cdot\beta}%
^{+}\})$, where
\begin{equation}
\left[
\begin{array}
[c]{c}%
\mathbf{\eta}\\
\mathbf{\eta}^{+}%
\end{array}
\right]  =\left[
\begin{array}
[c]{cc}%
U^{\dag} & -V^{T}\\
-V^{\dag} & U^{T}%
\end{array}
\right]  \times\left[
\begin{array}
[c]{c}%
\mathbf{\xi}\\
\mathbf{\xi}^{+}%
\end{array}
\right]  \label{Eq.RelnCharacFnVariableBogolToStandard}%
\end{equation}
As the differentials transform as
\begin{equation}
d^{2}\mathbf{\gamma\,}d^{2}\mathbf{\gamma}^{+}\mathbf{\,=}\left\vert \left[
\begin{array}
[c]{cc}%
U & -V\\
-V^{\ast} & U^{\ast}%
\end{array}
\right]  \right\vert \times d^{2}\beta\mathbf{\,}d^{2}\mathbf{\beta}%
^{+}\mathbf{\,} \label{Eq.TransDiffls}%
\end{equation}
we can then use the result (\ref{Eq.MatrixEqnsUVMk3}) to show that the
determinant in (\ref{Eq.TransDiffls}) is merely a constant of magnitude one.

Since the standard phase space and characteristic function variables are
linear functions of the corresponding Bogoliubov variables and also have
$(\exp(i\{\widehat{\mathbf{b}}\cdot\mathbf{\eta}^{+}+\eta\mathbf{\cdot
}\widehat{\mathbf{b}}^{\dag}\})=(\exp(i\{\widehat{\mathbf{c}}\cdot\mathbf{\xi
}^{+}+\mathbf{\xi\cdot}\widehat{\mathbf{c}}^{\dag}\})$ we thus can replace Eq.
(\ref{Eq.WignerCharReln}) by
\begin{align}
\chi_{NC}(\mathbf{\eta,\eta}^{+})  &  =Tr_{NC}(\exp(i\{\widehat{\mathbf{b}%
}\cdot\mathbf{\eta}^{+}+\eta\mathbf{\cdot}\widehat{\mathbf{b}}^{\dag
}\})\,\widehat{\rho}_{NC})\label{Eq.DefnCharFnBogolModes}\\
&  =%
{\textstyle\int}
d^{2}\beta\mathbf{\,}d^{2}\mathbf{\beta}^{+}\mathbf{\,}\exp(i\{\mathbf{\beta
}\cdot\mathbf{\eta}^{+}+\mathbf{\eta\cdot\beta}^{+}\}\,W_{NC}(\mathbf{\beta
},\mathbf{\beta}^{+}) \label{Eq.WignerCharRelnBogolModes}%
\end{align}
showing that a Wigner distribution function can be defined in terms of
Bogoliubov modes, even though these do not satisfy standard orthogonality
conditions.\smallskip

\subsection{Non-Condensate Modes and Squeezed Vacuum State}

It is of some interest to calculate the stochastic averages for the stochastic
phase space variables associated with the standard modes $\widehat{c}%
_{i},\widehat{c}_{i}^{\dag}$ associated with Eq. (\ref{Eq.FluctFieldOpr3}). It
can be shown that the vacuum state for Bogoliubov modes is equivalent to a
squeezed vacuum state for the standard non-condensate modes.

For the first order QCF\ we have from Eq.
(\ref{Eq.RelnStochPSVStandardToBogolMk2})
\[
\overline{\left[
\begin{array}
[c]{c}%
\widetilde{\mathbf{\gamma}}\\
\widetilde{\mathbf{\gamma}}^{+}%
\end{array}
\right]  }=\left[
\begin{array}
[c]{cc}%
U & -V\\
-V^{\ast} & U^{\ast}%
\end{array}
\right]  \times\overline{\left[
\begin{array}
[c]{c}%
\widetilde{\mathbf{\beta}}\\
\widetilde{\mathbf{\beta}}^{+}%
\end{array}
\right]  }%
\]
so that as $\overline{\widetilde{\beta}_{k}}=\overline{\widetilde{\beta}%
_{k}^{+}}=0$ it follows that $\overline{\widetilde{\gamma}_{i}}=\overline
{\widetilde{\gamma}_{i}^{+}}=0$. There is no difference between the standard
and Bogoliubov non-condensate modes in this regard. However, this is not the
case for second order QCF. For the second order QCF we see that from Eq.
(\ref{Eq.RelnStochPSVStandardToBogolMk2}) that
\begin{align}
&  \overline{\left[
\begin{array}
[c]{c}%
\widetilde{\mathbf{\gamma}}^{+}\\
\widetilde{\mathbf{\gamma}}%
\end{array}
\right]  \times\left[
\begin{array}
[c]{cc}%
\widetilde{\mathbf{\gamma}} & \widetilde{\mathbf{\gamma}}^{+}%
\end{array}
\right]  }\nonumber\\
&  =\left[
\begin{array}
[c]{cc}%
U^{\ast} & -V^{\ast}\\
-V & U
\end{array}
\right]  \times\overline{\left[
\begin{array}
[c]{c}%
\widetilde{\mathbf{\beta}}^{+}\\
\widetilde{\mathbf{\beta}}%
\end{array}
\right]  \times\left[
\begin{array}
[c]{cc}%
\widetilde{\mathbf{\beta}} & \widetilde{\mathbf{\beta}}^{+}%
\end{array}
\right]  }\times\left[
\begin{array}
[c]{cc}%
U^{T} & -V^{\dag}\\
-V^{T} & U^{\dag}%
\end{array}
\right]  \label{Eq.RelnStochAverSecondOrder}%
\end{align}
where the stochastic averages of both sides have been taken. But from
(\ref{Eq.SecondOrderStochAverNonCondModes}) we have
\begin{equation}
\overline{\left[
\begin{array}
[c]{c}%
\widetilde{\mathbf{\beta}}^{+}\\
\widetilde{\mathbf{\beta}}%
\end{array}
\right]  \times\left[
\begin{array}
[c]{cc}%
\widetilde{\mathbf{\beta}} & \widetilde{\mathbf{\beta}}^{+}%
\end{array}
\right]  }=\frac{1}{2}\left[
\begin{array}
[c]{cc}%
E & 0\\
0 & E
\end{array}
\right]  \label{Eq.StochAverSecondOrderBogol}%
\end{equation}
so that
\begin{equation}
\overline{\left[
\begin{array}
[c]{c}%
\widetilde{\mathbf{\gamma}}^{+}\\
\widetilde{\mathbf{\gamma}}%
\end{array}
\right]  \times\left[
\begin{array}
[c]{cc}%
\widetilde{\mathbf{\gamma}} & \widetilde{\mathbf{\gamma}}^{+}%
\end{array}
\right]  }=\frac{1}{2}\left[
\begin{array}
[c]{cc}%
(UU^{\dag}+VV^{\dag})^{\ast} & (-VU^{T}-UV^{T})^{\ast}\\
-VU^{T}-UV^{T} & (UU^{\dag}+VV^{\dag})
\end{array}
\right]  \label{Eq.Result}%
\end{equation}
Evaluating these sub-matrices using (\ref{Eq.MatrixElemUandV}) and
(\ref{Eq.CondUVMkA}) gives
\begin{align}
\frac{1}{2}(UU^{\dag}+VV^{\dag})_{i,j}  &  =\frac{1}{2}\delta_{i,j}+(VV^{\dag
})_{i,j}\nonumber\\
-\frac{1}{2}(VU^{T}+UV^{T})_{i,j}  &  =-(VU^{T})_{i,j}
\label{Eq.DetailedExpns}%
\end{align}
so that $\overline{\widetilde{\gamma}_{i}^{+}\widetilde{\gamma}_{i}}=\frac
{1}{2}+(VV^{\dag})_{i,i}$, which is always greater than $\frac{1}{2}$. This
shows that the state with all Bogoliubov modes in the vacuum state is not a
state where the standard non-condensate modes are in the vacuum state. Also
$\overline{\widetilde{\gamma}_{i}\widetilde{\gamma}_{i}}=-(VU^{T})_{i,i}$
which is no longer zero as for the Bogoliubov modes. Described in terms of the
standard non-condensate modes, the non-condensate state is actually a
multi-mode squeezed vacuum.\pagebreak\ 

\section{Appendix S7 - No Driving Case}

\label{Appendix - No Driving}

To confirm that both driving and interactions must both be present for
DTC\ creation, we consider Figs. 14, 15 for the PPD and OBP for the case of
interactions $(gN=-0.01)$\ but no driving, and Figs. 4, 9 for the case of no
interaction $(gN=0)$\ but with driving. In the latter case the PPD shows a
mixture of $T,2T$\ periodicity corresponding to transfer of bosons back and
forth between the two Wannier modes, so no simple $2T$ periodicity occurs. In
the former case, the PPD shows an irregular behaviour after a transient
interval where a $2T$\ periodicity (associated with t$_{bounce}=2T$) is
initially seen, but which rapidly disappears. The corresponding OBP and its FT
shows that there is no regular periodicity. Unlike cases where both driving
and interactions are present and a DTC occurs, a DTC is not present in
situations where one factor is absent. Of course just having both factors
present does not guarantee DTC behaviour. The interaction needs to be
sufficiently strong to allow DTC behaviour.%
\begin{figure}[ptb]%
\centering
\includegraphics[
height=2.9369in,
width=2.8176in
]%
{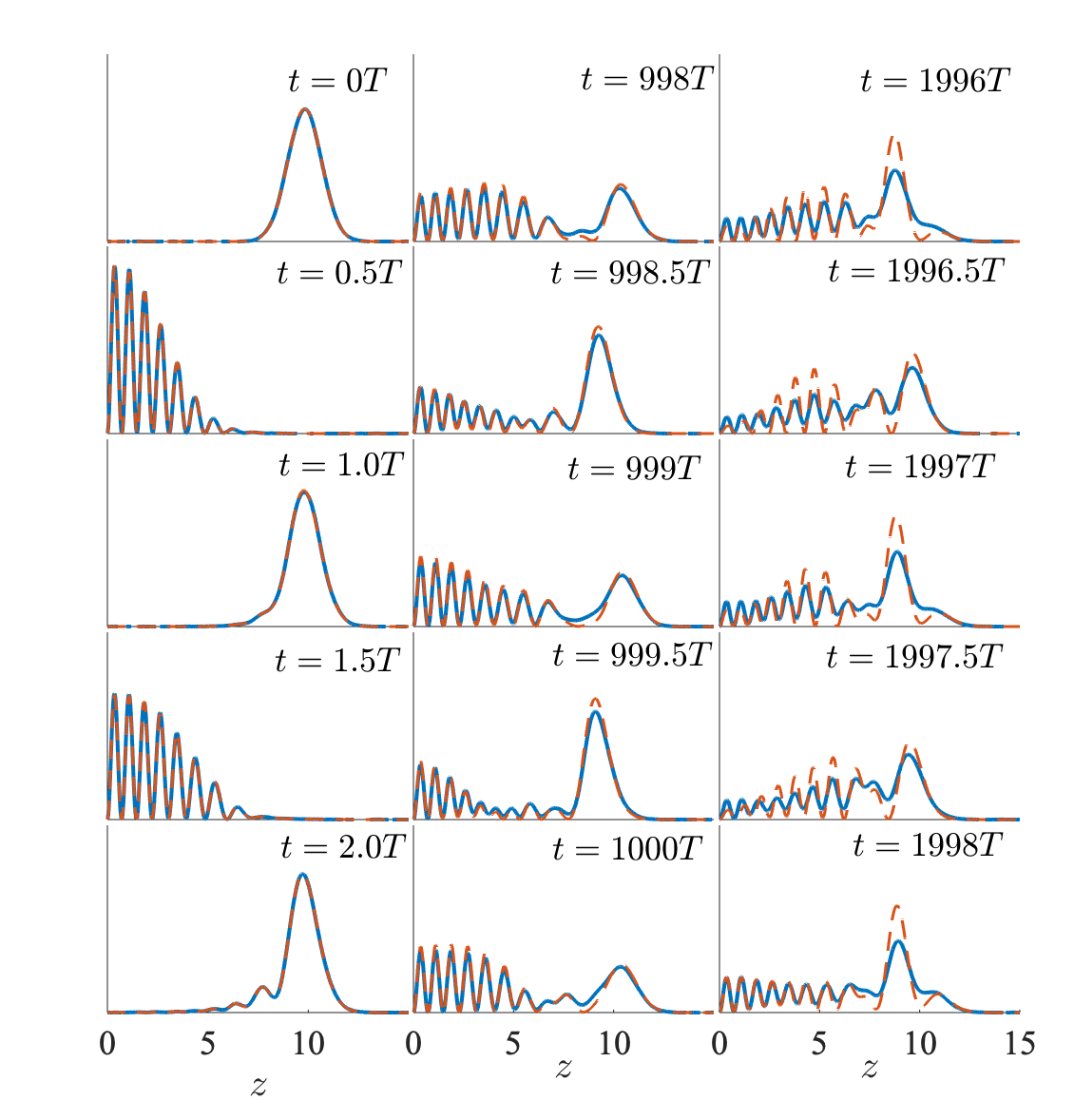}%
\caption{PPD for harmonic trap initial conditions and static mirror case as a
function of $z$ and $t$ for $gN=-0.1$ and $N=600$. The blue solid (red dashed)
curves are calculated using the TWA (GPE) approach.}%
\end{figure}
\begin{figure}[ptb]%
\centering
\includegraphics[
height=2.9369in,
width=3.9055in
]%
{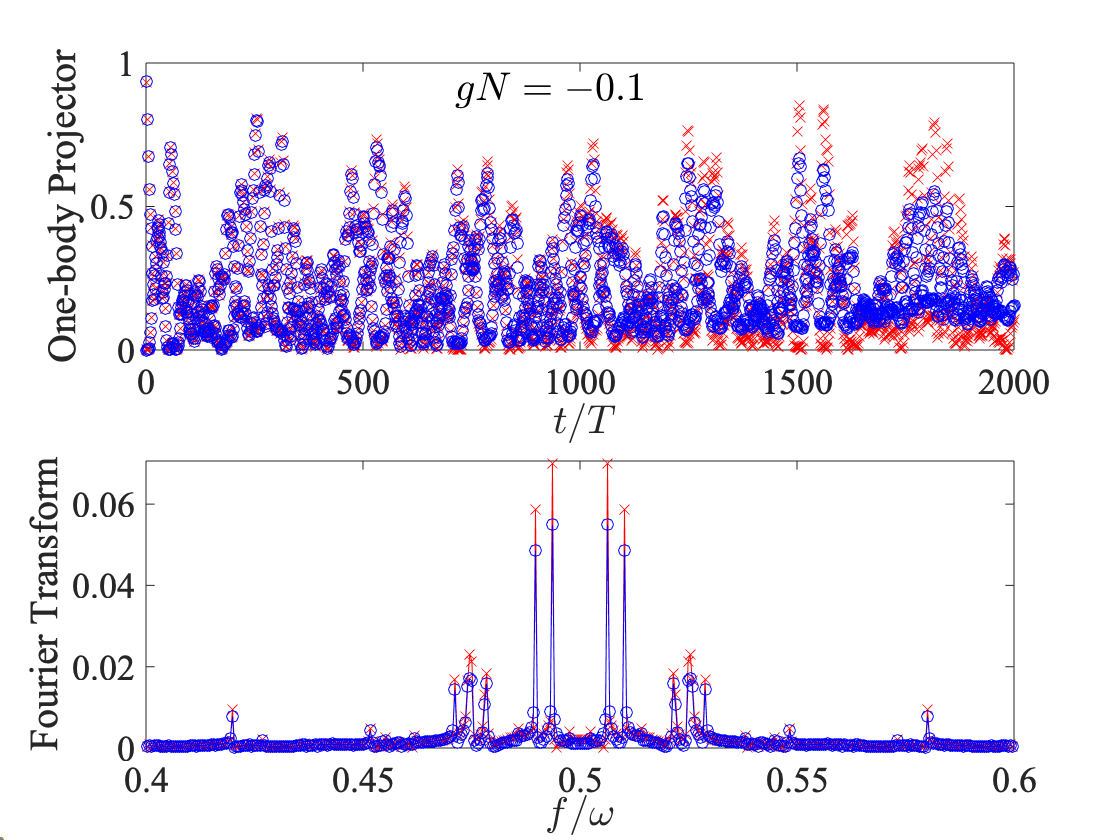}%
\caption{OBP and the corresponding FT for harmonic trap initial conditions and
static mirror case with $gN=-0.1$ and $N=600$. The blue circle (red cross)
symbols show the TWA (GPE) results.}%
\end{figure}

\pagebreak


\begin{thebibliography}{99}                                                                                               %


\bibitem {Sacha15a}K. Sacha. Modeling spontaneous breaking of time-translation
symmetry. \textit{Phys. Rev. A }\textbf{91}, 033617 (2015).

\bibitem {Khemani16a}V. Khemani, A. Lazarides, R. Moessner and S. L. Sondhi.
Phase structure of driven quantum systems. \textit{Phys. Rev. Lett.
}\textbf{116}, 250401 (2016).

\bibitem {Else16a}D. V. Else, B. Bauer and C. Nayak. Floquet time crystals.
\textit{Phys. Rev. Lett.} \textbf{117}, 090402 (2016).

\bibitem {Yao17a}N. Y. Yao, A. C. Potter, I. D. Potirniche and A. Vishwanath.
Discrete time crystals. rigidity, criticality, and realizations. \textit{Phys.
Rev. Lett.} \textbf{118}, 030401 (2017).

\bibitem {Sacha20a}K. Sacha. \textit{Time Crystals} (Berlin: Springer, 2020).

\bibitem {Zhang17a}J. Zhang, P. Hess, A. Kyprianidis, P. Becker, A. Lee, J.
Smith, G. Pagano, I. D. Potirniche, A. C. Potter, A. Vishwanath et al..
Observation of a discrete time crystal. \textit{Nature} \textbf{543}, 217 (2017).

\bibitem {Choi17a}S. Choi, J. Choi, R. Landig, G. Kucsko, H. Zhou, J. Isoya,
F. Jelezko, S. Onoda, H. Sumiya, V. Khemani et al.. Observation of dicrete
time-crystalline order in a disordered dipolar many body system.
\textit{Nature} \textbf{543}, 221 (2017).

\bibitem {Pal18a}S. Pal, N. Nishad, T. S. Mahesh and G. J. Sreejith. Temporal
order in periodically driven spins in star-shaped clusters. \textit{Phys. Rev.
Lett.} \textbf{120}, 180602 (2018).

\bibitem {Rovny18a}J. Rovny, R L Blum and S E Barrett. Observation of
discrete-time-crystal signatures in an ordered dipolar many-body system.
\textit{Phys. Rev. Lett. }\textbf{120}, 180603 (2018).

\bibitem {Rovny18b}J. Rovny, R. L. Blum and S. E. Barrett. $^{31}P$ NMR study
of discrete time-crystalline signatures in an ordered crystal of ammonium
dihydrogen phosphate. \textit{Phys. Rev. B} \textbf{97}, 184301 (2018).

\bibitem {Smits18a}J. Smits, L. Liao, H. T. C. Stoof and P. van der Straten.
Observation of a space-time crystal in a superfluid quantum gas. \textit{Phys.
Rev. Lett.} \textbf{121}, 185301 (2018).

\bibitem {Liao18a}L. Liao, J. Smits, P. van der Straten and H. T. C. Stoof.
Dynamics of a space-time crystal in an atomic Bose-Einsten condensate.
\textit{Phys. Rev. A }\textbf{99}, 013625 (2018).

\bibitem {Smits20a}J. Smits, H. T. C. Stoof and P. van der Straten. On the
long-term stability of space-time crystals. \textit{ArXiv}:2007.07038 (2020).

\bibitem {Sacha18b}K. Sacha and J. Zakrzewski. Time crystals. \textit{Rep.
Prog. Phys}. \textbf{81}, 016401 (2018).

\bibitem {Yao18a}N. Y. Yao and C. Nayak. Time crystals in periodically driven
systems. \textit{Phys. Today }Sept Issue 44 (2018).

\bibitem {Khemani19a}V. Khemani, R. Moessner and S. L. Sondhi. A brief history
of time crystals. \textit{ArXiv }1910.10745 (2019).

\bibitem {Giergel18a}K. Giergiel, A. Kosior, P. Hannaford and K. Sacha. Time
crystals: analysis of experimental conditions. \textit{Phys. Rev. A}
\textbf{98}, 013613 (2018).

\bibitem {Giergiel20a}K. Giergiel, T. Tran, A. Zaheer, A. Singh, A. Sidorov,
K. Sacha and P. Hannaford. Creating big time crystals with ultracold atoms,
\textit{New J. Phys. }\textbf{22}, 085004 (2020).

\bibitem {Kuros20a}A. Kuro\'{s}, R. Mukherjee, W. Golletz, F. Sauvage, K.
Giergiel, F. Mintert and K. Sacha. Phase diagram and optimal control for
n-tupling discrete time crystal. \textit{New J. Phys.} \textbf{22}, 095001. (2020).

\bibitem {Steel98}M. J. Steel, M. K. Olsen, L. I. Plinek, P. D. Drummond, S.
M. Tan, M. J. Collet, D. F. Walls and R. Graham. Dynamical quantum noise in
trapped Bose-Einstein condensates, \textit{Phys. Rev. A} \textbf{58}, 4824 (1998).

\bibitem {Blakie08}P. B. Blakie, A. S. Bradley, M. J. Davis, R. J. Ballagh and
C. W. Gardiner. Dynamics and statistical mechanics of ultra-cold Bose gases
using c-field techniques, \textit{Adv. Phys.} \textbf{57}, 363 (2008).

\bibitem {Dalton15a}B. J. Dalton, J. Jeffers and S. M. Barnett. \textit{Phase
Space Methods for Degenerate Quantum Gases} (Oxford: Oxford University Press, 2015).

\bibitem {Gardiner17a}C. W. Gardiner and P. Zoller. \textit{Quantum World of
Ultra-Cold Atoms and Light: Book 3 - Ultra-Cold Atoms }(Singapore: World
Scientific, 2017).

\bibitem {King19}K. L. Ng, R. Polkinghorne, B. Opanchuk, and P. D. Drummond.
Phase-space representations of thermal Bose-Einstein condensates, \textit{J.
Phys. A: Math. Theor.} \textbf{52}, 035302 (2019).

\bibitem {Morgan00}S. A. Morgan. A gapless theory of Bose-Einstein
condensation in dilute gases at finite temperature, \textit{J. Phys B: At.
Mol. Opt. Phys. }\textbf{33}, 3847 (2000).

\bibitem {Lewenstein}M. Lewenstein and L. You. Quantum phase diffusion of a
Bose-Einstein condensate, \textit{Phys. Rev. Lett.} \textbf{77}, 3489 (1996);
P. Villain, M. Lewenstein, R. Dum, Y. Casti, L. You, A. Imamoglu and T. A. B.
Kennedy. Quantum dynamics of the phase of a Bose-Einstein condensate.
\textit{J. Mod. Opt. }\textbf{44, }1775 \ (1997).

\bibitem {Proukakis08}N. P. Proukakis and B. Jackson. Finite-temperature
models of Bose-Einstein condensation. \textit{J. Phys. B: At. Mol. Opt. Phys.}
\textbf{41}, 203002 (2008).

\bibitem {Leggett01a}A. J. Leggett. Bose-Einstein condensation in the alkali
gases: Some fundamental concepts. \textit{Rev. Mod. Phys. }\textbf{73, }307 (2001).

\bibitem {Floquet}J. H. Shirley. Solution of the Schr\"{o}dinger equation with
a Hamitonian periodic in time. \textit{Phys. Rev. }\textbf{138, }B979 (1965).

\bibitem {Sacha15c}K. Sacha \ Anderson localisation and Mott insulator phase
in the time domain. \textit{Sci. Rep. }\textbf{5 }10787\textit{\ }(2015).

\bibitem {Bogoliubov}N. Bogoliubov. On the theory of superfluidity. \textit{J.
Phys. (USSR) }\textbf{11, }23 (1947).

\bibitem {Fetter72}A. Fetter. Non uniform states of an imperfect Bose gas.
\textit{Ann. Phys. }\textbf{70 }67 (1972).

\bibitem {Wick}G. C. Wick. The evolution of the collision matrix.
\textit{Phys. Rev. }\textbf{80, }268 (1950).

\bibitem {Olsen04a}M. K. Olsen, A. S. Bradley and S. B. Cavalcanti. Fock-state
dynamics in Raman photoassociation of Bose-Einstein condensates. \textit{Phys
\ Rev \ A }\textbf{70, }033611 (2004).

\bibitem {Mierzejewski17a}M. Mierzejewski, K. Giergiel and K. Sacha. Many-body
localization caused by temporal disorder. \textit{Phys. Rev. B} \textbf{96},
140201 (2017).

\bibitem {Giergiel19a}K. Giergiel, A. Kuros and K. Sacha. Discrete time
quasicrystals. \textit{Phys. Rev. B} \textbf{99}, 220303 (2019).

\bibitem {Giergiel19b}K. Giergiel, A. Dauphin, M. Lewenstein, J. Zakrzewski
and K. Sacha. Topological time crystals, \textit{New J. Phys}. \textbf{21},
052003 (2019).

\bibitem {Giergiel18b}K. Giergiel, A. Miroszewski and K. Sacha. Time crystal
platform: From quasicrystal structures in time to systems with exotic
interactions, \textit{Phys. Rev. Lett}. \textbf{120}, 140401 (2018).

\bibitem {Kosier18a}A. Kosior and K. Sacha. Dynamical quantum phase
transitions in discrete time crystals. \textit{Phys. Rev. A} \textbf{97},
053621 (2018).

\bibitem {Roberts01a}J. L. Roberts, N. R. Claussen, S. L. Cornish, E. A.
Donley, E. A. Cornell and C. E. Wieman. Controlled collapse of a Bose-Einstein
condensate, \textit{Phys. Rev. Lett}. \textbf{86}, 4211 (2001).

\bibitem {Shirley65a}J Shirley. Solution of the Schrodinger Equation with a
Hamiltonian Periodic in Time. \textit{Phys. Rev}. \textbf{165}, B480 (1965).
\end{thebibliography}
\end{document}